\documentclass[journal]{IEEEtran-v17}

\pdfoutput=1

\newlength{\figwidth}
\usepackage[nosort]{cite}
\usepackage{url}
\usepackage[cmex10]{amsmath} 
\usepackage{bbm}
\usepackage{graphicx}
\usepackage{subfigure}
\usepackage{ifpdf}
\usepackage[stretch=16,shrink=16,step=4]{microtype}

\makeatletter
\def\@IEEEinterspaceratioM{0.265}
\def\@IEEEinterspaceMINratioM{0.1651}
\def\@IEEEinterspaceMAXratioM{0.38}

\def\@IEEEinterspaceratioB{0.31}
\def\@IEEEinterspaceMINratioB{0.19}
\def\@IEEEinterspaceMAXratioB{0.38}
\@IEEEtunefonts
\makeatother

\usepackage{amssymb}
\usepackage{bbm}
\usepackage{mathrsfs}
\usepackage{xspace}

\newenvironment{textbmatrix}{	\setlength{\arraycolsep}{2.5pt}%
								\big[\begin{matrix}}{\end{matrix}\big]%
								\raisebox{0.08ex}{\vphantom{M}}}

\def\be{\begin{equation}}
\def\ee{\end{equation}}
\def\benn{\begin{equation*}}
\def\eenn{\end{equation*}}
\def\een{\nonumber \end{equation}}
\def\mat{\begin{bmatrix}}
\def\emat{\end{bmatrix}}
\def\btm{\begin{textbmatrix}}
\def\etm{\end{textbmatrix}}

\def\ba#1\ea{\begin{align}#1\end{align}}
\def\bann#1\eann{\begin{align*}#1\end{align*}}
\def\bs#1\es{\begin{split}#1\end{split}} 
\def\bmult#1\emult{\begin{multline}#1\end{multline}} 
\def\bmnn#1\emnn{\begin{multline*}#1\end{multline*}} 
\def\bg#1\eg{\begin{gather}#1\end{gather}} 
\def\bi#1\ei{\begin{itemize}#1\end{itemize}} 
\def\bsa#1\esa{\begin{IEEEeqnarray}{rCl}#1\end{IEEEeqnarray}}
\def\bsann#1\esann{\begin{IEEEeqnarray*}{rCl}#1\end{IEEEeqnarray*}}

\newcommand{\safemath}[2]{\newcommand{#1}{\ensuremath{#2}\xspace}}

\newcommand{\lefto}{\mathopen{}\left}

\DeclareMathOperator{\Prob}{\mathbb{P}}		
\DeclareMathOperator{\Exop}{\mathbb{E}}		
\DeclareMathOperator{\Varop}{\mathbb{V}\!\mathrm{ar}} 

\newcommand{\simplifiedmathchoice}[2]{\mathchoice{#1}{#2}{#2}{#2}}
\newcommand{\Ex}[1]{\simplifiedmathchoice{\ensuremath{\Exop\lefto[#1\right]}}{\ensuremath{\Exop\bigl[#1\bigr]}}} 	
\newcommand{\Var}[1]{\simplifiedmathchoice{\ensuremath{\Varop\lefto[#1\right]}}{\ensuremath{\Varop\bigl[#1\bigr]}}} 
\safemath{\normal}{\mathcal{N}}				
\safemath{\circnorm}{\mathcal{CN}}			
\safemath{\uniform}{\mathcal{U}}				
\newcommand{\abs}[1]{\simplifiedmathchoice{\left\lvert#1\right\rvert}{\bigl\lvert#1\bigr\rvert}}		
\newcommand{\card}[1]{\lvert#1\rvert}			
\newcommand{\Union}{\bigcup}
\newcommand{\Intersect}{\bigcap}
\safemath{\interior}{\mathrm{Int}}			 
\newcommand{\vecnorm}[1]{\left\lVert#1\right\rVert}		
\newcommand{\conj}[1]{\ensuremath{#1^*}} 
\newcommand{\tp}[1]{\ensuremath{#1^{T}}} 		
\newcommand{\herm}[1]{\ensuremath{#1^{H}}} 	

\newcommand{\ent}[1]{\ensuremath{\left[#1\right]}} 
\safemath{\dfn}{:=}							
\newcommand{\natseg}[3]{#1\in[#2\!:\!#3]} 

\safemath{\SNR}{\text{\sc snr}} 				
\safemath{\No}{N_0}							
\safemath{\Es}{E_s}							
\safemath{\Eb}{E_b}							
\safemath{\EbNo}{\frac{\Eb}{\No}}
\safemath{\EsNo}{\frac{\Es}{\No}}

\providecommand{\Hop}{\ensuremath{\mathbb{H}}} 
\safemath{\LH}{L_{\Hop}}						
\safemath{\SH}{S_\Hop}						
\safemath{\HH}{H_{\Hop}}						
\safemath{\CH}{C_\Hop}						
\safemath{\RH}{R_\Hop}						
\safemath{\Rh}{R_h}							

\newcommand{\iid}{i.i.d.\@\xspace}
\newcommand{\as}{a.s.\@\xspace}

\newcommand{\pdf}[1]{f_{#1}}			
\newcommand{\cdf}[1]{F_{#1}} 			

\safemath{\dB}{\,\mathrm{dB}}
\safemath{\dBm}{\,\mathrm{dBm}}
\safemath{\Hz}{\,\mathrm{Hz}}
\safemath{\kHz}{\,\mathrm{kHz}}
\safemath{\MHz}{\,\mathrm{MHz}}
\safemath{\GHz}{\,\mathrm{GHz}}
\safemath{\s}{\,\mathrm{s}}
\safemath{\ms}{\,\mathrm{ms}}
\safemath{\mus}{\,\mathrm{\mu s}}
\safemath{\ns}{\,\mathrm{ns}}
\safemath{\meter}{\,\mathrm{m}}
\safemath{\mm}{\,\mathrm{mm}}
\safemath{\cm}{\,\mathrm{cm}}
\safemath{\m}{\,\mathrm{m}}
\safemath{\W}{\,\mathrm{W}}
\safemath{\J}{\,\mathrm{J}}
\safemath{\K}{\,\mathrm{K}}
\safemath{\bit}{\,\mathrm{bit}}

\safemath{\define}{\triangleq}			

\newcommand{\given}{\,\vert\,}				
\newcommand{\setgiven}{\,\middle|\,}				
\safemath{\equivalent}{\sim}
\safemath{\distas}{\sim}					

\safemath{\reals}{\mathbb{R}}
\safemath{\positivereals}{\mathbb{R}^{+}}
\safemath{\integers}{\mathbb{Z}}
\safemath{\posint}{\mathbb{Z}_{+}}
\safemath{\naturals}{\mathbb{N}}
\safemath{\complexset}{\mathbb{C}}

\safemath{\setA}{\mathcal{A}}
\safemath{\setB}{\mathcal{B}}
\safemath{\setC}{\mathcal{C}}
\safemath{\setD}{\mathcal{D}}
\safemath{\setE}{\mathcal{E}}
\safemath{\setF}{\mathcal{F}}
\safemath{\setG}{\mathcal{G}}
\safemath{\setH}{\mathcal{H}}
\safemath{\setI}{\mathcal{I}}
\safemath{\setJ}{\mathcal{J}}
\safemath{\setK}{\mathcal{K}}
\safemath{\setL}{\mathcal{L}}
\safemath{\setM}{\mathcal{M}}
\safemath{\setN}{\mathcal{N}}
\safemath{\setO}{\mathcal{O}}
\safemath{\setP}{\mathcal{P}}
\safemath{\setQ}{\mathcal{Q}}
\safemath{\setR}{\mathcal{R}}
\safemath{\setS}{\mathcal{S}}
\safemath{\setT}{\mathcal{T}}
\safemath{\setU}{\mathcal{U}}
\safemath{\setV}{\mathcal{V}}
\safemath{\setW}{\mathcal{W}}
\safemath{\setX}{\mathcal{X}}
\safemath{\setY}{\mathcal{Y}}
\safemath{\setZ}{\mathcal{Z}}
\safemath{\emptySet}{\varnothing}

\safemath{\bma}{\mathbf{a}}
\safemath{\bmb}{\mathbf{b}}
\safemath{\bmc}{\mathbf{c}}
\safemath{\bmd}{\mathbf{d}}
\safemath{\bme}{\mathbf{e}}
\safemath{\bmf}{\mathbf{f}}
\safemath{\bmg}{\mathbf{g}}
\safemath{\bmh}{\mathbf{h}}
\safemath{\bmi}{\mathbf{i}}
\safemath{\bmj}{\mathbf{j}}
\safemath{\bmk}{\mathbf{k}}
\safemath{\bml}{\mathbf{l}}
\safemath{\bmm}{\mathbf{m}}
\safemath{\bmn}{\mathbf{n}}
\safemath{\bmo}{\mathbf{o}}
\safemath{\bmp}{\mathbf{p}}
\safemath{\bmq}{\mathbf{q}}
\safemath{\bmr}{\mathbf{r}}
\safemath{\bms}{\mathbf{s}}
\safemath{\bmt}{\mathbf{t}}
\safemath{\bmu}{\mathbf{u}}
\safemath{\bmv}{\mathbf{v}}
\safemath{\bmw}{\mathbf{w}}
\safemath{\bmx}{\mathbf{x}}
\safemath{\bmy}{\mathbf{y}}
\safemath{\bmz}{\mathbf{z}}

\safemath{\bmxi}{\boldsymbol{\xi}}
\safemath{\bmlambda}{\mathbf{\lambda}}
\safemath{\bmmu}{\mathbf{\mu}}
\safemath{\bmtheta}{\boldsymbol{\theta}}
\safemath{\bmphi}{\boldsymbol{\phi}}

\safemath{\bA}{\mathbf{A}}
\safemath{\bB}{\mathbf{B}}
\safemath{\bC}{\mathbf{C}}
\safemath{\bD}{\mathbf{D}}
\safemath{\bE}{\mathbf{E}}
\safemath{\bF}{\mathbf{F}}
\safemath{\bG}{\mathbf{G}}
\safemath{\bH}{\mathbf{H}}
\safemath{\bI}{\mathbf{I}}
\safemath{\bJ}{\mathbf{J}}
\safemath{\bK}{\mathbf{K}}
\safemath{\bL}{\mathbf{L}}
\safemath{\bM}{\mathbf{M}}
\safemath{\bN}{\mathbf{N}}
\safemath{\bO}{\mathbf{O}}
\safemath{\bP}{\mathbf{P}}
\safemath{\bQ}{\mathbf{Q}}
\safemath{\bR}{\mathbf{R}}
\safemath{\bS}{\mathbf{S}}
\safemath{\bT}{\mathbf{T}}
\safemath{\bU}{\mathbf{U}}
\safemath{\bV}{\mathbf{V}}
\safemath{\bW}{\mathbf{W}}
\safemath{\bX}{\mathbf{X}}
\safemath{\bY}{\mathbf{Y}}
\safemath{\bZ}{\mathbf{Z}}

\safemath{\bDelta}{\mathbf{\Delta}}
\safemath{\bLambda}{\mathbf{\Lambda}}
\safemath{\bPhi}{\mathbf{\Phi}}
\safemath{\bSigma}{\mathbf{\Sigma}}
\safemath{\bOmega}{\mathbf{\Omega}}
\safemath{\bTheta}{\mathbf{\Theta}}

\safemath{\bZero}{\mathbf{0}}

\safemath{\veca}{\bma}
\safemath{\vecb}{\bmb}
\safemath{\vecc}{\bmc}
\safemath{\vecd}{\bmd}
\safemath{\vece}{\bme}
\safemath{\vecf}{\bmf}
\safemath{\vecg}{\bmg}
\safemath{\vech}{\bmh}
\safemath{\veci}{\bmi}
\safemath{\vecj}{\bmj}
\safemath{\veck}{\bmk}
\safemath{\vecl}{\bml}
\safemath{\vecm}{\bmm}
\safemath{\vecn}{\bmn}
\safemath{\veco}{\bmo}
\safemath{\vecp}{\bmp}
\safemath{\vecq}{\bmq}
\safemath{\vecr}{\bmr}
\safemath{\vecs}{\bms}
\safemath{\vect}{\bmt}
\safemath{\vecu}{\bmu}
\safemath{\vecv}{\bmv}
\safemath{\vecw}{\bmw}
\safemath{\vecx}{\bmx}
\safemath{\vecy}{\bmy}
\safemath{\vecz}{\bmz}
\safemath{\vecZero}{\bZero}

\safemath{\vecxi}{\bmxi}
\safemath{\veclambda}{\bmlambda}
\safemath{\vecmu}{\bmmu}
\safemath{\vectheta}{\bmtheta}
\safemath{\vecphi}{\bmphi}

\safemath{\matA}{\bA}
\safemath{\matB}{\bB}
\safemath{\matC}{\bC}
\safemath{\matD}{\bD}
\safemath{\matE}{\bE}
\safemath{\matF}{\bF}
\safemath{\matG}{\bG}
\safemath{\matH}{\bH}
\safemath{\matI}{\bI}
\safemath{\matJ}{\bJ}
\safemath{\matK}{\bK}
\safemath{\matL}{\bL}
\safemath{\matM}{\bM}
\safemath{\matN}{\bN}
\safemath{\matO}{\bO}
\safemath{\matP}{\bP}
\safemath{\matQ}{\bQ}
\safemath{\matR}{\bR}
\safemath{\matS}{\bS}
\safemath{\matT}{\bT}
\safemath{\matU}{\bU}
\safemath{\matV}{\bV}
\safemath{\matW}{\bW}
\safemath{\matX}{\bX}
\safemath{\matY}{\bY}
\safemath{\matZ}{\bZ}
\safemath{\matZero}{\bZero}

\safemath{\matDelta}{\bDelta}
\safemath{\matLambda}{\bLambda}
\safemath{\matPhi}{\bPhi}
\safemath{\matSigma}{\bSigma}
\safemath{\matOmega}{\bOmega}
\safemath{\matTheta}{\bTheta}

\newtheorem{theorem}{Theorem}
\newtheorem{lemma}{Lemma}
\newtheorem{corollary}{Corollary}

\newtheorem{definition} {Definition}

\newcounter{MYtempeqncnt}
\newcommand\relphantom[1]{\mathrel{\phantom{#1}}}
\newlength\myboxwidth
\newcommand{\algobox}[2]{\begin{center}\setlength\fboxsep{0.6em}
\noindent\framebox[0.93\linewidth][l]{%
\setlength\myboxwidth{0.93\linewidth}%
\addtolength\myboxwidth{-\labelindent}%
\addtolength\myboxwidth{-2\fboxsep}%
\noindent%
\begin{minipage}[]{\myboxwidth}\textit{#1}\par{}#2\end{minipage}
}
\end{center}}

\begin{document}

\IEEEoverridecommandlockouts
\ifpdf
\pdfinfo{
	/Title		(Crystallization in Large Wireless Networks)
	/Author		(Veniamin I. Morgenshtern, Helmut~Boelcskei)
	/Keywords	(Large wireless networks, capacity scaling, interference relay network, distributed orthogonalization, amplify-and-forward, large-deviations theory, large random matrices, crystallization.)
}
\fi

\title{Crystallization in Large Wireless Networks}
%
%

\author{Veniamin~I.~Morgenshtern and
			Helmut~B\"{o}lcskei,~\IEEEmembership{Senior~Member,~IEEE}%
\thanks{This paper was presented in part at IEEE ISIT 2005, Adelaide, Australia, Sept.~2005 and at the 
Allerton Conference on Communications, Control and Computing, Monticello, IL, USA, Oct.~2005. 
This research was supported by Nokia Research Center
Helsinki, Finland and by the STREP project No. IST-027310 (MEMBRANE) within the Sixth Framework Programme of the European Commission.}%
\thanks{The authors are with the Communication  Technology Laboratory,
ETH Zurich, 8092 Zurich, Switzerland (e-mail: \{vmorgens,boelcskei\}@nari.ee.ethz.ch).}}%

\maketitle
\begin{abstract}
We analyze fading interference relay networks where~$M$ single-antenna source-destination terminal pairs communicate 
concurrently and in the same frequency band through 
a set of~$K$ single-antenna  relays using half-duplex two-{}hop relaying. Assuming that the relays have channel state information (CSI), it is shown that in the large-$M$ limit, provided~$K$ grows fast enough as a function of~$M$, the network ``decouples'' in the sense that the individual source-destination terminal pair capacities are strictly positive. The corresponding required rate of growth of $K$ as a function of $M$ is found to be sufficient to also make the individual source-destination fading links converge to nonfading links. 
 We say that the network  {\em ``crystallizes''}
as it breaks up into a set of effectively isolated {\em ``wires in the air''}. A large-deviations analysis is performed 
to characterize the ``crystallization'' rate, i.e., the rate (as a function of~$M,K$) at which the decoupled links converge to
nonfading links. In the course of this analysis, we develop a new technique for characterizing the large-deviations behavior of
certain sums of dependent random variables.
For the case of no CSI at the relay level, assuming amplify-and-forward relaying, we compute the per source-destination terminal pair capacity for $M,K\to\infty$, with $K/M\to\beta$ fixed,
using tools from large random matrix theory. 
\end{abstract}

\section{Introduction}
\IEEEPARstart{T}{he capacity} of the relay channel~\cite{meulen71-1 ,cover79-09} is still unknown in the general case. Recently, 
the problem has attracted significant attention, with progress being made on several aspects~\cite{kramer05-09}.  
Sparked by~\cite{gupta02-03,gastpar05-03}, analysis of the capacity\footnote{Throughout the paper, when we talk about capacity, we mean the capacity induced by the considered protocols, not the capacity of the network itself.} scaling behavior of large wireless (relay) networks has emerged as an 
interesting tool~\cite{gupta03-08,grossglauser02-10,leveque05-03,xie04-05,jovicic04-10,franceschetti-,wang06-09,dana03-,bolcskei06-}, which 
often allows to make stronger statements than a finite-number-of-nodes analysis. 
In parallel, the design of distributed space-time codes~\cite{laneman03-10,laneman04-12,nabar04-08}, 
the area of network coding~\cite{ahlswede00-07, koetter03-10}, and the 
understanding of the impact of relaying protocols and multiple-antenna terminals on network capacity~\cite{nabar04-08,azarian05-12,wang05-01} 
have seen remarkable activity.  

\begin{figure}
\centering
\includegraphics[width=\figwidth]{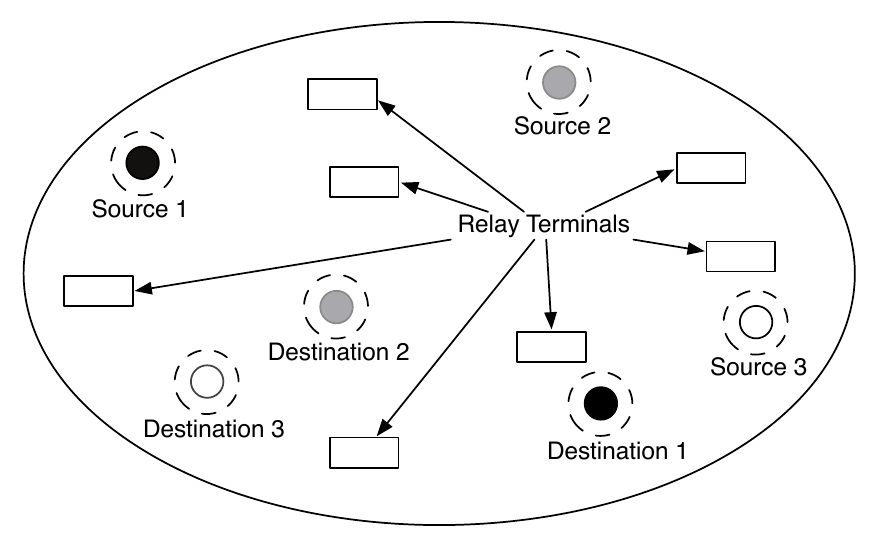}
\caption{Dense wireless interference relay network with dead-zones around source and destination terminals. Each terminal employs one antenna.}
\label{fig:mu_setup}
\end{figure}

This paper deals with interference fading relay networks where $M$ single-antenna source-destination terminal 
pairs communicate concurrently and in the same frequency band through half-duplex two-hop relaying over a common set 
of $K$ single-antenna relay terminals (see Fig.~\ref{fig:mu_setup}). Two setups are considered, i) the {\em coherent} case,
where the relays have channel state information (CSI), perform matched-filtering, and the destination terminals cannot cooperate, 
and ii) the {\em noncoherent} case, where the relays do not have CSI,
perform amplify-and-forward (AF) relaying, and the destination terminals can cooperate. In the coherent case, the network operates
in a completely distributed fashion, i.e., with no cooperation between any of the terminals whereas in the
noncoherent case the destination terminals can cooperate and perform joint decoding. 

\subsection{Contributions and Relation to Previous Work} 

Our main contributions for the coherent case can be summarized as follows:
\begin{itemize}
\item
We consider two different protocols, P1 introduced (for the finite-$M$ case)
in~\cite{bolcskei06-} and P2 introduced in~\cite{dana03-}. P1 relies on the idea of relay partitioning (i.e., each relay is assigned to one source-destination
terminal pair) and requires each relay terminal to know its assigned backward (source to relay) and forward (relay to destination) channel only.
The relays perform matched-filtering with respect to (w.r.t.) their assigned backward and forward channels.
P2 does not use relay partitioning, requires each relay terminal to know all $M$ backward and all $M$ forward channels, and
performs matched-filtering w.r.t.\ all $M$ backward and $M$ forward links.

Previous work for the coherent case has established the power efficiency scaling of P2 for $M\to\infty$ with $K=M^{2}$~\cite{dana03-};
in~\cite{bolcskei06-} it was shown that for P1 with $M$ fixed, in the $K\to\infty$ limit, network capacity scales as $C=(M/2)\log(K)+O(1)$. 
The results in~\cite{bolcskei06-} and the corresponding proof techniques, however, rely heavily on $M$ being fixed 
when $K\to\infty$. When $M,K\to\infty$, the amount of interference (at each destination terminal) grows with $M$. Establishing
the corresponding network capacity scaling behavior, therefore, requires fundamentally new techniques, which are developed in this paper. In particular, we derive the network (ergodic) capacity scaling behavior for $M,K\to\infty$ for P1 and P2 by computing a lower and an upper bound
on the per source-destination terminal pair capacity, and by showing that the bounds exhibit the same scaling (in $M,K$)
behavior. The technique used to establish the lower bound is based on a result found in a completely different context in~\cite{medard00-05} and applied in~\cite{dana03-}  
to derive the power efficiency scaling of P2. For our purposes, we need a slight generalization of the result in~\cite{medard00-05}, which follows, in a straightforward fashion, from a result on nearest-neighbor decoding reported in~\cite{lapidoth02-05}. For the sake of completeness, we state,  in Appendix~\ref{appendix:Medard}, the relevant inequality in the form needed in the context of this paper.
The matching upper bound on the per source-destination terminal pair capacity poses significantly more technical challenges and is based on a large-deviations analysis of the individual link SINR (signal to interference plus noise ratio) random variables (RVs).
In summary, we prove that in the large-$M$ limit, provided the number of relay terminals $K$ grows fast enough as a function of $M$,
under both protocols P1 and P2 the network ``decouples'' in the sense that the individual source-destination terminal pair (ergodic) capacities are strictly positive. 
The corresponding minimum rates of growth are $K \propto M^{3}$ for P1 and~$K \propto M^{2}$ for P2, with the per source-destination terminal pair capacity scaling (for~$M,K\to\infty$) given by $C_{\mathrm{P1}}=(1/2)\log\lefto(1+\Theta\lefto(K/M^{3}\right)\right)$ and $C_{\mathrm{P2}}=(1/2)\log\lefto(1+\Theta\lefto(K/M^{2}\right)\right)$, respectively.
The protocols P1 and P2 thus trade off CSI at the relays for the required (for the network to decouple) rate of growth of the number of relays. We hasten to add that an ergodic-capacity lower bound for P2 was previously established in~\cite{dana03-}; this bound is restated
(and reproved under slightly different assumptions) in this paper for the sake of completeness. It appears, however, that~\cite{dana03-} does not establish the minimum rate of growth of the number of relays for the network to decouple. 

\item
We analyze the network outage capacity behavior induced by P1 and P2 using a large-deviations approach. More specifically,
we show that the growth rates $K\propto M^{3}$ in P1 and $K \propto M^{2}$ in P2 are sufficient
to not only make the network decouple, but also to make to the individual source-destination fading links
converge to nonfading links. We say that the {\em network ``crystallizes''}
as it breaks up into a set of {\em effectively isolated ``wires in the air''}. Each of the decoupled links experiences distributed spatial diversity (or relay diversity),
with the corresponding diversity order going to infinity as $M\to\infty$.
Consequently, in the large-$M$ limit, time diversity
(achieved by coding over a sufficiently long time horizon) is not needed to achieve ergodic capacity.
We obtain bounds on the outage capacity of the individual source-destination links, which allow to characterize
the ``crystallization'' rate (more precisely a guaranteed ``crystallization'' rate as we do not know whether our bounds are tight), 
i.e., the rate (as a function of $M,K$) at which the decoupled links converge to nonfading links. In the course of this analysis, we develop a new technique for characterizing the large-deviations behavior of
certain sums of dependent RVs. This technique builds on the well-known truncation approach and is reported in Appendix~\ref{sec:Trunc}.

\item
For P1 and P2, we establish the impact of cooperation at the relay level on network (ergodic) capacity scaling. More specifically, it is shown
that, asymptotically in $M$ and $K$, cooperation (realized by vector matched filtering) in groups of $L$ relays leads to an $L$-fold reduction in the
total number of relays needed to achieve a given per source-destination terminal pair capacity.

\end{itemize}

Previous work for the noncoherent (AF) case~\cite{bolcskei06-} demonstrated that for~$M$ fixed and $K\to\infty$,
AF relaying turns the fading interference relay network into a fading point-to-point  multiple-input multiple-output (MIMO) link, showing that the use
of relays as active scatterers can recover spatial multiplexing gain in poor scattering environments. Our main
contributions for the noncoherent (AF) case are as follows:
\begin{itemize}
\item
Like in the coherent case, the proof techniques for the noncoherent (AF) case in~\cite{bolcskei06-} rely heavily on $M$ being finite. 
Building on results reported in~\cite{silverstein95-11}, we compute the $M,K\to\infty$ (with $K/M\to\beta$ fixed) per source-destination terminal pair capacity 
using tools from large-random-matrix theory~\cite{tulino04,muller03-10}. The limiting eigenvalue density function of the effective MIMO channel matrix between the source and destination terminals is characterized in terms of its Stieltjes transform
as the unique solution of a fixed-point equation, which can be transformed into a fourth-order equation. Upon solving this
fourth-order equation and applying the inverse Stieltjes transform, the remaining steps to computing the limiting eigenvalue density function, and based on that the asymptotic network capacity, need to be
carried out numerically. We show that this can be accomplished in a straightforward fashion and provide a corresponding algorithm.
\item
We show that for $\beta\to\infty$, the fading AF relay network is turned into a fading point-to-point MIMO link (in a sense to be made precise in Section~\ref{sec:ncoh}), thus
establishing the large-$M,K$ analog of the result found previously for the finite-$M$ and $K\to\infty$ case in~\cite{bolcskei06-}.
\end{itemize}

\subsection{Notation}

The superscripts~$\tp{}$,~$\herm{}$, and~$\conj{}$ stand for transposition, conjugate transpose, and element-wise conjugation, respectively. $|\setX|$ is the cardinality of the set $\setX$. $\log(x)$ stands for the logarithm to the base~$2$, and~$\ln(x)$ is the natural logarithm. $I[x]=1$ if $x$ is $\mathrm{true}$ and $I[x]=0$ if~$x$ is $\mathrm{false}$. $\delta[k]=1$ for~$k=0$ and~$0$ otherwise.  The unit step function $u(x)=0$ for $x<0$ and $u(x)=1$ for~$x\ge 0$. $\Exop$ and~$\Varop$ denote the expectation and variance operator, respectively. $\lceil x\rceil$ stands for the smallest integer greater than or equal to~$x$. $\arg(x)$ stands for the argument
of $x\in \complexset$. A circularly symmetric zero-mean complex Gaussian RV is a RV $Z = X+j\,Y \sim \circnorm(0,\sigma^2)$, 
where~$X$ and~$Y$ are independent identically distributed (\iid ) $\normal(0,\sigma^2/2)$. An exponentially distributed RV with parameter~$\lambda$ is a real-valued RV~$X$ with probability density function (pdf) given by~$\pdf{X}(x)=\lambda \exp(-\lambda x) u(x)$. A Rayleigh-distributed RV with parameter~$\alpha^{2}$ is a real-valued RV~$X$ with pdf~$\pdf{X}(x)=(x/\alpha^{2}) \exp\lefto(-x^{2}/(2\alpha^{2})\right)\!u(x)$. $\uniform\left(a,b\right)$ denotes the uniform distribution over the interval~$[a,b]$.  $\delta(x)$ is the Dirac delta distribution. The moment-generating function (MGF) of a RV~$X$ is defined as $M_{X}(s)\define \int_{-\infty}^{\infty} e^{s x} \pdf{X}(x) dx$. $(x)^{+}=x$ for~$x>0$ and~$0$ otherwise. For two functions~$f(x)$ and~$g(x)$, the notation~$f(x)=O(g(x))$ means that 
$|f(x)/g(x)|$ remains bounded as $x\to\infty$. We write~$g(x)=\Theta (f(x))$ to denote that $f(x)=O(g(x))$ and~$g(x)=O(f(x))$.  For two functions~$f(x)$ and~$g(x)$, the notation $f(x)=o(g(x))$ means that 
$|f(x)/g(x)|\to 0$ as $x\to\infty$. Matrices and vectors (both deterministic and random) are denoted by uppercase and lowercase, respectively, boldface letters. The element of a matrix~$\matX$ in the $n$th row and $m$th column and the $n$th element of a vector~$\vecx$ are denoted as~$[\matX]_{n,m}$ and~$[\vecx]_{n}$, respectively. $\lambda_{i}(\matX)$, $\lambda_{\min}(\matX)$, and $\lambda_{\max}(\matX)$ stand for the $i$th, the minimum, and the maximum eigenvalue of a matrix~$\matX$, respectively. $\matX\circ\matY$ is the Hadamard (or element-wise) product of the  matrices~$\matX$ and~$\matY$.  $\vecnorm{\vecx}$ denotes the $\ell^{2}$-norm of the vector~$\vecx$. $\Re z$ and~$\Im z$ designate the real and imaginary part of $z\in\complexset$, respectively. $\complexset^{+}\define\left\{z\in\complexset \given \Im{z}>0\right\}$. For any $n,m\in\naturals$, $m\ge n$,~$[n\!:\! m]$ denotes the natural numbers $\left\{n,n+1,\ldots,m\right\}$.

\subsection{Organization of the Paper}

The rest of this paper is organized as follows. Section~\ref{sec:model} describes the general channel model and the parts of the signal model\footnote{The motivation for the channel model considered in this paper can be found in~\cite{bolcskei06-}.} 
that pertain to both the coherent and the noncoherent case. Sections~\ref{sec:coh} and~\ref{sec:ergodic} focus on the coherent case exclusively: Section~\ref{sec:coh} contains the large-deviations analysis of the individual link SINRs for P1 and P2.
In Section~\ref{sec:ergodic}, we present our ergodic-capacity
scaling results, discuss the ``crystallization'' phenomenon, and study the impact of cooperation at the relay level. In Section~\ref{sec:ncoh}, we
present our results on the asymptotic network capacity for the noncoherent (AF) case. We conclude in Section~\ref{sec:conclusion}.
The new technique to establish the large-deviations behavior of certain sums of dependent RVs is presented in Appendix~\ref{sec:Trunc}. Appendix~\ref{sec:union} summarizes a set of (union) bounds used heavily throughout the paper.
Appendices~\ref{appendix:P1ConvProof}
and~\ref{appendix:ErgP1LThProof} contain the proofs of Theorems~\ref{thm:P1Conv}
and~\ref{thm:P1ergL}, respectively. 
The result from~\cite{lapidoth02-05} needed for the proof of
the ergodic capacity lower bounds for P1 and P2 is summarized in Appendix~\ref{appendix:Medard}. Appendix~\ref{appendix:RM} contains some essentials from large-random-matrix theory
needed in Section~\ref{sec:ncoh}. In Appendix~\ref{appendix:integral}, we detail part of the solution of the fixed-point equation underlying the main result in Section~\ref{sec:ncoh}. 

\section{Channel and Signal Model}
\label{sec:model}
In this section, we present the channel and signal model and additional basic assumptions. We restrict ourselves to the aspects that apply to both coherent and noncoherent networks and to both protocols considered in the coherent case. Relevant specifics for the coherent case will be provided in Sections~\ref{sec:P1} and~\ref{sec:P2} and for the noncoherent case in Section~\ref{sec:ncoh}.

\subsection{General Assumptions}
\begin{figure}
\centering
\includegraphics[width=\figwidth]{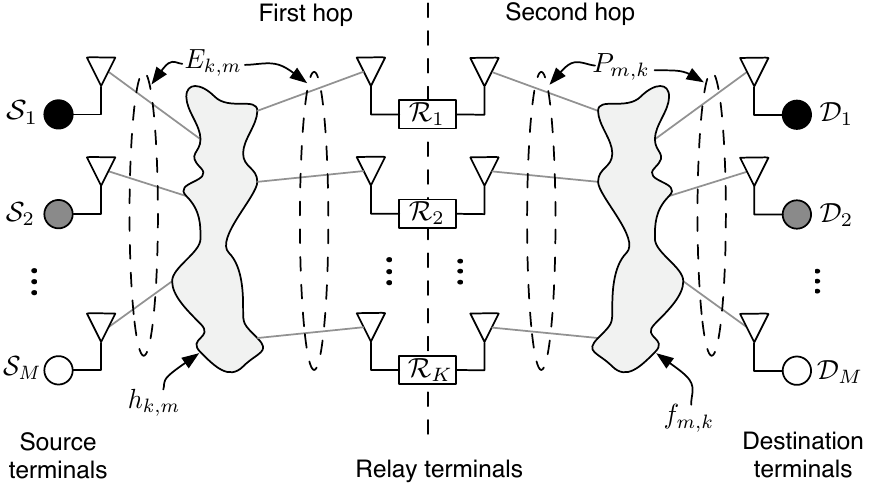}
\caption{Two-hop wireless relay network setup.}
\label{fig:CM}
\end{figure}

We consider an interference relay network (see Figs.~\ref{fig:mu_setup} and~\ref{fig:CM}) 
consisting of $K+2M$ single-antenna terminals 
with $M$ designated source-destination terminal pairs $\{\mathcal{S}_m, \mathcal{D}_m\}$ $\left(\natseg{m}{1}{M}\right)$ and $K$ relays $\mathcal{R}_k$ $\left(\natseg{k}{1}{K}\right)$. We assume a ``dead-zone'' of non-zero radius, free of relays, around each of the source and destination terminals, no direct link between the individual source-destination terminal pairs (e.g., due to large separation), and a domain of fixed area (i.e., dense network assumption). Transmission takes place in half-duplex fashion (the terminals cannot transmit and receive simultaneously)  in two hops (a.k.a.\ two-hop relaying) over two disjoint time slots. In the first time slot, the source terminals simultaneously broadcast their information to all the relay terminals (i.e., each relay terminal receives a superposition of all source signals). After processing the received signals, the relay terminals simultaneously broadcast the processed data to all the destination terminals during the second time slot. Our setup can be considered as an interference channel~\cite{carleial78-01} with dedicated relays, hence the terminology {\it interference relay network}.

\subsection{Channel and Signal Model}
Throughout the paper, frequency-flat fading over the bandwidth of interest as well as perfectly synchronized transmission and reception between the terminals is assumed. For the finite-$M$ and $K\to\infty$ case it has been shown in~\cite{nabar04-10} that the perfect-synchronization assumption can be relaxed, under quite general conditions on the synchronization errors, without impact on the capacity scaling laws. The input-output (I-O) relation for the link between the source terminals and the relay terminals during the first time slot is given by
\be
\label{eq:IO1}
\vecr=\left(\matE\circ\matH\right) \vecs+\vecz
\ee
where $\vecr=\tp{[r_1, r_2,\ldots, r_{K}]}$ with $r_{k}$ denoting the signal received at the $k$th relay terminal, $\matE\in \reals^{K\times M}$ with $[\matE]_{k,m}=\sqrt{E_{k,m}}$ where $E_{k,m}$ denotes the average energy received at $\mathcal{R}_k$ through the $\mathcal{S}_m\rightarrow \mathcal{R}_k$ link\footnote{$\mathcal{A}\rightarrow\mathcal{B}$ signifies communication from terminal $\mathcal{A}$ to terminal $\mathcal{B}$.}
(having accounted for path loss and shadowing in the $\mathcal{S}_m\rightarrow \mathcal{R}_k$ link), $\matH\in \complexset^{K\times M}$ with $[\matH]_{k,m}=h_{k,m}$ $(\natseg{k}{1}{K}$, $\natseg{m}{1}{M})$ where $h_{k,m}\sim\circnorm(0,1)$  denotes the \iid complex-valued channel gains corresponding to the $\mathcal{S}_m\rightarrow \mathcal{R}_k$ links, $\vecs=\tp{[s_1, s_2,\ldots, s_{M}]}$ where $s_m$ is the zero-mean Gaussian signal transmitted by $\mathcal{S}_m$ and the vector~$\vecs$ is i.i.d. temporally and spatially (across source terminals). Finally, $\vecz=\tp{[z_1, z_2,\ldots, z_{K}]}$ where $z_k\sim\circnorm(0,\sigma^2)$ is temporally and spatially (across relay terminals) white noise.
The $k$th relay terminal processes its received signal~$r_{k}$ to produce the output signal~$t_{k}$. The collection of output signals~$t_{k}$, organized in the vector $\vect=\tp{[t_1, t_2,\ldots, t_{K}]}$, is then broadcast to the destination terminals during the second time slot, while the source terminals are silent. The $m$th destination terminal receives the signal $y_{m}$ with $\vecy=\tp{[y_{1},y_{2},\ldots,y_{M}]}$ given by 
\be
\label{eq:IO2}
\vecy=\left(\matP\circ\matF\right) \vect+\vecw
\ee
where $\matP\in \reals^{M\times K}$ with $[\matP]_{m,k}=\sqrt{P_{m,k}}$ and $P_{m,k}$ denotes the average energy received at $\mathcal{D}_m$ through the $\mathcal{R}_k\rightarrow \mathcal{D}_m$ link
(having accounted for path loss and shadowing in the $\mathcal{R}_k\rightarrow \mathcal{D}_m$ link). Furthermore, $\matF\in \complexset^{M\times K}$ with $[\matF]_{m,k}=f_{m,k}$ $(\natseg{m}{1}{M}$, $\natseg{k}{1}{K})$ where $f_{m,k}\sim\circnorm(0,1)$ denotes the \iid complex-valued channel gains corresponding to the $\mathcal{R}_k\rightarrow \mathcal{D}_m$ links, and $\vecw=\tp{[w_1, w_2,\ldots, w_{M}]}$ with $w_m\sim\circnorm(0,\sigma^2)$ being temporally and spatially (across destination terminals) white noise. Throughout the paper, we impose a per-source-terminal power constraint $\Ex{\abs{s_{m}}^{2}}\le 1/M$ $(\natseg{m}{1}{M})$, which results in the total transmit power trivially satisfying $\Ex{\vecnorm{\vecs}^{2}}\le 1$. Furthermore, we impose a per-relay-terminal power constraint $\Ex{\abs{t_{k}}^{2}}\le P_{\mathrm{rel}}/K$ $\left(\natseg{k}{1}{K}\right)$, which results in the total power transmitted by the relay terminals satisfying $\Ex{\vecnorm{\vect}^{2}}\le P_{\mathrm{rel}}$.
As already mentioned above, path loss and shadowing are accounted for through the $E_{k,m}$ $(\natseg{k}{1}{K}$, $\natseg{m}{1}{M})$ (for the first hop) and the $P_{m,k}$ $(\natseg{m}{1}{M}$, $\natseg{k}{1}{K})$ (for the second hop). We assume that these parameters are deterministic, uniformly bounded from above (follows from the dead-zone assumption) and below (follows from considering a domain of fixed area) so that for all $k,m$
\be
\label{EPbounds}
0<\underline{E}\le E_{k,m}{}\le \overline{E}<\infty\quad\ 0<\underline{P}\le P_{m,k}{}\le \overline{P}<\infty.%
\ee
Throughout the paper, we assume that the source terminals $\mathcal{S}_m$ $(\natseg{m}{1}{M})$ do not have CSI. The assumptions on CSI at the relays and the destination terminals depend on the setup (coherent or noncoherent case) and the protocol (in the coherent case) and will be made specific when needed.

A discussion of the motivation for the two scenarios analyzed in this paper can be found in~\cite{bolcskei06-}.

\section{The Coherent Case}
\label{sec:coh}
\begin{figure*}[!t]
\normalsize
\setcounter{MYtempeqncnt}{\value{equation}}
\setcounter{equation}{14}
\be
\label{eq:SNIRP1}
\mathrm{SINR}^{\mathrm{P1}}_{m}\define \left.\Bigl|\sum_{k=1}^{K} a^{m,m}_{k}\Bigr|^{2}\right/\Biggl(\sum_{\hat{m}\ne m} \Bigl|\sum_{k=1}^{K} a^{m,\hat{m}}_{k}\Bigr|^{2}+\sigma^{2}  M\sum_{k=1}^{K}\abs{b_{k}^{m}}^{2}+K M \sigma^{2}\Biggr)
\ee
\setcounter{equation}{\value{MYtempeqncnt}}
\hrulefill
\vspace*{-5pt}
\end{figure*}
In this section, we describe the two protocols P1 and P2 and derive the corresponding SINR concentration results along with the resulting bounds on the individual source-destination link outage probability induced by P1 and P2. Note that the results in this section do not require ergodicity of~$\matH$ and~$\matF$. 
\subsection{Protocol 1 (P1)}
\label{sec:P1}
The basic setup was introduced in Section~\ref{sec:model}. We shall next describe the specifics of P1. 
The~$K$~relay terminals
are partitioned 
into~$M$ subsets $\mathcal{M}_m$ $\left(\natseg{m}{1}{M}\right)$ with\footnote{For simplicity, we assume that~$K$ is an integer multiple of~$M$. Moreover, in the remainder of the paper all results pertaining to P1 implicitly assume $K\ge M$.}
$\card{\mathcal{M}_m}=K/M$. The relays in~$\mathcal{M}_m$ are assumed to assist the $m$th source-destination terminal pair $\{\mathcal{S}_m, \mathcal{D}_m\}$.
This assignment is succinctly described through the relay partitioning function $p: [1,K]\rightarrow [1,M]$ defined as
\benn
p(k)\define m \Leftrightarrow \mathcal{R}_k \in \mathcal{M}_m.
\eenn
We assume that the $k$th relay terminal has perfect knowledge of the phases~$\arg(h_{k,p(k)})$ and  $\arg(f_{p(k),k})$  of the single-input single-output (SISO) backward (from the perspective of the relay) channel $\mathcal{S}_{p(k)}\rightarrow \mathcal{R}_k$ and the corresponding forward channel $\mathcal{R}_k\rightarrow \mathcal{D}_{p(k)}$, respectively. We furthermore define~$\tilde{h}_{k,p(k)}\define \exp\lefto(j\arg(h_{k,p(k)})\right)$ and~$\tilde{f}_{p(k),k}\define \exp\lefto(j\arg(f_{p(k),k})\right)$. 
The signal $r_k$ received at the $k$th relay terminal is first cophased w.r.t.\ the assigned backward channel followed by an energy normalization so that
\be
u_k=d_{\mathrm{P1}, k}\, \conj{\tilde{h}_{k,p(k)}}\,r_k
\label{eq:RelayProc1P1}
\ee
where 
\be
d_{\mathrm{P1},k}\define \sqrt{P_{\mathrm{rel}}} \left[\frac{K}{M}\sum_{m=1}^M E_{k,m}+ K \sigma^2 \right]^{-1/2}
\label{eq:dP1L1}
\ee
ensures that the per-relay power constraint $\Ex{\abs{u_k}^2}=P_{\mathrm{rel}}/K$ is met. The relay terminal $\mathcal{R}_k$ 
then computes the transmit signal $t_k$ by cophasing w.r.t.\ its assigned forward channel, i.e.,
\be
t_k=\conj{\tilde{f}_{p(k),k}}\,u_k
\label{eq:RelayTrP1}
\ee
which, obviously, satisfies~$\Ex{\abs{{t}_k}^2}\le P_{\mathrm{rel}}/K$ with equality and hence meets the total power constraint (across relays)~$\Ex{\vecnorm{\vect}^2}=\sum_{k=1}^{K}\Ex{\abs{{t}_k}^2}=P_{\mathrm{rel}}$. In summary, P1 ensures that the relays $\mathcal{R}_{k}\in\mathcal{M}_m$ forward the 
signal intended for $\mathcal{D}_m$, namely, the signal transmitted by~$\mathcal{S}_{m}$, in a \textquotedblleft doubly coherent\textquotedblright~(w.r.t.\ backward and forward channels) fashion, whereas the signals transmitted 
by the source terminals $\mathcal{S}_{\hat{m}}$ with $\hat{m}\, \neq\, m$ are forwarded to $\mathcal{D}_{m}$ in a \textquotedblleft noncoherent\textquotedblright~fashion  (i.e., phase 
incoherence occurs either on the backward  or the forward link or on both links). The idea underlying P1 has originally been introduced in~\cite{bolcskei06-} (for the finite-$M$ case).

We shall next derive the I-O relation for the SISO channels $\mathcal{S}_m\to \mathcal{D}_m$ $\left(\natseg{m}{1}{M}\right)$. The destination terminal $\mathcal{D}_m$ receives doubly (backward and forward link) coherently combined contributions corresponding to the signal $s_m$, with interfering terms containing contributions from the signals $s_{\hat{m}}$ with $\hat{m}\ne m$ as well as noise, forwarded by the relays. Combining~\eqref{eq:IO1}, \eqref{eq:RelayProc1P1}, \eqref{eq:RelayTrP1}, and~\eqref{eq:IO2}, it follows (after some straightforward algebra) that the signal received at $\mathcal{D}_m\  (\natseg{m}{1}{M})$ is given by\footnote{The notation $\sum_{\hat{m}\ne m}$ stands for the summation over $\natseg{\hat{m}}{1}{M}$ s.t.\ $\hat{m}\ne m$. If not specified, the upper limit of the summation is clear from the context.} 
\bmult
y_m=s_m\underbrace{\frac{1}{\sqrt{K}}\sum_{k=1}^{K} a_k^{m,m}}_{\text{effective channel gain}}\\
{+}\:\underbrace{\sum_{\hat{m}\ne m} s_{\hat{m}}\,\frac{1}{\sqrt{K}}\sum_{k=1}^K a_k^{m,\hat{m}}}_{\text{interference}}+
\underbrace{\frac{1}{\sqrt{K}}\sum_{k=1}^{K} b_k^{m}  z_k+w_m}_{\text{noise}}
\label{eq:P1SISO}
\emult
where
\ba
\label{eq:P1aDef}
a_k^{m,\hat{m}}&\define C_{\mathrm{P1}, k}^{m,\hat{m}}\, \conj{\tilde{f}}_{p(k) ,k}\, f_{m,k}\, \conj{\tilde{h}}_{k,p(k)}\, h_{k,\hat{m}}\\
\label{eq:P1bDef}
b_k^{m}&\define C_{\mathrm{P1},k}^{m}\, \conj{\tilde{f}}_{p(k) ,k}\, f_{m,k}\, \conj{\tilde{h}}_{k,p(k)}\\
\intertext{with}
\label{eq:BigCP1}
C_{\mathrm{P1},k}^{m,\hat{m}}&= \sqrt{K} d_{\mathrm{P1}, k}  \sqrt{P_{m,k} E_{k,\hat{m}}}\\
\label{eq:SmallCP1}
C_{\mathrm{P1},k}^{m}&= \sqrt{K} d_{\mathrm{P1}, k} \sqrt{P_{m,k}}.%
\ea
The normalization factor~$\sqrt{K}$ in~\eqref{eq:P1SISO}, \eqref{eq:BigCP1}, and~\eqref{eq:SmallCP1} is introduced for convenience of exposition. Using~\eqref{EPbounds}, it now follows that
\bsa
\label{eq:P1Cbound}
\underline{C}\define\sqrt{\frac{\underline{P}\,\underline{E}P_{\mathrm{rel}}}{\overline{E}+\sigma^{2}}}&{}\le C_{\mathrm{P1},k}^{m,\hat{m}} &{}\le \sqrt{\frac{\overline{P}\,\overline{E}P_{\mathrm{rel}}}{\underline{E}+\sigma^{2}}}\define\overline{C}\\
\label{eq:P1cbound}
\underline{c}\define\sqrt{\frac{\underline{P}P_{\mathrm{rel}}}{\overline{E}+\sigma^{2}}}&{}\le  C_{\mathrm{P1},k}^{m} &{}\le \sqrt{\frac{\overline{P}P_{\mathrm{rel}}}{\underline{E}+\sigma^{2}}}\define\overline{c}%
\esa
for all $\natseg{k}{1}{K}$, $\natseg{m}{1}{M}$, and $\natseg{\hat{m}}{1}{M}$. In the following, it will be essential that the constants $\underline{C}$, $\underline{c}$, $\overline{C}$, and~$\overline{c}$ do not depend on~$M,K$.

Since we assumed that the destination terminals $\mathcal{D}_{m}$ $\left(\natseg{m}{1}{M}\right)$ cannot cooperate, the~$\mathcal{D}_{m}$ cannot perform joint decoding so that the network can be viewed as a collection of $M$ SISO channels $\mathcal{S}_{m}\to\mathcal{D}_{m}$, i.e., as an interference channel with dedicated relays.  We can see from~\eqref{eq:P1SISO} that 
each of these
SISO channels consists of a fading effective channel, fading interference, caused by the source signals not intended for
a given destination terminal, and finally a noise term incorporating thermal noise forwarded by the relays and thermal noise added at the destination terminals. 
In the remainder of this section, we make the conceptual assumption that each of the destination terminals~$\mathcal{D}_{m}$ has perfect knowledge of the fading and path loss and shadowing coefficients in the entire network,  i.e.,~$\mathcal{D}_{m}\,\left(\natseg{m}{1}{M}\right)$ knows~$\matH, \matF, \matE$ and~$\matP$ perfectly. An immediate consequence of this assumption is that~$\mathcal{D}_{m}\,\left(\natseg{m}{1}{M}\right)$ has perfect knowledge of the effective channel gain~$(1/\sqrt{K})\sum_{k=1}^{K} a_k^{m,m}$, the interference channel gains~$(1/\sqrt{K})\sum_{k=1}^K a_k^{m,\hat{m}}\ (\hat{m}\ne m)$, and the quantity~$(1/\sqrt{K})\sum_{k=1}^{K} b_k^{m}$.  
Conditioned on~$\matH$ and~$\matF$, both the interference and the noise term in~\eqref{eq:P1SISO} are Gaussian, so that the mutual information for the ${\cal S}_{m}\,\rightarrow\,{\cal D}_{m}$ link is given by
\be
\label{eq:MIP1}
I\lefto(y_{m}; s_{m}\given \matH,\matF\right)=\frac{1}{2}\log\lefto(1+\mathrm{SINR}^{\mathrm{P1}}_{m}\right)
\ee
where $\mathrm{SINR}^{\mathrm{P1}}_{m}$, defined in \eqref{eq:SNIRP1} at the top of the page, is the effective SINR in the SISO channel $\mathcal{S}_{m}\to\mathcal{D}_{m}$.
\addtocounter{equation}{1}
\begin{figure*}[!t]
\normalsize
\setcounter{MYtempeqncnt}{\value{equation}}
\setcounter{equation}{20}
\be
\label{eq:SNIRP2}
\mathrm{SINR}^{\mathrm{P2}}_{m}\define 
\left.\Bigl|\sum_{k=1}^{K} \sum_{\tilde{m}=1}^{M}a^{m,m,\tilde{m}}_{k}\Bigr|^{2}\right/\Biggl(\sum_{\hat{m}\ne m} \Bigl|\sum_{k=1}^{K} \sum_{\tilde{m}=1}^{M}a^{m,\hat{m},\tilde{m}}_{k}\Bigr|^{2}+\sigma^{2}M \sum_{k=1}^{K}\Bigl|\sum_{\tilde{m}=1}^{M}b_{k}^{m,\tilde{m}}\Bigr|^{2}+K M^{2}\sigma^{2}\Biggr)
\ee
\setcounter{equation}{\value{MYtempeqncnt}}
\hrulefill
\vspace*{-5pt}
\end{figure*}

We conclude by noting that the large-deviations results in Section~\ref{sec:convergence} rely heavily on the assumption that~$\mathcal{D}_{m}\,\left(\natseg{m}{1}{M}\right)$ knows~$\matH, \matF, \matE$, and~$\matP$ perfectly. The ergodic capacity-scaling results in Section~\ref{sec:ergodic} will, however, be seen to require significantly less channel knowledge at the destination terminals. 
\subsection{Protocol 2 (P2)}
\label{sec:P2}
The only difference between P1 and P2 is in the processing at the relays. Whereas in P1 the $K$ relay terminals are partitioned into $M$ clusters (of equal size) with each of these clusters assisting one particular source-destination terminal pair, in P2 each relay assists all source-destination terminal pairs so that relay partitioning is not needed. 
In turn, P2 requires that each relay knows
the phases of all its $M$~backward and $M$~forward channels, i.e., $\mathcal{R}_k$ needs knowledge of $\tilde{h}_{k,m}$ and $\tilde{f}_{m,k}$, respectively,
for $\natseg{m}{1}{M}$. Consequently, P2 requires significantly more CSI at the relays than P1. The relay processing stage in P2 computes
\ba
\label{eq:RelayTrP2}
t_k&=d_{\mathrm{P2},k}\!\left(\sum_{m=1}^M \conj{\tilde{h}_{k,m}}\, \conj{\tilde{f}_{m,k}} \right) r_k\\
\intertext{where}
d_{\mathrm{P2},k}&\define \sqrt{P_{\mathrm{rel}}}\lefto[K\sum_{m=1}^M E_{k,m}+ M K\sigma^2 \right]^{-1/2}\nonumber
\ea
ensures that 
the power constraint~$\Ex{\abs{t_k}^2}=P_{\mathrm{rel}}/K$ and hence~$\Ex{\vecnorm{\vect}^2}=\sum_{k=1}^{K}\Ex{\abs{t_k}^2}=P_{\mathrm{rel}}$ is met.

Again, we start by deriving the I-O relation for the SISO channels $\mathcal{S}_m\to \mathcal{D}_m$ $\left(\natseg{m}{1}{M}\right)$. Like in P1, the destination terminal $\mathcal{D}_m$ receives doubly (backward and forward link) coherently combined contributions corresponding to the signal $s_m$, interfering terms containing contributions from the signals $s_{\hat{m}}$ with $\hat{m}\ne m$, as well as noise forwarded by the relays. Combining~\eqref{eq:IO1}, \eqref{eq:RelayTrP2}, and~\eqref{eq:IO2}, it follows that the signal received at $\mathcal{D}_m\, \left(\natseg{m}{1}{M}\right)$ is given by
\bmult
y_m=s_m\underbrace{\frac{1}{\sqrt{K M}}\sum_{k=1}^{K} \sum_{\tilde{m}=1}^{M} a_k^{m,m,\tilde{m}}}_{\text{effective channel gain}}\\
{+}\:\underbrace{\sum_{\hat{m}\ne m} s_{\hat{m}}\, \frac{1}{\sqrt{K M}}\sum_{k=1}^K \sum_{\tilde{m}=1}^{M} a_k^{m,\hat{m},\tilde{m}}}_{\text{interference}}\\
{+}\:\underbrace{\frac{1}{\sqrt{K M}}\sum_{k=1}^{K} \sum_{\tilde{m}=1}^{M} b_k^{m,\tilde{m}} z_k+w_m}_{\text{noise}}
\label{eq:P2SISO}
\emult
where
\ba
a_k^{m,\hat{m},\tilde{m}}&\define C_{\mathrm{P2}, k}^{m,\hat{m}}\, \conj{\tilde{f}}_{\tilde{m},k}\, f_{m,k}\, \conj{\tilde{h}}_{k,\tilde{m}}\, h_{k,\hat{m}}\nonumber\\
b_k^{m,\tilde{m}}&\define C_{\mathrm{P2},k}^{m}\, \conj{\tilde{f}}_{\tilde{m},k}\, f_{m,k}\, \conj{\tilde{h}}_{k,\tilde{m}}\nonumber\\
\intertext{with}
\label{eq:BigCP2}
C_{\mathrm{P2},k}^{m,\hat{m}}&\define \sqrt{K M} d_{\mathrm{P2}, k}  \sqrt{P_{m,k} E_{k,\hat{m}}}\\
\label{eq:SmallCP2}
C_{\mathrm{P2},k}^{m}&\define \sqrt{K M} d_{\mathrm{P2}, k} \sqrt{P_{m,k}}.%
\ea
Again, the normalization~$\sqrt{KM}$ in~\eqref{eq:P2SISO}, \eqref{eq:BigCP2} and~\eqref{eq:SmallCP2} is introduced for convenience of exposition and
\benn
\underline{C}\le  C_{\mathrm{P2},k}^{m,\hat{m}} \le \overline{C},\qquad 
\underline{c}\le C_{\mathrm{P2},k}^{m} \le \overline{c}
\eenn
for all~$\natseg{k}{1}{K}$, $\natseg{m}{1}{M}$, and $\natseg{\hat{m}}{1}{M}$ with the constants~$\underline{C}$, $\underline{c}$, $\overline{C}$, and $\overline{c}$ not depending on~$M, K$.

Recalling that we assume perfect knowledge of~$\matH, \matF, \matE$, and~$\matP$ at each of the destination terminals, ${\cal D}_{m}$, the mutual information for the ${\cal S}_{m}\,\rightarrow\,{\cal D}_{m}$ link in P2 is given by
\be
\label{eq:MIP2}
I\lefto(y_{m}; s_{m}\given \matH,\matF\right)=\frac{1}{2}\log\lefto(1+\mathrm{SINR}^{\mathrm{P2}}_{m}\right)
\ee%
\addtocounter{equation}{1}%
where $\mathrm{SINR}^{\mathrm{P2}}_{m}$, defined in \eqref{eq:SNIRP2} at the top of the page, is the effective SINR in the SISO channel $\mathcal{S}_{m}\to\mathcal{D}_{m}$.

\subsection{Large-Deviations Analysis of SINR}
\label{sec:convergence}
Our goal in this section is to prove that $\mathrm{SINR}^{\mathrm{P1}}_{m}$ and $\mathrm{SINR}^{\mathrm{P2}}_{m}$ for~$\natseg{m}{1}{M}$ (and, thus, the corresponding mutual information quantities~\eqref{eq:MIP1} and~\eqref{eq:MIP2}) lie within ``narrow intervals'' around their mean values with\footnote{The precise meaning of ``narrow intervals'' and ``high probability'' is explained in the formulation of Theorems~\ref{thm:P1Conv} and~\ref{thm:P2Conv} in Section~\ref{sec:concP1P2}.} ``high probability'' when $M,K\to\infty$. The technique we use to prove these {\em concentration results} is based on a large-deviations analysis and can be summarized as follows:
\renewcommand{\theenumi}{\roman{enumi}}
\begin{enumerate}
\item
\label{step:sum1}
Consider each sum in the numerator and denominator of~\eqref{eq:SNIRP1} and~\eqref{eq:SNIRP2} separately.
\item
\label{step:sum2}
Represent the considered sum as a sum of independent RVs or as a sum of dependent complex-valued RVs with independent phases. 
\item
\label{step:sum3}
Find the mean value of the considered sum.
\item
\label{step:sum4}
Employ a large-deviations analysis to prove that the considered sum lies within a narrow interval around its mean with high probability, i.e., establish a concentration result. 
\item
\label{step:sum5}
Combine the concentration results for the separate sums using the union bounds summarized in Appendix~\ref{sec:union} to obtain concentration results for $\mathrm{SINR}^{\mathrm{P1}}_{m}$ and $\mathrm{SINR}^{\mathrm{P2}}_{m}$.
\end{enumerate}
\renewcommand{\theenumi}{\arabic{enumi}}
\subsubsection{Chernoff bounds}
\label{sec:chernoff}
Before embarking on a detailed discussion of the individual Steps~\ref{step:sum1}--\ref{step:sum5} above, we note that a well-known technique to establish large-deviations results for sums of RVs (as required in Step~\ref{step:sum4} above) is based on Chernoff bounds. This method, which yields the precise exponential behavior for the tails of the distributions under question, can, unfortunately, not be applied to all the sums in~\eqref{eq:SNIRP1} and~\eqref{eq:SNIRP2}. 
To solve this problem, we develop a new technique, which allows to establish large-deviations results for certain sums of dependent complex-valued RVs with independent phases
where the RVs occurring in the sum are s.t.\ their MGF does not need to be known. The new technique is based on the well-known idea of truncation of RVs and will, therefore, be called truncation technique. Even though truncation of RVs is a standard concept in probability theory, and in particular in large-deviations analysis, we could not find the specific approach developed in this paper in the literature. We therefore decided to present the truncation technique as a stand-alone concept and summarized the main results in Appendix~\ref{sec:Trunc}. Before proceeding, we note that even though the truncation technique has wider applicability than Chernoff bounds, it yields weaker exponents for the tails of the distributions under question.  

Although the proofs of the main concentration results, Theorems~\ref{thm:P1Conv} and~\ref{thm:P2Conv} in Section~\ref{sec:concP1P2}, are entirely based on the truncation technique, we still discuss the results of the application of Chernoff bounds (without giving all the details) in the following, restricting our attention to P1, to motivate the development of the truncation technique and to provide a reference for the quality (in terms of tightness of the bounds) of the results in Theorems~\ref{thm:P1Conv} and~\ref{thm:P2Conv}. Moreover, the developments below introduce some of the key elements of the proofs of  Theorems~\ref{thm:P1Conv} and~\ref{thm:P2Conv}.  
 
Following the approach outlined in Steps~\ref{step:sum1}--\ref{step:sum5} above, we start by writing~$\mathrm{SINR}^{\mathrm{P1}}_{m}$ as
\be
\mathrm{SINR}^{\mathrm{P1}}_{m}=\frac{\abs{S^{(1)}+S^{(2)}}^{2}}{S^{(3)}+\sigma^{2} M S^{(4)}+K M \sigma^{2}}
\label{eq:SINRP1Sums}
\ee
and establishing bounds on the probability of large deviations of
\ba
\label{eq:S1def}
S^{(1)}&\define\sum_{k: p(k)=m}  C_{\mathrm{P1}, k}^{m,m} \abs{f_{m,k}}\abs{h_{k,m}}\\
\label{eq:S2def}
S^{(2)}&\define\sum_{k:p(k)\ne m} C_{\mathrm{P1}, k}^{m,m}\, \conj{\tilde{f}_{p(k),k}}\, f_{m,k}\,  \conj{\tilde{h}_{k,p(k)}}\,  h_{k,m} \\
\label{eq:S3def}
S^{(3)}&\define\sum_{\hat{m}\ne m}\Bigl| \sum_{k=1}^K C_{\mathrm{P1}, k}^{m,\hat{m}}\, \conj{\tilde{f}_{p(k),k}}\, f_{m,k}\,  \conj{\tilde{h}_{k,p(k)}}\, h_{k,\hat{m}}\Bigr|^{2}\\
\label{eq:S4def}
S^{(4)}&\define\sum_{k=1}^K \left(C_{\mathrm{P1}, k}^{m}\right)^{2}\abs{f_{m,k}}^{\! 2}.
\ea
We shall see in the following that the pdfs of the terms in~$S^{(1)}, S^{(2)}$, and~$S^{(4)}$ have a structure that is simple enough for Chernoff bounds to be applicable.
We start with the analysis of the simplest term, namely~$S^{(4)}$. 
To avoid unnecessary technical details and to simplify the exposition, we assume (only in the ensuing analysis of the large deviations behavior of~$S^{(4)}$) that
\be
\label{eq:COne}
C_{\mathrm{P1}, k}^{m,\hat{m}}=C_{\mathrm{P1}, k}^{m}=1 
\ee
for all $\natseg{m,\hat{m}}{1}{M}$, $\natseg{k}{1}{K}$.
Defining\footnote{For notational convenience, we shall omit the index~$m$ in what follows.} $X_{k}\define\abs{f_{m,k}}^{2}$, we have
\benn
S^{(4)}=\sum_{k=1}^{K} X_{k}
\eenn
where the~$X_{k}$ are \iid exponentially distributed with parameter~$\lambda=1$, i.e., $\pdf{X_{k}}(x)=\exp\lefto(-x\right) u(x)$ and hence~$\Ex{X_{k}}=1$.
For convenience, we centralize~$X_{k}$ and define~$Z_{k}\define X_{k}-1$. The MGF of~$Z_{k}$ is given by 
\be
M_{Z_{k}}(s)=\int_{0}^{\infty} e^{s(x-1)}e^{-x} dx=\frac{e^{-s}}{1-s},\qquad \Re s\le 1.
\label{eq:MGFS4}
\ee
Since the RVs~$Z_{k}$ are independent, we obtain, using the standard Chernoff bound (see, for example,~\cite[Section~5.4]{gallager68-}), for~$x>0$
\ba
\Prob\lefto\{\sum_{k=1}^{K} Z_{k}\ge x\right\}&\le \min_{0\le s \le 1} \left(M_{Z_{k}}(s)\right)^{K} e^{-s x}\nonumber\\
&=\min_{0\le s \le 1} e^{-Ks-K\ln(1-s)-s x}.
\label{eq:S4Chernoff1}
\ea
Because $\left(M_{Z_{k}}(s)\right)^{K} \exp\lefto(-s x\right)$ is convex in~$s$~\cite[Section~5.4]{gallager68-}, the minimum in~\eqref{eq:S4Chernoff1} can easily be seen to be taken on for~$s=x/(x+K)$, which gives
\ba
\Prob\lefto\{\sum_{k=1}^{K} Z_{k}\ge x\right\}&\le e^{K\ln(x+K)-K\ln(K)-x}.
\label{eq:S4Chernoff2}
\ea 
The corresponding relation for negative deviations ($x<0$) is
\ba
&\Prob\lefto\{\sum_{k=1}^{K} Z_{k}\le x\right\}
\le
\begin{cases}
e^{K\ln(x+K)-K\ln(K)-x},\ x>-K\\
0,\ x<-K.
\end{cases}
\label{eq:S4Chernoff3}
\ea
Finally, 
setting~$x=\sqrt{K}t$, we get the desired concentration result for the sum~$S^{(4)}$ as
\ba
&\Prob\lefto\{S^{(4)}-K\ge \sqrt{K} t\right\}
\le 
e^{K\ln\lefto(1+t/\sqrt{K}\right)- \sqrt{K}t},\ t\ge 0
\label{eq:S4Chernoff4}
\ea  
\bmult
\Prob\lefto\{S^{(4)}-K\le \sqrt{K} t\right\}
\\\le 
\begin{cases}
e^{K\ln\lefto(1+t/\sqrt{K}\right)- \sqrt{K} t},\ -\sqrt{K}< t\le 0\\
0,\ t\le -\sqrt{K}.
\end{cases}
\label{eq:S4Chernoff41}
\emult  
We now consider the case when~$K$ is large and $t=o\bigl(\sqrt{K}\bigr)$ so that
\be
\ln\lefto(1+\frac{t}{\sqrt{K}}\right)=\frac{t}{\sqrt{K}}-\frac{t^{2}}{2 K}+O\lefto(\left(\frac{t}{\sqrt{K}}\right)^{3}\right).
\label{eq:LogSeries}
\ee
If we omit higher (than second) order terms in~\eqref{eq:LogSeries}, the bound in~\eqref{eq:S4Chernoff4} and~\eqref{eq:S4Chernoff41} can be compactly written as
\be
\Prob\lefto\{\abs{S^{(4)}-K}\ge \sqrt{K} t\right\}\le 2 e^{-t^{2}/2}.
\label{eq:S4Chernoff}
\ee  
We can, therefore, conclude that
the probability of large deviations of~$S^{(4)}$ decays exponentially. 

Similar concentration results, using Chernoff bounds, can be established for~$S^{(1)}$ and~$S^{(2)}$. The derivation is somewhat involved (as it requires establishing upper bounds on the MGF), does not provide insights into the problem and will, therefore, be omitted.
Unfortunately, the simple technique used above to establish concentration results for~$S^{(4)}$ (and applicable to~$S^{(1)}$ and $S^{(2)}$) does not seem to be applicable to~$S^{(3)}$. To see this, we start by noting that~$S^{(3)}$ contains two classes of terms (in the sense of the properties of their pdf), i.e.,
\be
S^{(3)}=S^{(31)}+S^{(32)}
\label{eq:S3toS31S32}
\ee 
with
\ba
\label{eq:S31}
S^{(31)}&\define \sum_{\hat{m}\ne m}\sum_{k=1}^{K}\left(C_{\mathrm{P1}, k}^{m,\hat{m}}\right)^{2}\abs{f_{m,k}}^{2}\abs{h_{k,\hat{m}}}^{2}\displaybreak\\
\label{eq:S32}
S^{(32)}&\define \sum_{\hat{m}\ne m}\sum_{k=1}^{K}\sum_{\hat{k}\ne k}C_{\mathrm{P1}, k}^{m,\hat{m}}\, \conj{\tilde{f}_{p(k) ,k}}\, f_{m,k}\,  \conj{\tilde{h}_{k,p(k)}}\, h_{k,\hat{m}}\nonumber\\
&\relphantom{\define}\phantom{\sum_{\hat{m}\ne m}\sum_{k=1}^{K}} {}\times 
C_{\mathrm{P1}, \hat{k}}^{m,\hat{m}}\, \tilde{f}_{p(\hat{k}),\hat{k}}\, \conj{f_{m,\hat{k}}}\, \tilde{h}_{\hat{k},p(\hat{k})}\, \conj{h_{\hat{k},\hat{m}}}.
\ea
Now, there are two problems in applying the technique we have used so far to~$S^{(3)}$:   
First, it seems very difficult to compute the MGFs for the individual terms in~$S^{(31)}$ and $S^{(32)}$;
second, the individual terms in $S^{(31)}$ and $S^{(32)}$ are not {\em jointly}\footnote{We write ``jointly independent'', as opposed to ``pairwise independent'' here and in what follows to stress the fact that the joint pdf of the RVs under consideration can be factored into a product of the marginal pdfs. In several places throughout the paper we will deal with sets of RVs that turn out to be pairwise independent, but not jointly independent.} independent across the summation indices.
The first problem can probably be resolved using bounds on the exact MGFs (as can be done in the analysis of~$S^{(1)}$ and~$S^{(2)}$). The second problem, however, seems more fundamental. 
In particular, the individual terms in~$S^{(31)}$ are independent across $k$ but not across $\hat{m}$. In $S^{(32)}$, the individual terms are independent across $k$ but not across $\hat{k}$ and $\hat{m}$. Assuming that the problem of computing (or properly bounding) the MGFs is resolved, a natural way to overcome the second problem mentioned above would be to establish concentration results for the sums over~$k$, i.e., for
\ba
\hat{S}^{(31)}_{\hat{m}}&\define\sum_{k=1}^{K}\left(C_{\mathrm{P1}, k}^{m,\hat{m}}\right)^{2}\abs{f_{m,k}}^{2}\abs{h_{k,\hat{m}}}^{2}\\
\hat{S}^{(32)}_{\hat{m}, \hat{k}}&\define\sum_{k=1}^{K}C_{\mathrm{P1}, k}^{m,\hat{m}}\, \conj{\tilde{f}_{p(k),k}}\, f_{m,k}\,  \conj{\tilde{h}_{k,p(k)}}\, h_{k,\hat{m}}\,\nonumber\\
&\relphantom{\define}\phantom{\sum_{k=1}^{K}C_{\mathrm{P1}, k}^{m,\hat{m}}\,} {}\times 
C_{\mathrm{P1}, \hat{k}}^{m,\hat{m}}\, \tilde{f}_{p(\hat{k}),\hat{k}}\, \conj{f_{m,\hat{k}}}\, \tilde{h}_{\hat{k},p(\hat{k})}\, \conj{h_{\hat{k},\hat{m}}}
\label{eq:S321}
\ea
and to employ the union bound for sums (Lemmas~\ref{lemma:UBsums} and~\ref{lemma:UBsumsMixed} in Appendix~\ref{sec:union}) to obtain concentration results for $S^{(31)}$ and $S^{(32)}$. Unfortunately, this method, although applicable, yields results that are very loose in the sense of not reflecting the correct ``order-of-magnitude behavior'' of the typical deviations. To understand why this is the case, we perform an order-of-magnitude analysis as follows. For simplicity, we again assume that the condition~\eqref{eq:COne} is satisfied.
Note that for any $\natseg{\hat{k},k}{1}{K}$ s.t.\ $\hat{k}\ne k$ and any $\natseg{\hat{m}}{1}{M}$ s.t.\ $\hat{m}\ne m$, we have
\benn
\Ex{\conj{\tilde{f}_{p(k) ,k}}\, f_{m,k}\,  \conj{\tilde{h}_{k,p(k)}}\, h_{k,\hat{m}}\,\tilde{f}_{p(\hat{k}) ,\hat{k}}\, \conj{f_{m,\hat{k}}}\, \tilde{h}_{\hat{k},p(\hat{k})}\, \conj{h_{\hat{k},\hat{m}}}}=0.
\eenn 
Chernoff bounding $\hat{S}^{(32)}_{\hat{m}, \hat{k}}$ would, therefore, yield that 
\benn
\Prob\lefto\{\abs{\hat{S}^{(32)}_{\hat{m}, \hat{k}}}\ge \sqrt{K} t\right\}
\eenn
decays exponentially\footnote{We do not specify the exponent here.} in $t$. Then, applying the union bound for sums (Lemma~\ref{lemma:UBsums}) to $S^{(32)}=\sum_{\hat{m}\ne m} \sum_{\hat{k}\ne k} \hat{S}^{(32)}_{\hat{m}, \hat{k}}$, we would conclude that
\be
\Prob\lefto\{\abs{S^{(32)}}\ge (M-1)(K-1)\sqrt{K} t\right\}
\label{eq:S32L}
\ee
decays exponentially in~$t$. 
Even though the terms in~$S^{(32)}$ are not completely independent across~$\hat{k}$ and~$\hat{m}$, we will see in Section~\ref{sec:AppTrunc} that there is still enough independence between them for the truncation technique to reveal that
\be
\Prob\lefto\{\abs{S^{(32)}}\ge \sqrt{(M-1)(K-1)K} t\right\}
\label{eq:S32T}
\ee
decays exponentially in~$t$, which is a much stronger concentration result than~\eqref{eq:S32L}. The importance of the difference between~\eqref{eq:S32T} and~\eqref{eq:S32L} becomes clear if we consider~$S^{(31)}$. 
Since~$\hat{S}^{(31)}_{\hat{m}}$ is a sum over~$K$ independent terms, each of which satisfies~$\Ex{\abs{f_{m,k}}^{2}\abs{h_{k,\hat{m}}}^{2}}=1$, Chernoff bounding would yield that 
\benn
\Prob\lefto\{\abs{\hat{S}^{(31)}_{\hat{m}}-K}\ge \sqrt{K}t \right\}
\eenn 
decays exponentially in~$t$. Applying the union bound to~$S^{(31)}=\sum_{\hat{m}\ne m}\hat{S}^{(31)}_{\hat{m}}$, one can then show that
\be
\Prob\lefto\{\abs{S^{(31)}-K (M-1)}\ge (M-1)\sqrt{K}t\right\}
\label{eq:S31disc}
\ee 
decays exponentially in~$t$.  
When~$M$ and~$K$ are large, we would now conclude  from~\eqref{eq:S32L} and~\eqref{eq:S31disc}  that~$S^{(3)}=S^{(31)}+S^{(32)}$ deviates around~$KM$ with a typical deviation of order~$MK\sqrt{K}$. Since the typical deviations are larger (by a factor of~$\sqrt{K}$) than the mean, the corresponding deviation result is useless. On the other hand, if we use the bound~\eqref{eq:S32T} combined with~\eqref{eq:S31disc}, again assuming that~$M$ and~$K$ are large, we can conclude that~$S^{(3)}$ deviates around~$KM$ with a typical deviation of order~$\sqrt{M}K+M\sqrt{K}$, which is an order of magnitude smaller than the mean.
As already mentioned, the truncation technique allows us to establish useful concentration results for sums with dependent terms such as that in~\eqref{eq:S321}. 

\subsubsection{Application of the truncation technique}
\label{sec:AppTrunc}
In this section, we demonstrate how the desired concentration results for~$S^{(31)}$ and~$S^{(32)}$, defined in~\eqref{eq:S31} and~\eqref{eq:S32}, respectively, can be obtained by application of the truncation technique. The following results will be used in the proof of Theorem~\ref{thm:P1Conv} and will, therefore, be formulated for general~$C_{\mathrm{P1}, k}^{m,\hat{m}}$ and~$C_{\mathrm{P1}, k}^{m}$.  

\paragraph*{Analysis of~$S^{(31)}$} 
Consider~$\hat{S}^{(31)}_{\hat{m}}$. The variables~$X_{k}\define\abs{f_{m,k}}^{2}$ and~$Y_{k,\hat{m}}\define\abs{h_{k,\hat{m}}}^{2}$ are exponentially distributed with parameter~$\lambda=1$. Therefore, we have 
\benn
\Prob\mathopen{}\Bigl\{X_{k}\ge x\Bigr\}=\Prob\mathopen{}\Bigl\{Y_{k,\hat{m}}\ge x\Bigr\}
\le e^{-x},\quad x\ge 0,\ \text{for all } k,\hat{m}.
\eenn
Define $Z_{k,\hat{m}}\define X_{k}Y_{k,\hat{m}}$. From the union bound for products it follows that
\benn
\Prob\mathopen{}\Bigl\{Z_{k,\hat{m}}\ge x^{2}\Bigr\}=\Prob\mathopen{}\Bigl\{X_{k}Y_{k,\hat{m}}\ge x^{2}\Bigr\}\le 2 e^{-x}
\eenn
which yields
\benn
\Prob\mathopen{}\Bigl\{Z_{k,\hat{m}}\ge x\Bigr\} \le 2e^{-\sqrt{x}}.
\eenn
Next, using~$\Ex{Z_{k,\hat{m}}}=1$ and $\Ex{\left(Z_{k,\hat{m}}\right)^{2}}=4$ for all $k, \hat{m}\ne m$ and the independence of the RVs~$Z_{k,\hat{m}}$ across $\natseg{k}{1}{K}$, it follows from Corollary~\ref{cor:trunc1}, taking into account~\eqref{eq:P1Cbound}, that for $K\ge 2$
\benn
\Prob\lefto\{\abs{\hat{S}^{(31)}_{\hat{m}}-\sum_{k=1}^{K}\left(C_{\mathrm{P1}, k}^{m,\hat{m}}\right)^{2}}\ge \sqrt{K} x\right\}\le 6 K e^{-\Delta^{(31)} x^{2/5}} 
\eenn
where~$\Delta^{\!(31)}\define\min\bigl[1, (1/8)\overline{C}^{-4}\bigr]$. Applying the union bound for sums (see Lemma~\ref{lemma:UBsums}) and using~\eqref{eq:P1Cbound}, we finally obtain the desired\footnote{We note that we do not avoid using the union bound on~$S^{(31)}$. It is important, however, that we do not use it when analyzing~$S^{(32)}$.} concentration result for~$S^{(31)}$ as
\bmult
\label{eq:S31Boundu}
\Prob\lefto\{S^{(31)}\ge (M-1)K\overline{C}^{2}+(M-1)\sqrt{K} x\right\}\\
\le 
6 (M-1) K e^{-\Delta^{\!(31)} x^{2/5}}
\emult
and
\bmult
\label{eq:S31Boundl}
\Prob\lefto\{S^{(31)}\le (M-1)K\underline{C}^{2}-(M-1)\sqrt{K} x\right\}\\
\le 
6 (M-1) K e^{-\Delta^{\!(31)} x^{2/5}}.
\emult

\paragraph*{Analysis of~$S^{(32)}$}
We start by rewriting~\eqref{eq:S32} as
\bmult
S^{(32)}=\sqrt{K-1}\\{}\times\sum_{\hat{m}\ne m}\sum_{k=1}^{K}C_{\mathrm{P1}, k}^{m,\hat{m}}\, \conj{\tilde{f}_{p(k),k}}\, f_{m,k}\, \conj{\tilde{h}_{k,p(k)}}\, h_{k,\hat{m}}\, T^{(32)}_{\hat{m},k}
\label{eq:S32withT}
\emult
where~$T^{(32)}_{\hat{m},k}$ is defined as
\benn
T^{(32)}_{\hat{m},k}\define\frac{1}{\sqrt{K-1}}\sum_{\hat{k}\ne k}  C_{\mathrm{P1}, \hat{k}}^{m,\hat{m}}\, \tilde{f}_{p(\hat{k}),\hat{k}}\, \conj{f_{m,\hat{k}}}\, \tilde{h}_{\hat{k},p(\hat{k})}\, \conj{h_{\hat{k},\hat{m}}}.
\eenn

The concentration result for~$S^{(32)}$ (and other similar sums occurring in the proofs of Theorems~\ref{thm:P1Conv} and~\ref{thm:P2Conv}) will be established by applying (one or multiple times) the following general steps:
\begin{itemize}
\item
Establish a concentration result for~$T^{(32)}_{\hat{m},k}$.
\item
Represent the terms on the right-hand side (RHS) of~\eqref{eq:S32withT} in the form~$C_{\mathrm{P1}, k}^{m,\hat{m}}\, Z_{\hat{m},k} \exp(j \hat{\phi}_{k,\hat{m}})$ where
\benn
Z_{\hat{m},k}\define T^{(32)}_{\hat{m},k} \abs{f_{m,k}} \abs{h_{k,\hat{m}}}
\eenn
and 
\benn
\hat{\phi}_{k,\hat{m}}\define\arg\lefto(\conj{\tilde{f}_{p(k),k}}\, f_{m,k}\,  \conj{\tilde{h}_{k,p(k)}}\, h_{k,\hat{m}}\right)
\eenn
so that the sum~$S^{(32)}$ can be written as
\benn
S^{(32)}\define\sqrt{K-1}\sum_{\hat{m}\ne m}\sum_{k=1}^{K} C_{\mathrm{P1}, k}^{m,\hat{m}}\, Z_{\hat{m},k}\, e^{j \hat{\phi}_{k,\hat{m}}}.
\eenn
\item
Use the concentration result for~$T^{(32)}_{\hat{m},k}$ together with the union bound for products (see Lemma~\ref{lemma:UBprods}) to establish bounds on the tail behavior of~$Z_{\hat{m},k}$ and verify condition~\eqref{eq:DecayCond} in Theorem~\ref{lemma:trunc}.
\item
If needed, split up the sum~$S^{(32)}$ into several sums, so that the phases $\exp(j \hat{\phi}_{k,\hat{m}})$ are jointly independent in each of these sums and Theorem~\ref{lemma:trunc} can be applied (to each of these sums separately). 
\item
Finally, apply Theorem~\ref{lemma:trunc} to each of the sums resulting in the previous step separately and use the union bound for sums to establish the desired concentration result for~$S^{(32)}$.
\end{itemize}
    
Following this procedure, we start by deriving a concentration result for~$T^{(32)}_{\hat{m},k}$. Since~$T^{(32)}_{\hat{m},k}$ is of the same nature as~$S^{(2)}$, we could, in principle, use Chernoff bounds. This would, however, lead to an exponent with a complicated dependence on~$t$, which can be simplified only under certain assumptions on~$t$, such as e.g. $t=o\bigl(\sqrt{K}\bigr)$ in~\eqref{eq:LogSeries}. What we need is a simple universal bound for~$\Prob\bigl\{\abs{T^{(32)}_{\hat{m},k}}\ge x\bigr\}$, which is valid for all~$x$ and allows to verify condition~\eqref{eq:DecayCond} in Theorem~\ref{lemma:trunc} for~$Z_{\hat{m},k}$. Such a bound can be obtained by applying the truncation technique to~$T^{(32)}_{\hat{m},k}$ as follows.
Define~$X_{\hat{k}}\define\abs{f_{m,\hat{k}}}$, $Y_{\hat{k},\hat{m}}\define\abs{h_{\hat{k},\hat{m}}}$ and 
\benn
\phi_{\hat{k},\hat{m}}\define\arg\lefto(\tilde{f}_{p(\hat{k}),\hat{k}}\,\conj{f_{m,\hat{k}}}\,  \tilde{h}_{\hat{k},p(\hat{k})}\, \conj{h_{\hat{k},\hat{m}}}\right)
\eenn
so that 
\benn
T^{(32)}_{\hat{m},k}=\frac{1}{\sqrt{K-1}}\sum_{\hat{k}\ne k}  C_{\mathrm{P1}, \hat{k}}^{m,\hat{m}} X_{\hat{k}} Y_{\hat{k},\hat{m}} e^{j \phi_{\hat{k},\hat{m}}}.
\eenn
The RVs~$X_{\hat{k}}$ and~$Y_{\hat{k},\hat{m}}$ (for all $\hat{k}, \hat{m}$) are Rayleigh distributed with parameter~$\alpha^{2}=1/2$. Therefore, we have
\benn
\Prob\mathopen{}\Bigl\{X_{\hat{k}}\ge x\Bigr\}=\Prob\lefto\{Y_{\hat{k},\hat{m}}\ge x\right\}\le e^{-x^{2}}, \qquad x\ge 0
\eenn
and the union bound for products yields
\be
\Prob\mathopen{}\Bigl\{X_{\hat{k}}Y_{\hat{k},\hat{m}}\ge x\Bigr\}\le 2 e^{-x}, \qquad x\ge 0
\label{eq:XYBound1}
\ee
which shows that condition~\eqref{eq:DecayCondCor} in Corollary~\ref{cor:trunc} is satisfied.
Next, rewrite~$\phi_{\hat{k},\hat{m}}$ as
\bmult
\phi_{\hat{k},\hat{m}}=\arg\lefto(\tilde{f}_{p(\hat{k}),\hat{k}}\right)\oplus \arg\lefto(\conj{f_{m,\hat{k}}}\right)\\
\oplus  \arg\lefto(\tilde{h}_{\hat{k},p(\hat{k})}\right)\oplus \arg\lefto(\conj{h_{\hat{k},\hat{m}}}\right)
\label{eq:phasesum}
\emult
where~$\oplus$ stands for addition modulo $2\pi$. Because the $f$'s and the $h$'s in~\eqref{eq:phasesum} are independent across $\natseg{\hat{k}}{1}{K}$, it follows that the phases $\phi_{\hat{k},\hat{m}}$ are also independent across $\natseg{\hat{k}}{1}{K}$, which is precisely what we need for the truncation technique to be applicable. Recalling that $m\ne \hat{m}$, and, therefore, either $p(\hat{k})\ne m$ or $p(\hat{k})\ne \hat{m}$, \eqref{eq:phasesum} implies that $\phi_{\hat{k},\hat{m}}\sim\uniform(-\pi,\pi)$ and hence $\Ex{\exp(j \phi_{\hat{k},\hat{m}})}=0$ for all $\hat{k},\hat{m}$. Since $\phi_{\hat{k},\hat{m}}$ is independent of $X_{\hat{k}}$ and $Y_{\hat{k},\hat{m}}$, we have $\Ex{\exp(j\phi_{\hat{k},\hat{m}})X_{\hat{k}}Y_{\hat{k},\hat{m}}}=0$ for all $\hat{k},\hat{m}$ and hence $\Ex{T^{(32)}_{\hat{m},k}}=0$ for all $\hat{m},k$. 
Finally, applying Corollary~\ref{cor:trunc} to $T^{(32)}_{\hat{m},k}$, taking into account~\eqref{eq:P1Cbound}, we get for $K\ge 2$ and $x\ge 0$ that
\be
\Prob\lefto\{\abs{T^{(32)}_{\hat{m},k}}\ge x\right\} \le 8 (K-1) e^{-\Delta^{\!(T)} x^{2/3}}
\label{eq:T32Bound}
\ee
with $\Delta^{\!(T)}\define 2^{-1/3}\min\lefto[1, (1/2)\overline{C}^{-2}\right]$.

\begin{figure*}[!t]
\normalsize
\setcounter{MYtempeqncnt}{\value{equation}}
\setcounter{equation}{53}
\ba
\label{eq:LP1}
&L_{\mathrm{P1}}(x)\define \frac{\pi^{2}}{16}\frac{\underline{C}^{2}}{\overline{C}_{\mathrm{SN}}^{2}}\frac{K}{M^{3}} \frac{\max\lefto[0,1-\frac{8}{\underline{C} \pi}\frac{M}{\sqrt{K}}x\right]^{2}}{\frac{\overline{C}^{2}}{\overline{C}_{\mathrm{SN}}^{2}}+
\frac{3}{\overline{C}_{\mathrm{SN}}^{2}}\frac{x}{\sqrt{M}}+\frac{\sigma^{2}}{\overline{C}_{\mathrm{SN}}^{2}}\left(\overline{c}^{2}+\frac{x}{\sqrt{K}}\right)+\frac{\sigma^{2}}{\overline{C}^{2}_{\mathrm{SN}}}}\\
\label{eq:UP1}
&U_{\mathrm{P1}}(x)\define \frac{\pi^{2}}{16}\frac{\overline{C}^{2}}{\underline{C}_{\mathrm{SN}}^{2}}\frac{K}{M^{3}} \frac{\left(1+\frac{8}{\overline{C} \pi}\frac{M}{\sqrt{K}}x\right)^{2}}{\max\lefto[0,\frac{\underline{C}^{2}}{\underline{C}_{\mathrm{SN}}^{2}}\frac{M-1}{M}-
\frac{3}{\underline{C}_{\mathrm{SN}}^{2}}\frac{x}{\sqrt{M}}\right]+\max\lefto[0,\frac{\sigma^{2}}{\underline{C}_{\mathrm{SN}}^{2}}\left(\overline{c}^{2}-\frac{x}{\sqrt{K}}\right)\right]+\frac{\sigma^{2}}{\underline{C}^{2}_{\mathrm{SN}}}}\\
\setcounter{equation}{56}
\label{eq:LP2}
&L_{\mathrm{P2}}(x)\define \frac{\pi^{2}}{16}\frac{\underline{C}^{2}}{\overline{C}_{\mathrm{SN}}^{2}}\frac{K}{M^{2}} \frac{\max\lefto[0,1-\frac{8}{\underline{C} \pi}\sqrt{\frac{M}{K}}x\right]^{2}}{\frac{\overline{C}^{2}}{\overline{C}_{\mathrm{SN}}^{2}}+
\frac{4}{\overline{C}_{\mathrm{SN}}^{2}}\frac{x}{\min\left[\sqrt{M},\sqrt{K}\right]}+\frac{\sigma^{2}}{\overline{C}_{\mathrm{SN}}^{2}}\left(\overline{c}^{2}+2 \frac{x}{\sqrt{K}}\right)+\frac{\sigma^{2}}{\overline{C}^{2}_{\mathrm{SN}}}}\\
\label{eq:UP2}
&U_{\mathrm{P2}}(x)\define \frac{\pi^{2}}{16}\frac{\overline{C}^{2}}{\underline{C}_{\mathrm{SN}}^{2}}\frac{K}{M^{3}} \frac{\left(1+\frac{8}{\overline{C} \pi}\sqrt{\frac{M}{K}}x\right)^{2}}{\max\lefto[0,\frac{\underline{C}^{2}}{\underline{C}_{\mathrm{SN}}^{2}}\frac{M-1}{M}-
\frac{4}{\underline{C}_{\mathrm{SN}}^{2}}\frac{x}{\min\left[\sqrt{M},\sqrt{K}\right]}\right]+\max\lefto[0,\frac{\sigma^{2}}{\underline{C}_{\mathrm{SN}}^{2}}\left(\underline{c}^{2}-2\frac{x}{\sqrt{K}}\right)\right]+\frac{\sigma^{2}}{\underline{C}^{2}_{\mathrm{SN}}}}
\ea
\setcounter{equation}{\value{MYtempeqncnt}}
\hrulefill
\vspace*{-5pt}
\end{figure*}

We are now ready to establish the concentration result for~$S^{(32)}$. First, rewrite~$\hat{\phi}_{k,\hat{m}}$ as
\bmult
\hat{\phi}_{k,\hat{m}}\define\arg\lefto(\conj{\tilde{f}_{p(k),k}}\right)
\oplus\arg\lefto(f_{m,k}\right)\\
\oplus\arg\lefto(\conj{\tilde{h}_{k,p(k)}}\right)\oplus \arg\lefto(h_{k,\hat{m}}\right).
\emult
Similar to $\phi_{\hat{k},\hat{m}}$ in~\eqref{eq:phasesum}, because $\hat{m}\ne m$ we conclude that $\hat{\phi}_{k,\hat{m}}\sim \uniform(-\pi,\pi)$. Furthermore, because $\hat{k}\ne k$ the $\hat{\phi}_{k,\hat{m}}$ are independent of $T^{(32)}_{\hat{m},k}$, and therefore also of $Z_{\hat{m},k}$ (for all $k,\hat{m}$).
To apply Corollary~\ref{cor:trunc} to $S^{(32)}$, the $\hat{\phi}_{k,\hat{m}}$ are required to be jointly independent across $\natseg{\hat{m}}{1}{M}$ for $\hat{m}\ne m$ and $\natseg{k}{1}{K}$. 
It can be verified that this is not the case. There is, however, a simple way to resolve this problem by considering the two disjoint index sets
\bann
&I_{1}\define\mathopen{}\Bigl\{\left(\hat{m}, k\right) \Bigl| \nonumber\\&\qquad\qquad  \hat{m}\in[1:M],\ \hat{m}\ne m,\ k\in[1:K],\ p(k)\ne\hat{m}\Bigr\}\nonumber\\
&I_{2}\define\mathopen{}\Bigl\{\left(\hat{m}, k\right) \Bigl| \nonumber\\&\qquad\qquad  \hat{m}\in[1:M],\ \hat{m}\ne m,\ k\in[1:K],\ p(k)=\hat{m}\Bigr\}. 
\eann
It follows by inspection that within each of the  sets $\bigl\{\hat{\phi}_{k,\hat{m}}\bigr\}_{(k,\hat{m})\in I_{1}}$ and $\bigl\{\hat{\phi}_{k,\hat{m}}\bigr\}_{(k,\hat{m})\in I_{2}}$ the phases are jointly independent. Separating $S^{(32)}$ into two sums corresponding to the group of indices~$I_{1}$ and~$I_{2}$, we get
\be
S^{(32)}=S^{(321)}+S^{(322)}
\label{eq:S32toS321S322}
\ee
with
\bann
S^{(321)}&\define\sqrt{K-1}\sum_{\hat{m}\ne m}\sum_{k: p(k)\ne \hat{m}} C_{\mathrm{P1}, k}^{m,\hat{m}}\, Z_{\hat{m},k}\, e^{j \hat{\phi}_{k,\hat{m}}}\nonumber\\
S^{(322)}&\define\sqrt{K-1}\sum_{\hat{m}\ne m}\sum_{k: p(k)=\hat{m}} C_{\mathrm{P1}, k}^{m,\hat{m}}\, Z_{\hat{m},k}\, e^{j \hat{\phi}_{k,\hat{m}}}.
\eann
Applying the union bound for products first to $\abs{f_{m,k}} \abs{h_{k,\hat{m}}}$ as in~\eqref{eq:XYBound1}, then to $Z_{\hat{m},k}$ using~\eqref{eq:T32Bound}, and using the simple bound
\benn
2 e^{-x}+8 (K-1)e^{-\Delta^{\!(T)} x^{1/3}} \le 16 (K-1)e^{-\Delta^{\!(T)} x^{1/3}}
\eenn
which is valid for~$x\ge 1$, we get
\benn
\Prob\lefto\{\abs{Z_{\hat{m},k}}\ge x\right\}\le 16 (K-1) e^{-\Delta^{\!(T)} x^{1/3}}
\eenn
for $K\ge 2$ and $x\ge 1$. Therefore, using $\Ex{Z_{\hat{m},k}\, \exp(j \hat{\phi}_{k,\hat{m}})}=0$ for all $k,\hat{m}\ne m$, applying Corollary~\ref{cor:trunc} to $S^{(321)}$ (which consists of $K(M-1)^{2}/M$ terms) and to $S^{(322)}$ (which consists of~$K(M-1)/M$ terms) separately, taking into account~\eqref{eq:P1Cbound}, we obtain that for $K\ge 2$, $M>2$, and $x\ge 1$
\bmult
\Prob\lefto\{\abs{S^{(321)}}\ge \sqrt{\frac{(K-1)K(M-1)^{2}}{M}}\, x\right\}\\
\le 64\, \frac{(K-1) K (M-1)^{2}}{M}\, e^{-\Delta^{\!(32)} x^{2/7}}
\label{eq:S321Bound}
\emult
and
\bmult
\Prob\lefto\{\abs{S^{(322)}}\ge \sqrt{\frac{(K-1)K(M-1)}{M}}\, x\right\}\\
\le 64\, \frac{(K-1) K (M-1)}{M}\, e^{-\Delta^{\!(32)} x^{2/7}}
\label{eq:S322Bound}
\emult
where $\Delta^{\!(32)}=2^{-10/21}\min\mathopen{}\bigl[1, (1/2)\overline{C}^{-2}\bigr]$.
Combining~\eqref{eq:S4Chernoff} (and similar bounds for $S^{(1)}$ and $S^{(2)}$), \eqref{eq:S321Bound}, \eqref{eq:S322Bound}, \eqref{eq:S32toS321S322}, \eqref{eq:S31Boundu}, \eqref{eq:S31Boundl}, and~\eqref{eq:S3toS31S32}, we can now state the final concentration result for $\mathrm{SINR}^{\mathrm{P1}}_{m}$ by carrying out Step~\ref{step:sum5} in the summary presented in the first paragraph of Section~\ref{sec:convergence}. Recall, however, that we used the classical Chernoff-bounding technique to establish the large-deviations behavior of $S^{(1)}$, $S^{(2)}$, and $S^{(4)}$, whereas we employed the truncation technique to analyze the large-deviations behavior of $S^{(3)}$. 
Even though the Chernoff bounds are tighter than the bounds obtained through the truncation technique, the tightness of the final bounds for the tail behavior of $\mathrm{SINR}^{\mathrm{P1}}_{m}$ and $\mathrm{SINR}^{\mathrm{P2}}_{m}$ is determined by the weakest exponent in the bounds for the individual terms $S^{(1)}, S^{(2)}, S^{(3)}$ and $S^{(4)}$. Therefore, employing Chernoff bounds for $S^{(1)}$, $S^{(2)}$, and $S^{(4)}$ and the truncation technique for~$S^{(3)}$ will not lead to a significantly tighter final result, compared to the case where the truncation technique is used throughout. Motivated by this observation and for simplicity of exposition, we therefore decided to state the concentration results in Section~\ref{sec:concP1P2} for $\mathrm{SINR}^{\mathrm{P1}}_{m}$ and $\mathrm{SINR}^{\mathrm{P2}}_{m}$ obtained by applying the truncation technique throughout.

\subsection{Concentration Results for P1 and P2}
\label{sec:concP1P2}
In Section~\ref{sec:convergence}, we outlined how the large-deviations behavior of the SINR (for P1 and P2) can be established based on the truncation technique and on union bounds. The resulting key statement, made precise in Theorems~\ref{thm:P1Conv} and~\ref{thm:P2Conv} below, is that the probability of the SINR falling outside a narrow interval around its mean is ``exponentially small''. We proceed with the formal statement of the results.
\label{sec:GenConv}
\begin{theorem}
\label{thm:P1Conv}
For any $K\ge 2$, $M\ge 2$, for any $x\ge 1$, the probability $P_{\mathrm{P1}}(x)$ of the event
\benn
\mathrm{SINR}^{\mathrm{P1}}_{m}\notin\left[L_{\mathrm{P1}}(x),U_{\mathrm{P1}}(x)\right],\qquad \natseg{m}{1}{M}
\eenn
where $L_{\mathrm{P1}}(x)$ and $U_{\mathrm{P1}}(x)$ are defined at the top of the page in \eqref{eq:LP1} and \eqref{eq:UP1}, respectively,
with the constants $\overline{C}_{\mathrm{SN}}$ and $\underline{C}_{\mathrm{SN}}$ given by
\addtocounter{equation}{2}
\benn
\overline{C}_{\mathrm{SN}}\define \sqrt{\overline{C}^{2}+\sigma^{2}\!\left(\overline{c}^{2}+1\right)}\qquad \underline{C}_{\mathrm{SN}}\define \sqrt{\underline{C}^{2}+\sigma^{2}\!\left(\underline{c}^{2}+1\right)}
\eenn
satisfies the following inequality
\be
\label{eq:POutP1}
P_{\mathrm{P1}}(x)\le 302\, K^{2} M e^{-\Delta_{\mathrm{P1}}\, x^{2/7}}
\ee
with 
$
\Delta_{\mathrm{P1}}\define \min\mathopen{}\Bigl[2^{-\frac{10}{21}}, 1/\big(2^{\frac{31}{21}} \overline{C}^{\,2}\big), 1/\big(8\, \overline{C}^{\, 4}\big), 1/\big(4\, \overline{c}^{\, 4}\big)
\Bigr].$

\begin{IEEEproof}
See Appendix~\ref{appendix:P1ConvProof}.
\end{IEEEproof}
\end{theorem}

\begin{theorem}
\label{thm:P2Conv}
For any $K\ge 2$, $M\ge 2$, for any $x\ge 1$,  the probability $P_{\mathrm{P2}}(x)$ of the event
\benn
\mathrm{SINR}^{\mathrm{P2}}_{m}\notin\left[L_{\mathrm{P2}}(x),U_{\mathrm{P2}}(x)\right], \qquad \natseg{m}{1}{M}
\eenn%
where $L_{\mathrm{P2}}(x)$ and $U_{\mathrm{P2}}(x)$ are defined at the top of the page in \eqref{eq:LP2} and \eqref{eq:UP2}, respectively,
satisfies the following inequality
\addtocounter{equation}{2}
\be
\label{eq:POutP2}
P_{\mathrm{P2}}(x)\le 814\, K^{2} M^{3} e^{-\Delta_{\mathrm{P2}}\, x^{2/9}}
\ee
with
$\Delta_{\mathrm{P2}}\define \min\mathopen{}\Bigl[2^{-\frac{11}{5}}, 1/\big(2^{\frac{61}{36}} \overline{C}^{\,2}\big), 1/\big(8\, \overline{C}^{\, 4}\big), 1/\big(4\, \overline{c}^{\, 4}\big)
\Bigr].$
\end{theorem}
\begin{IEEEproof}
The proof idea is the same as that underlying the proof of Theorem~\ref{thm:P1Conv} with large parts of the proof itself being very similar to the proof of Theorem~\ref{thm:P1Conv}. For the sake of brevity the details of the proof are therefore omitted.
\end{IEEEproof}

The concentration results in Theorems~\ref{thm:P1Conv} and~\ref{thm:P2Conv} form the basis for showing that, provided the rate of growth of~$K$ as a function of~$M$ is fast enough, the network ``decouples'' (see Theorems~\ref{thm:P1erg} and~\ref{thm:P2erg}) and ``crystallizes'' (see Theorem~\ref{thm:P1ConvOUT}). Moreover, as outlined in Theorem~\ref{thm:P1ConvOUT}, the outage capacity behavior of the $\mathcal{S}_{m}\to\mathcal{D}_{m}$ links can be inferred from~\eqref{eq:POutP1} and~\eqref{eq:POutP2}.

\section{Ergodic Capacity and Cooperation at the Relay Level}
\label{sec:ergodic}
The focus in the previous section was on establishing concentration results for the individual link SINRs for P1 and P2. Based on these results, in this section, we study the ergodic capacity realized by the two protocols and we establish the corresponding capacity scaling and outage capacity behavior. 
\subsection{Ergodic Capacity of P1 and P2}
Throughout this section, we assume that all channels in the network are ergodic.
The two main results are summarized as follows.
\begin{theorem}[Ergodic capacity of P1]
\label{thm:P1erg}
Suppose that destination terminal $\mathcal{D}_m\ \left(\natseg{m}{1}{M}\right)$ has perfect knowledge of the mean of the effective channel gain of the $\mathcal{S}_m\rightarrow\mathcal{D}_m$ link, given by 
$(\pi/(4\sqrt{K}))\!\sum_{k: p(k)=m} C_{\mathrm{P1},k}^{m,m}.$
Then, for any $\epsilon, \delta>0$ there exist $M_0, K_{0}>0$ s.t.\ for all $M\,\ge\,M_0$, $K\ge K_{0}$, the per source-destination terminal 
pair capacity achieved by P1 satisfies
\bmult
\label{eq:P1bounds}
\frac{1}{2}\log\lefto(1+\frac{\pi^{2}}{16}\frac{\underline{C}^{2}}{\overline{C}^{2}_{\mathrm{SN}}}\frac{K}{M^3}(1-\epsilon)\right)\le C_{\mathrm{P1}}\\
 \le 
 \frac{1}{2}\log\lefto(1+\frac{\pi^2}{16}\frac{\overline{C}^{2}}{\underline{C}^{2}_{\mathrm{SN}}}\,\frac{\max\lefto[K,M^{2+\delta}\right]}{M^3}(1+\epsilon)\right).
\emult
\end{theorem}
\begin{theorem}[Ergodic capacity of P2]
\label{thm:P2erg}
Suppose that destination terminal $\mathcal{D}_m\ \left(\natseg{m}{1}{M}\right)$ has perfect knowledge of the mean of the effective channel gain of the $\mathcal{S}_m\rightarrow\mathcal{D}_m$ link, given by $(\pi/(4\sqrt{K M}))\!\sum_{k=1}^{K} C_{\mathrm{P2},k}^{m,m}$. 
Then, for any $\epsilon, \delta>0$ there exist $M_0, K_0>0$, s.t.\ for all $M\,\ge\,M_0$, $K\,\ge\,K_0$, the per source-destination terminal pair capacity 
achieved by P2 satisfies
\bmult
\label{eq:P2bounds}
\frac{1}{2}\log\mathopen{}\left(1+\frac{\pi^2}{16}\frac{\underline{C}^{2}}{\overline{C}^{2}_{\mathrm{SN}}}\,\frac{K}{M^2}(1-\epsilon)\right)\le C_{\mathrm{P2}}\\ 
 \le 
 \frac{1}{2}\log\mathopen{}\left(1+\frac{\pi^2}{16}\frac{\overline{C}^{2}}{\underline{C}^{2}_{\mathrm{SN}}}\,\frac{\max\lefto[K,M^{1+\delta}\right]}{M^2}(1+\epsilon)\right).
\emult
\end{theorem}
 
The proofs of Theorems~\ref{thm:P1erg} and~\ref{thm:P2erg} are very similar. Below we present the proof of Theorem~\ref{thm:P1erg} only. The proof of Theorem~\ref{thm:P2erg} is omitted.

\begin{figure*}[!t]
\normalsize
\setcounter{MYtempeqncnt}{\value{equation}}
\setcounter{equation}{66}
\ba
\label{eq:BN}
B^{N}(M,K,x) &\define \frac{K}{\max[K,M^{2+\delta}]}\left(1+A_{1}\frac{M}{\sqrt{K}}x\right)^{2}\\
\label{eq:BD}
B^{D}(M,K,x) &\define \left.\Biggl(\underline{C}^{2}\max\lefto[0,\frac{M-1}{M}-\frac{A_{2} x}{\underline{C}^{2} \sqrt{M}}\right]+\underline{c}^{2}\sigma^{2}\max\lefto[0,1-\frac{x}{\underline{c}^{2}\sqrt{K}}\right]+\sigma^{2}\Biggr)\right/\underline{C}^{2}_{\mathrm{SN}}
\ea
\setcounter{equation}{\value{MYtempeqncnt}}
\hrulefill
\vspace*{-5pt}
\end{figure*}

\begin{IEEEproof}[Proof of Theorem~\ref{thm:P1erg}]
We start by establishing the lower bound in~\eqref{eq:P1bounds}, the proof of which uses the result summarized in Appendix~\ref{appendix:Medard}. To apply Lemma~\ref{lemma:Medard} in Appendix~\ref{appendix:Medard}, 
we start from~\eqref{eq:P1SISO} and define
\bann
\bar{F}_{m}&\define \frac{1}{\sqrt{K}}\sum_{k=1}^{K} \Ex{a_k^{m,m}}\\
\tilde{F}_{m}&\define \frac{1}{\sqrt{K}}\sum_{k=1}^{K} \left(a_k^{m,m}-\Ex{a_k^{m,m}}\right)\\
W_{m}&\define \sum_{\hat{m}\ne m} s_{\hat{m}}\frac{1}{\sqrt{K}}\sum_{k=1}^K a_k^{m,\hat{m}}+
\frac{1}{\sqrt{K}}\sum_{k=1}^{K} b_k^{m} z_k+w_m.
\eann
With these definitions, we can now rewrite~\eqref{eq:P1SISO} as
\benn
y_{m}=\left(\bar{F}_{m}+\tilde{F}_{m}\right) s_{m}+W_{m}.
\eenn
Straightforward, but tedious, manipulations yield
\bsann
\bar{F}_{m}&=&\frac{\pi}{4} \frac{1}{\sqrt{K}}\sum_{k: p(k)=m} C_{\mathrm{P1},k}^{m,m}\\
\Var{\tilde{F}_{m}}&=&\frac{1}{K}\left(\sum_{k=1}^{K} \left(C_{\mathrm{P1},k}^{m,m}\right)^{2}-\frac{\pi^{2}}{16}\sum_{k:p(k)=m} \left(C_{\mathrm{P1},k}^{m,m}\right)^{2}\right)\\
\Var{W_{m}}&=&\frac{1}{K M} \sum_{\hat{m}\ne m}\sum_{k=1}^{K} \left(C_{\mathrm{P1},k}^{m,\hat{m}}\right)^{2}
\\&&\qquad\qquad\qquad\qquad 
{+}\:\frac{\sigma^{2}}{K}\sum_{k=1}^{K} \left(C_{\mathrm{P1},k}^{m}\right)^{2}+\sigma^{2}.%
\esann
Next, we use~\eqref{eq:P1Cbound} and~\eqref{eq:P1cbound} to lower-bound~$\bar{F}_{m}$ and upper-bound~$\Var{\tilde{F}_{m}}$ and~$\Var{W_{m}}$, substitute the resulting bounds into~\eqref{eq:Medard}, and obtain\footnote{We note that this bound is valid for arbitrary~$M$ and~$K$ and is, therefore, somewhat stronger than the asymptotic bound we are actually seeking.}
\be
\label{eq:Th3MIbound}
I(y_{m};s_{m})\ge \frac{1}{2}\log\lefto(1+\frac{\pi^{2}}{16}\frac{\underline{C}^{2}}{(1/M)\overline{C}^{2}+\overline{C}^{2}_{\mathrm{SN}}}\frac{K}{M^{3}}\right).
\ee
Finally, fix~$\epsilon>0$ and set
\benn
M_{0}=\frac{1-\epsilon}{\epsilon} \frac{\overline{C}^{2}}{\overline{C}^{2}_{\mathrm{SN}}}.
\eenn
It then follows that for any~$M\ge M_{0}$, the inequality
\benn
\frac{\underline{C}^{2}}{(1/M)\overline{C}^{2}+\overline{C}^{2}_{\mathrm{SN}}} \ge \frac{\underline{C}^{2}}{\overline{C}^{2}_{\mathrm{SN}}} (1-\epsilon)
\eenn
is satisfied, which together with~\eqref{eq:Th3MIbound} completes the proof of the lower bound.

Proving the upper bound on~$C_{\mathrm{P1}}$ in~\eqref{eq:P1bounds} turns out to be significantly more challenging. The method we use to this end is based 
on the concentration result for~$\mathrm{SINR}^{\mathrm{P1}}_{m}$ in Theorem~\ref{thm:P1Conv}. 
We start by noting that the per-stream ergodic capacity can be upper-bounded by assuming that~$\mathcal{D}_{m}$ has perfect knowledge of~$\matH$ and~$\matF$, i.e., 
\bann
C^{\mathrm{P1}}_{m}&\le\frac{1}{2}\Exop_{\matH,\matF}\lefto[I(y_{m};s_{m}\given \matH,\matF)\right]\nonumber\\
&=\frac{1}{2}\Exop_{\matH,\matF}\lefto[\log\left(1+\mathrm{SINR}^{\mathrm{P1}}_{m}\right)\right]\nonumber\\
&\le \frac{1}{2}\log\lefto(1+\Exop_{\matH,\matF}\lefto[\mathrm{SINR}^{\mathrm{P1}}_{m}\right]\right)
\eann
where the last step follows from Jensen's inequality.

Now fix~$\epsilon>0$. To prove the upper bound in~\eqref{eq:P1bounds}, it suffices to show that there exist~$M_{0},K_{0}>0$ s.t.\ for all $M\ge M_{0}$ and $K\ge K_{0}$
\benn
\Exop_{\matH,\matF}\lefto[\mathrm{SINR}^{\mathrm{P1}}_{m}\right]\le A \frac{\max[K,M^{2+\delta}]}{M^{3}}(1+\epsilon)
\eenn
where we define
\benn
A\define \frac{\pi^{2}}{16}\frac{\overline{C}^{2}}{\underline{C}^{2}_{\mathrm{SN}}}.
\eenn
To simplify the exposition, we define
\benn
g(M,K)\define\frac{1}{A} \mathrm{SINR}^{\mathrm{P1}}_{m}(M,K)\frac{M^3}{\max[K,M^{2+\delta}]}.
\eenn
Note that we make the dependence of $\mathrm{SINR}^{\mathrm{P1}}_{m}$ on~$M$ and~$K$ explicit by using the notation $\mathrm{SINR}^{\mathrm{P1}}_{m}(M,K)$. 
In the remainder of the proof, we show that
\be
\label{eq:ToShowP1}
\Exop_{\matH,\matF}\lefto[g(M,K)\right]\le 1+\epsilon
\ee
for~$M$ and~$K$ large enough. 
Let $\pdf{g}(x)$ denote the pdf of $g(M,K)$. Then, the expectation $\Exop_{\matH,\matF}\lefto[g(M,K)\right]$ can be written as
\ba
\Exop_{\matH,\matF}\lefto[g(M,K)\right]&=\int_0^\infty t\, \pdf{g}(t)dt\nonumber
\\&=\int_0^{1+\epsilon_1} t\, \pdf{g}(t)dt+\int_{1+\epsilon_1}^\infty t\, \pdf{g}(t)dt
\label{eq:Into2Int}
\ea
where~$\epsilon_1>0$ is chosen s.t.\
\benn
1+\epsilon_1<1+\epsilon/3.
\eenn
Consequently, we have
\ba
\label{eq:B1ErgFinalP1}
\int_0^{1+\epsilon_1} t\, \pdf{g}(t)dt &\le \left(1+\epsilon_{1}\right) \int_0^{1+\epsilon_1} \pdf{g}(t)dt\nonumber\\
&\le 1+\epsilon_{1}< 1+\epsilon/3.
\ea

For bounding the second integral on the RHS of~\eqref{eq:Into2Int}, it is convenient to write the upper bound in Theorem~\ref{thm:P1Conv} in the following form:
there exist $\Delta>0$, $\delta_{1}>0$, $\delta_{2}>0$, and $A_{1}, A_{2}, A_{3}>0$ such that for any $x\ge 1$ and $M, K\ge 2$
\be
\label{eq:gProb}
\Prob\mathopen{}\Bigl\{g(M,K)\ge B(M,K,x)\Bigr\}\le A_{3} M^{\delta_{1}} K^{\delta_{2}} e^{-\Delta x^{2/7}}
\ee 
with
\benn
B(M,K,x)\define \frac{B^{N}(M,K,x)}{B^{D}(M,K,x)}
\eenn%
\addtocounter{equation}{2}%
where $B^{N}(M,K,x)$ and $B^{D}(M,K,x)$ are defined at the top of the page in \eqref{eq:BN} and \eqref{eq:BD}, respectively.
The second integral on the RHS of~\eqref{eq:Into2Int} will be shown, for~$M$ and~$K$ large enough, to be upper bounded by~$2\epsilon/3$ by splitting it up and proving that
\be
\int_{1+\epsilon_1}^{\lceil t_{0}\rceil} t\, \pdf{g}(t)dt\le \epsilon/3
\label{eq:int1P1}
\ee
and
\be
\int_{\lceil t_{0}\rceil }^\infty t\, \pdf{g}(t)dt\le \epsilon/3
\label{eq:int2P1}
\ee
where the parameter~$t_{0}>1+\epsilon_{1}$, independent of~$M,K$, will be chosen later. It will become clear later why we need to split up the second integral on the RHS of~\eqref{eq:Into2Int} according to~\eqref{eq:int1P1} and~\eqref{eq:int2P1}.
The integral in~\eqref{eq:int1P1} can be bounded as follows
\bann
\int_{1+\epsilon_1}^{\lceil t_{0}\rceil} t\, \pdf{g}(t)dt &\le \lceil t_{0}\rceil \int_{1+\epsilon_1}^{\lceil t_{0}\rceil } \pdf{g}(t)dt
\nonumber\\& 
\le \lceil t_{0}\rceil \Prob\mathopen{}\Bigl\{g(M,K)\ge 1+\epsilon_{1}\Bigr\}.
\eann
Set $x(M)=\bigl(\min\mathopen{}\bigl[\sqrt{M},M^{\delta}\bigr]\bigr)^{1/3}$. With this choice of $x(M)$, it is not difficult to show that
\bann
&\lim_{M,K\to\infty} A_{1}\frac{M\, x(M)}{\sqrt{\max\lefto[K,M^{2+\delta}\right]}} =0\nonumber\\
&\lim_{M,K\to\infty} A_{2}\frac{x(M)}{\underline{C}^{2}\sqrt{M}}=0\nonumber\\
&\lim_{M,K\to\infty} \frac{x}{\underline{c}^{2}\sqrt{K}}=0
\eann
which yields
\be
\lim_{M,K\to\infty} B^{N}(M,K,x(M))=\lim_{M,K\to\infty} \frac{K}{\max[K,M^{2+\delta}]}\le 1.
\label{eq:BNLim}
\ee
Using $\underline{C}^{2}_{\mathrm{SN}}=\underline{C}^{2}+\sigma^{2}\!\left(\underline{c}^{2}+1\right)$, we can furthermore conclude that
\benn
\lim_{M,K\to\infty} B^{D}(M,K,x(M))=1
\eenn
which, together with~\eqref{eq:BNLim}, implies that
\benn
\lim_{M,K\to\infty} B(M,K,x(M))\le 1.
\eenn
We can, therefore, conclude that there exist~$M_{0}^{(11)}, K_{0}^{(11)}>0$ s.t.\ for any~$M\ge M_{0}^{(11)}$ and $K\ge K_{0}^{(11)}$
\be
B(M,K,x(M))\le 1+\epsilon_{1}.
\label{eq:B1P1}
\ee
Trivially, we have
\benn
\lim_{M,K\to\infty} M^{\delta_{1}} K^{\delta_{2}} e^{-\Delta \left(x(M)\right)^{2/7}}=0
\eenn
and, therefore, there exist $M_{0}^{(12)}$, $K_{0}^{(12)}>0$ s.t.\ for any $M\ge M_{0}^{(12)}$ and $K\ge K_{0}^{(12)}$
\be
A_{3}M^{\delta_{1}} K^{\delta_{2}} e^{-\Delta \left(x(M)\right)^{2/7}}\le \frac{\epsilon}{3 \lceil t_{0}\rceil}.
\label{eq:ProbP1}
\ee
Combining~\eqref{eq:B1P1} and~\eqref{eq:ProbP1} and setting 
\benn
M_{0}^{(1)}=\max[M_{0}^{(11)},M_{0}^{(12)}],\quad K_{0}^{(1)}=\max[K_{0}^{(11)}\!,K_{0}^{(12)}]
\eenn
we get that for any $M\ge M_{0}^{(1)}$ and $K\ge K_{0}^{(1)}$
\be
\lceil t_{0}\rceil \Prob\mathopen{}\Bigl\{g(M,K)\ge 1+\epsilon_{1}\Bigr\}\le \epsilon/3
\label{eq:B2ErgFinalP1}
\ee
which concludes the proof of~\eqref{eq:int1P1}. 

To show~\eqref{eq:int2P1}, we note that
\be
\int_{\lceil t_{0}\rceil }^\infty t\, \pdf{g}(t)dt\le \sum_{n=\lceil t_{0}\rceil}^{\infty} (n+1) \Prob\mathopen{}\Bigl\{g(M,K)\ge n\Bigr\}\define S.
\label{eq:SumI2P1}
\ee
Expanding the square, upper-bounding $x$ by $x^{2}$ in $B^{N}(M,K,x)$ and substituting the $\max$ terms in $B^{D}(M,K,x)$ by~$0$, we obtain the bound
\bmult
\label{eq:P1BU}
B(M,K,x)\\
\le \frac{K}{\max[K,M^{2+\delta}]}\frac{\underline{C}^{2}_{\mathrm{SN}}}{\sigma^{2}} \left(1+\left(2 A_{1}\frac{M}{\sqrt{K}}+A_{1}^{2}\frac{M^{2}}{K}\right) x^{2}\right)
\\\define B_{1}(M,K,x^{2}).
\emult
Applying the change of variables~$y=x^{2}$ in~\eqref{eq:P1BU} and~\eqref{eq:gProb}, we finally get
\bmult
\Prob\Bigl\{g(M,K)\ge B_{1}(M,K,\sqrt{y})\Bigr\}\\
\le
\Prob\Bigl\{g(M,K)\ge B(M,K,\sqrt{y})\Bigr\}\\\le A_{3} M^{\delta_{1}} K^{\delta_{2}} e^{-\Delta y^{1/7}}.
\emult
Equating $B_{1}(M,K,y)$ with~$n$ and solving for~$y$, we find that
\benn
\Prob\mathopen{}\Bigl\{g(M,K)\ge n\Bigr\}\le A_{3} M^{\delta_{1}} K^{\delta_{2}}e^{-\Delta \left(y_{2}(n,M,K)\right)^{1/7}}
\eenn
with
\be
y_{2}(n,M,K)=\frac{\frac{\max[K,M^{2+\delta}]}{K}\left(\frac{\sigma^{2}}{\underline{C}^{2}_{\mathrm{SN}}}n-\frac{K}{\max[K,M^{2+\delta}]}\right)}{2 A_{1}\frac{M}{\sqrt{K}}+A_{1}^{2}\frac{M^{2}}{K}}.
\label{eq:y2}
\ee
Now,~$S$ defined in~\eqref{eq:SumI2P1} can be upper-bounded as
\ba
S&\le 2\sum_{n=\lceil t_{0}\rceil}^{\infty} n \Prob\lefto\{g(M,K)\ge n\right\}
\nonumber\\&\le
2 A_{3} M^{\delta_{1}} K^{\delta_{2}}\sum_{n=\lceil t_{0}\rceil}^{\infty} n e^{-\Delta \left(y_{2}(n,M,K)\right)^{1/7}}.
\label{ea:SumFromT}
\ea
If~$n$ is s.t.\ $\sigma^{2} n/\underline{C}^{2}_{\mathrm{SN}}>1$, then the expression in the parentheses in the numerator of~\eqref{eq:y2} is strictly positive and it follows that $\lim_{M,K\to\infty} y_{2}(n,M,K)=\infty$. Therefore, if~$t_{0}$ is chosen s.t.\ $\lceil t_{0}\rceil>\underline{C}^{2}_{\mathrm{SN}}/\sigma^{2}$, each term in the sum in~\eqref{ea:SumFromT} goes to zero exponentially fast in~$M,K$. Note that the split-up in~\eqref{eq:int1P1} and~\eqref{eq:int2P1} was needed to be able to choose~$t_{0}$ large enough here. To simplify the exposition in the following, we set $t_{0}=\left(2^{7}+1\right) \underline{C}^{2}_{\mathrm{SN}}/\sigma^{2}$, so that 
\benn
\left(\frac{\sigma^{2}}{\underline{C}^{2}_{\mathrm{SN}}}n-\frac{K}{\max[K,M^{2+\delta}]}\right)^{1/7}\ge 2
\eenn 
for $n\ge \lceil t_{0}\rceil$.
Next, we note that  
\benn
\lim_{M,K\to\infty} \frac{\max[K,M^{2+\delta}]}{K}\frac{1}{2 A_{1}\frac{M}{\sqrt{K}}+A_{1}^{2}\frac{M^{2}}{K}}=\infty
\eenn
so that there exist $M_{0}^{(2)}, K_{0}^{(2)}>0$ s.t.\ for any~$M\ge M_{0}^{(2)}$ and~$K\ge K_{0}^{(2)}$
\benn
\left(\frac{\max[K,M^{2+\delta}]}{K}\frac{1}{2 A_{1}\frac{M}{\sqrt{K}}+A_{1}^{2}\frac{M^{2}}{K}}\right)^{1/7}\ge 2.
\eenn
Now using that, trivially,
\benn
x y\ge x+y
\eenn
for $x,y \ge 2$, we have for any $M\ge M_{0}^{(2)}$, $K\ge K_{0}^{(2)}$ and $n\ge \lceil t_{0}\rceil$
\bmnn
\left(y_{2}(n,M,K)\right)^{1/7}\ge \left(\frac{\sigma^{2}}{\underline{C}^{2}_{\mathrm{SN}}}n-\frac{K}{\max[K,M^{2+\delta}]}\right)^{1/7}\\
+\left(\frac{\max[K,M^{2+\delta}]}{K}\frac{1}{2 A_{1}\frac{M}{\sqrt{K}}+A_{1}^{2}\frac{M^{2}}{K}}\right)^{1/7}
\emnn
which yields 
\bann
S\le 2 A_{3} M^{\delta_{1}} K^{\delta_{2}} 
e^{-\Delta\left(\frac{2A_{1}M\sqrt{K}+A_{1}^{2} M^{2}}{\max[K,M^{2+\delta}]}\right)^{-1/7}}\!\!
\sum_{n=\lceil t_{0}\rceil}^{\infty} h(n)
\eann
with
\benn
h(n)\define n \exp\lefto(-\Delta \left(\frac{\sigma^{2}}{\underline{C}^{2}_{\mathrm{SN}}} n-1\right)^{1/7}\right).
\eenn
Clearly,~$h(n)$ decays fast enough for~$
\sum_{n=\lceil t_{0}\rceil}^{\infty} h(n)$ to converge to a finite limit, in other words, there exists a constant~$C<\infty$ (independent of~$M, K$) s.t.\
\be
\label{eq:SumBoundP1}
\sum_{n=\lceil t_{0}\rceil}^{\infty} h(n)\le C.
\ee
Moreover, it is easily seen that
\benn
\lim_{M,K\to\infty} M^{\delta_{1}} K^{\delta_{2}} e^{-\Delta\left(\frac{2A_{1}M\sqrt{K}+A_{1}^{2} M^{2}}{\max[K,M^{2+\delta}]}\right)^{-1/7}}=0
\eenn
which, together with~\eqref{eq:SumBoundP1}, shows that~$S$ can be made arbitrarily small by choosing~$M$ and~$K$ large enough. More specifically, there exist~$M_{0}^{(3)}, K_{0}^{(3)}>0$ s.t.\ for any~$M\ge M_{0}^{(3)}$ and~$K\ge K_{0}^{(3)}$ 
\be
S\le \epsilon/3. 
\label{eq:B3ErgFinalP1}
\ee
Taking 
\bann
M_{0}&\define \max\mathopen{}\bigl[M_{0}^{(1)}, M_{0}^{(2)}, M_{0}^{(3)}\bigr]\\
K_{0}&\define \max\mathopen{}\bigl[K_{0}^{(1)}, K_{0}^{(2)}, K_{0}^{(3)}\bigr]
\eann
and combining~\eqref{eq:B1ErgFinalP1}, \eqref{eq:B2ErgFinalP1}, and~\eqref{eq:B3ErgFinalP1}, we have shown~\eqref{eq:ToShowP1}, which completes the proof.
\end{IEEEproof}

\subsection{The ``Crystallization'' Phenomenon}
\label{sec:crystallization}
As pointed out in the introduction, the ``crystallization'' phenomenon occurs for $M,K\,\rightarrow\,\infty$, provided that~$K$ scales fast enough as a function of~$M$, and manifests itself in two effects, namely, the {\em decoupling} of the 
individual ${\cal S}_{m}\,\rightarrow\,{\cal D}_{m}$ links
and the {\em convergence of each of the resulting SISO links to a nonfading link}.  

\subsubsection{Decoupling of the network} 
Theorems~\ref{thm:P1erg} and~\ref{thm:P2erg} show that in the $M,K\to\infty$ limit, the per-source destination terminal pair capacity scales 
as $C_{\mathrm{P1}}=(1/2)\log\lefto(1+\Theta\lefto(K/M^{3}\right)\right)$ in P1 
and $C_{\mathrm{P2}}=(1/2)\log\lefto(1+\Theta\lefto(K/M^{2}\right)\right)$
in P2. We can, therefore, conclude that if $K\propto M^{3+\alpha}$ in P1 and $K\propto M^{2+\alpha}$ in P2 with $\alpha\ge 0$,   
apart from the factor $1/2$, which is due to the use of two time slots, P1 and P2 achieve full spatial multiplexing gain~\cite{tse05} (i.e., full sum-capacity pre-log) without any cooperation of the terminals in the network, not even the destination terminals. The corresponding distributed array gain (i.e., the factor inside the log) is given by~$M^{\alpha}$ in both cases. 

The fact that the per source-destination terminal pair capacity is strictly positive when $K$ scales at least
as fast as~$M^{3}$ in P1 and at least as fast as $M^{2}$ in P2 shows that the individual ${\cal S}_{m}\,\rightarrow\,{\cal D}_{m}$ links in the network 
``decouple'' in the sense that the SINR is strictly positive for each of the links.
Note that this does not imply that the interference at the ${\cal D}_{m}$ (created by $s_{\hat{m}}$ with $\hat{m}\,\neq\,m$) vanishes. 
Rather, if~$K$ scales fast enough, the signal power starts dominating the interference (plus noise) power. 
Since both upper and lower bounds in Theorems~\ref{thm:P1erg} and~\ref{thm:P2erg} exhibit the same scaling behavior, the $K\propto M^{3}$ and $K\propto M^{2}$, respectively, 
thresholds are fundamental in the sense of defining the critical scaling rate by delineating the regime where interference dominates over the signal and hence drives the per source-destination terminal
pair capacity to zero from the regime where the signal dominates the interference and the per source-destination terminal pair capacity is strictly
positive. Further inspection of the upper and lower bounds in~\eqref{eq:P1bounds} and~\eqref{eq:P2bounds} reveals that, for fixed~$\epsilon>0$, unless all path-loss and shadowing coefficients $E_{k,m}$ and $P_{m,k}$ $\left(\natseg{k}{1}{K}, \natseg{m}{1}{M}\right)$ are equal and hence $\overline{C}^{2}=\underline{C}^{2}$ and $\overline{C}_{\mathrm{SN}}^{2}=\underline{C}_{\mathrm{SN}}^{2}$, there is a gap (apart from that due to~$\epsilon>0$) between the bounds.  

The order-of-magnitude reduction in the threshold for critical scaling in P2, when compared with P1, comes at the cost of each relay having to know all $M$ backward and $M$ forward channels. 
We can, therefore, conclude that P1 and P2 trade off the number of relay terminals for channel knowledge at the relays.

Finally, it is worthwhile to 
point out that in contrast to the finite-$M$ results for P1 in~\cite{bolcskei06-}, the destination terminals $\mathcal{D}_m$ do not need
knowledge of the fading coefficients $h_{k,m}$ and $f_{m,k}$. This can be seen by noting that the quantity $\bigl(\pi/(4\sqrt{K})\bigr)\sum_{k:p(k)=m} C^{m,m}_{\mathrm{P1},k}$,
which has to be known at ${\cal D}_{m}$, depends on $E_{k,m}$, $P_{m,k}$, $K$, and $M$ only. Moreover, the coefficient $\bigl(\pi/(4\sqrt{K})\bigr)\sum_{k:p(k)=m} C^{m,m}_{\mathrm{P1},k}$ can easily be acquired 
through training. 

\subsubsection{Convergence to nonfading links and ``crystallization''}
\label{sec:ConvInterpretation}
When the network decouples, it is interesting to ask how the
decoupled SISO links behave (in terms of their fading statistics) when $M$ and $K$ grow large. The answer to this question follows from the
concentration results in Theorems~\ref{thm:P1Conv} and~\ref{thm:P2Conv}, which can be reformulated to establish upper bounds on the outage probability for the individual $\mathcal{S}_{m}\to\mathcal{D}_{m}$ links.
For the sake of brevity, we focus on P1 in what follows. The goal is to arrive at a statement regarding
\bann
P_{\mathrm{out},\mathrm{P1}}(R)&=\Prob\lefto\{\frac{1}{2}\log\lefto(1+\mathrm{SINR}^{\mathrm{P1}}_{m}\right)\le R\right\}\nonumber\\
&=\Prob\lefto\{\mathrm{SINR}^{\mathrm{P1}}_{m}\le 2^{2 R}-1\right\}.
\eann 
The corresponding result is summarized in
\begin{theorem}[Outage probability for P1]\strut
\label{thm:P1ConvOUT}
\begin{enumerate}
\item
Assume that $K\ge 2$, $M\ge 2$, and $R\ge 0$ are s.t.\
\ba
x(R)&=\frac{1-e_{\mathrm{P1}}(M,K,R)}{\frac{16}{\underline{C}\pi}\frac{M}{\sqrt{K}}+e_{\mathrm{P1}}(M,K,R)\!\!\left(\!\frac{3}{\overline{C}_{\mathrm{SN}}^{2}}\frac{1}{\sqrt{M}}+\frac{\sigma^{2}}{\overline{C}_{\mathrm{SN}}^{2}}\frac{1}{\sqrt{K}}\!\right)}\nonumber\\
&\ge 1
\label{eq:XRDef}
\ea
where
\benn
e_{\mathrm{P1}}(M,K,R)=\frac{16}{\pi^{2}}\frac{\overline{C}_{\mathrm{SN}}^{2}}{\underline{C}^{2} }\frac{M^{3}}{K}\left(2^{2R}-1\right).
\eenn
Then, the individual link outage probability is upper-bounded as 
\be
P_{\mathrm{out},\mathrm{P1}}(R)\le 151\, K^{2} M e^{-\Delta_\mathrm{P1}\, x(R)^{2/7}}.
\label{eq:crystallrate}
\ee
\item
Under the same conditions on~$K,M$ and~$R$ as in 1), for any~$\epsilon, \delta>0$, $K\ge M^{3+\delta}$, and
\be
R\le \frac{1}{2}\log\lefto(1+\frac{\pi^{2}}{16}\frac{\underline{C}^{2}}{\overline{C}_{\mathrm{SN}}^{2}}\frac{K}{M^{3}} (1-\epsilon)\right),
\label{eq:RBoundErg}
\ee
we have
\benn
P_{\mathrm{out},\mathrm{P1}}(R)
\le\lim_{M,K\to\infty} 151\, K^{2} M e^{-\Delta_\mathrm{P1}\, x(R)^{2/7}}=0.
\eenn
\end{enumerate}
\end{theorem}
\begin{IEEEproof}
We start with the proof of statement~1).
Recall that Theorem~\ref{thm:P1Conv} provides us with a parametric upper bound on $\Prob\lefto\{\mathrm{SINR}^{\mathrm{P1}}_{m}\le L_{\mathrm{P1}}(x)\right\}$ 
with~$L_{\mathrm{P1}}(x)$ defined in~\eqref{eq:LP1}.
Assuming that
\be
x\le \frac{\underline{C}\pi \sqrt{K}}{16 M}
\label{eq:Xbound}
\ee
and using $\overline{C}_{\mathrm{SN}}^{2}=\overline{C}^{2}+\sigma^{2}\!\left(\overline{c}^{2}+1\right)$, we can lower-bound $L_{\mathrm{P1}}(x)$ as 
\benn
L_{\mathrm{P1}}(x)\ge \frac{\pi^{2}}{16}\frac{\underline{C}^{2}}{\overline{C}_{\mathrm{SN}}^{2}}\frac{K}{M^{3}} \frac{1-\frac{16}{\underline{C} \pi}\frac{M}{\sqrt{K}}x}{1+
\frac{3}{\overline{C}_{\mathrm{SN}}^{2}}\frac{x}{\sqrt{M}}+\frac{\sigma^{2}}{\overline{C}_{\mathrm{SN}}^{2}}\frac{x}{\sqrt{K}}}\define L_{\mathrm{P1}}'(x).
\eenn
Solving
\be
2^{2 R}-1=L_{\mathrm{P1}}'(x)
\label{eq:solutionXR}
\ee
for~$x(R)$ yields~\eqref{eq:XRDef}, which, by assumption, satisfies~$x(R)\ge 1$.
With
\benn
\Prob\mathopen{}\Bigl\{\mathrm{SINR}^{\mathrm{P1}}_{m}\le L_{\mathrm{P1}}'(x)\Bigr\}\le \Prob\mathopen{}\Bigl\{\mathrm{SINR}^{\mathrm{P1}}_{m}\le L_{\mathrm{P1}}(x)\Bigr\}
\eenn
we can now apply\footnote{Strictly speaking, one needs to use the upper bounds on $\Prob\lefto\{\mathrm{SINR}^{\mathrm{P1}}_{m}\le L_{\mathrm{P1}}(x)\right\}$ derived in the last paragraph of Appendix~\ref{appendix:P1ConvProof}.}  Theorem~\ref{thm:P1Conv} to obtain
\be
P_{\mathrm{out},\mathrm{P1}}(R)\le 151\, K^{2} M e^{-\Delta_{\mathrm{P1}}\, x(R)^{2/7}}.
\label{eq:PoutBB}
\ee

Finally, we note that~$x(R)$ in~\eqref{eq:XRDef} is trivially seen to satisfy~\eqref{eq:Xbound}. This concludes the proof of statement 1).

The proof of statement 2) is obtained by establishing a sufficient condition on~$x(R)$, for any~$R\ge 0$, to grow with increasing~$M$ (and by~$K\ge M^{3+\delta}$ with increasing~$K$). Using~\eqref{eq:XRDef}, it is easily verified that guaranteeing  
\benn
0\le e_{\mathrm{P1}}(M,K,R)\le 1-\epsilon
\eenn
for some~$0<\epsilon<1$ (independent of~$M,K$) provides such a condition. The final result is now obtained by solving 
\benn
e_{\mathrm{P1}}(M,K,R)=\frac{16}{\pi^{2}}\frac{\overline{C}_{\mathrm{SN}}^{2}}{\underline{C}^{2} }\frac{M^{3}}{K}\left(2^{2R}-1\right)\le 1-\epsilon
\eenn 
for~$R$.
\end{IEEEproof}

The implications of Theorem~\ref{thm:P1ConvOUT} are significant: For any transmission rate~$R$ less than the ergodic capacity (in the case $E_{k,m}=P_{m,k}$ for all $k,m$) or the ergodic capacity lower bound in Theorem~\ref{thm:P1erg} (in the case of general $E_{k,m}$ and $P_{m,k}$), the outage probability of each of the decoupled links goes to zero exponentially fast in the number of nodes in the network, provided~$K$ scales supercritically in~$M$. We have thus shown that choosing the rate of growth of~$K$ fast enough for the network to decouple automatically guarantees that the decoupled SISO links converge to nonfading links. 
Equivalently, we can say that each of the decoupled links experiences a distributed spatial diversity (or, more precisely, relay diversity) order that goes to infinity as $M\,\rightarrow\,\infty$.
Consequently, in the large-$M$ limit time diversity (achieved by coding over a sufficiently long time horizon) is not needed
to achieve ergodic capacity.
We say that the network ``crystallizes'' as it breaks up into a set of effectively isolated ``wires in the air''. From~\eqref{eq:crystallrate}, we can furthermore infer the ``crystallization'' rate, i.e., the rate (as a function of~$M$ and~$K$) at which the individual $\mathcal{S}_{m}\to\mathcal{D}_{m}$ links converge to nonfading links. We note, however, that the exponent~$2/7$ (and~$2/9$ for P2) is unlikely to be fundamental as it is probably a consequence of the application of the truncation technique. In this sense, we can only specify a guaranteed crystallization rate. We conclude by noting that the upper bound~\eqref{eq:PoutBB} (as well as the corresponding result for P2) tend to be rather loose. This is probably a consequence of the truncation technique and the use of union bounds to characterize the large-deviations behavior of the individual link SINR RVs.

\begin{figure}
\centering
\includegraphics[width=\figwidth]{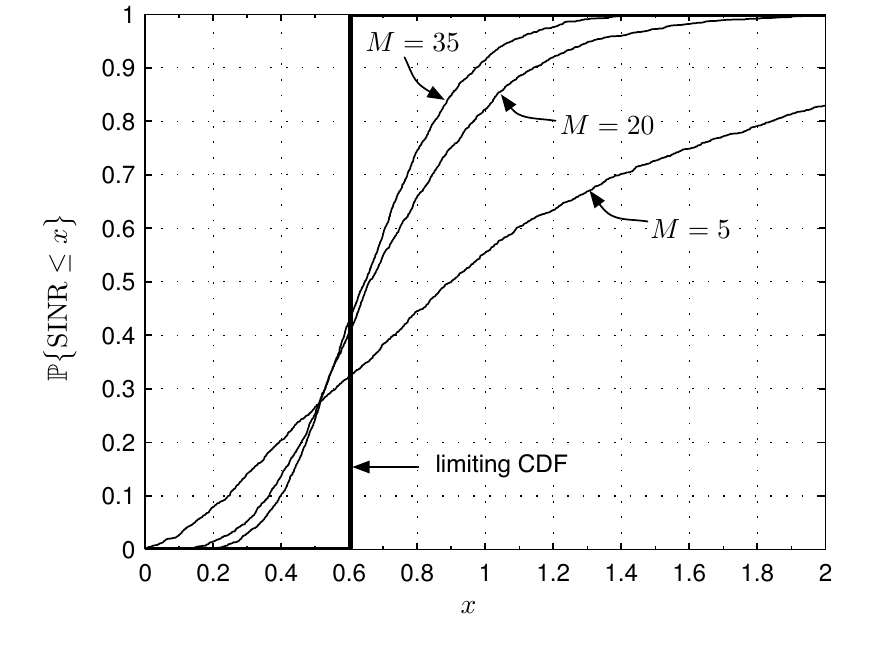}
 \includegraphics[width=\figwidth]{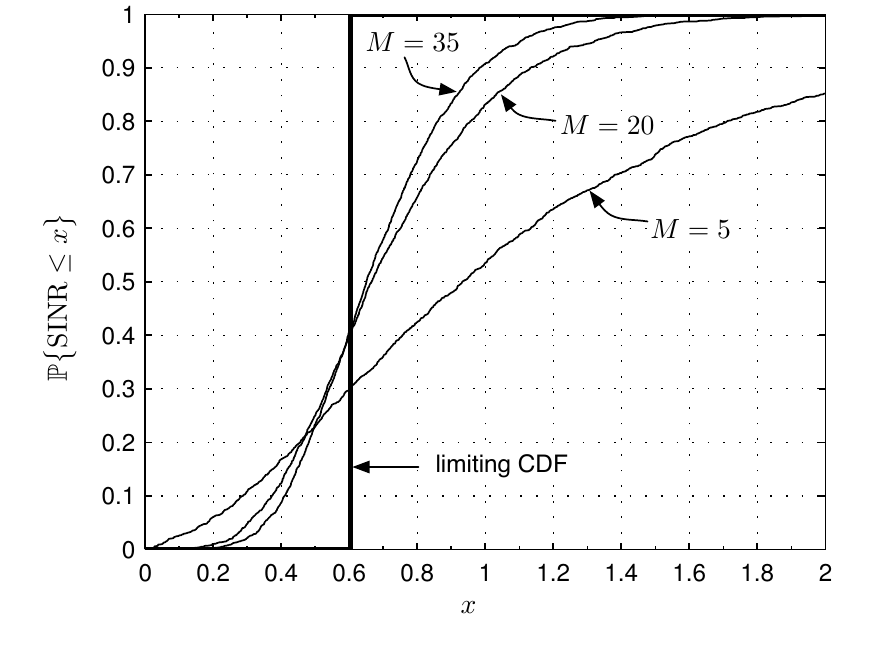}
     \caption{Simulated (Monte-Carlo) SINR CDFs for different values of $M$ for (a) $K=M^{3}$ in P1 and (b) $K=M^{2}$ in P2.}
\label{fig:convergenceP1}
\end{figure}

\paragraph*{Numerical results} We shall finally provide numerical results quantifying the outage behavior of P1 and P2.
For simplicity, we set $E_{k,m}=P_{m,k}=1$ for all $m,k$ and $\sigma^{2}=0.01$ in both simulation examples. This choice for the path loss and shadowing parameters, although not representative of a real-world propagation scenario, isolates the
dependence of our results on the network geometry. Moreover, it ensures that the distribution of the different SINR RVs for a given protocol is identical for all links so that it suffices to analyze the behavior of only
one SINR RV for each of the two protocols.
For $K=M^3$ in P1 and $K=M^2$ in P2, Fig.~\ref{fig:convergenceP1} shows the cumulative distribution functions (CDFs) (obtained through Monte-Carlo simulation) of $\mathrm{SINR}^{\mathrm{P1}}$ and $\mathrm{SINR}^{\mathrm{P2}}$, respectively, for different values of $M$. 
We observe that, for increasing $M$, the CDFs approach a step function at the corresponding mean values, i.e., 
the SINR RVs, indeed, converge to a deterministic quantity, and, consequently, the underlying fading channel converges to a nonfading channel.
The limiting mean values are given by the lower and upper bounds (which coincide in the case $E_{k,m}=P_{m,k}=1$ for all $m,k$) in~\eqref{eq:P1bounds} and~\eqref{eq:P2bounds} for P1 and P2, respectively. 
We can furthermore see that for fixed $M$ the CDFs are very similar for P1 and P2 (recall, however, that $K=M^3$ in P1 and $K=M^2$ in P2), suggesting that the convergence behavior is similar for the two protocols. The difference in the theoretically predicted convergence exponents (2/7 for P1 and 2/9 for P2) therefore does not seem to be fundamental to the two protocols and may, indeed, be a consequence of our proof technique as already pointed out above.

\subsection{Cooperation at the Relay Level}
\label{sec:coop}

The analysis carried out so far was based on the assumption that the relays cannot cooperate.
 The purpose of this section is to 
investigate the impact of cooperation (in fact, a specific form of cooperation) at the relay level on the ergodic-capacity scaling behavior in the coherent case.
Note that we continue to assume that the destination terminals cannot cooperate.
Before proceeding, we would like to mention that concentration results and an outage analysis along the
lines of the discussion in Sections~\ref{sec:coh} and~\ref{sec:crystallization}
are possible, but will be omitted for brevity of exposition.

Cooperation at the relay level will be accounted for by grouping the~$K$ single-antenna relay terminals into~$Q$ groups
\benn
\mathcal{G}_{q}\define\Bigl\{\mathcal{R}_{(q-1)L+1}, \mathcal{R}_{(q-1)L+2},\ldots, \mathcal{R}_{qL}\Bigr\},\ \ \ \natseg{q}{1}{Q}\nonumber 
\eenn
with $L$ relays in each group\footnote{For simplicity, we assume that $Q$ divides $K$ so that $K=Q L$.} and by assuming that the relays in each group can fully cooperate, 
but cooperation across groups is not possible. In order to simplify the exposition, in the remainder of this section, we think of a 
group $\mathcal{G}_{q}$ $\left(\natseg{q}{1}{Q}\right)$ as a single relay element with $L$ antenna elements and use
the term ``vector-relay (v-relay)'' terminal to address the $L$-antenna relays $\mathcal{G}_{1}, \mathcal{G}_{2},\ldots, \mathcal{G}_{Q}$. For $\natseg{q}{1}{Q}$ and $\natseg{m}{1}{M}$, the following notation will be used: 
{\allowdisplaybreaks
\bann
\vecr_{q}&\define\tp{[r_{(q-1)L+1},r_{(q-1)L+2},\ldots, r_{qL}]}\nonumber\\ 
\vect_{q}&\define\tp{[t_{(q-1)L+1},t_{(q-1)L+2},\ldots, t_{qL}]}\nonumber\\
\vecz_{q}&\define\tp{[z_{(q-1)L+1},z_{(q-1)L+2},\ldots, z_{qL}]}\nonumber\\
\vech_{q, m}&\define\tp{[h_{(q-1)L+1, m},h_{(q-1)L+2, m},\ldots, h_{qL, m}]}\nonumber\\
\vecf_{m, q}&\define\tp{[f_{m, (q-1)L+1},f_{m, (q-1)L+2},\ldots, f_{m, qL}]}\nonumber%
\eann}%
where $\vecr_{q}$ and $\vect_{q}$ are the ($L$-dimensional) vector-valued signals  received and transmitted by the $q$th v-relay, respectively, $\vecz_{q}$ is additive noise at the $q$th v-relay, $\vech_{q, m}$ contains the channel gains for the $\mathcal{S}_{m}\to \mathcal{G}_{q}$ link, and $\vecf_{m, q}$ contains the channel gains for the $\mathcal{G}_{q}\to \mathcal{D}_{m}$ link. Additionally, for simplicity, we assume that relays belonging to a given group $q$ are located close to each other so that
\bann
\hat{E}_{q, m}&\define E_{(q-1)L+1, m}=E_{(q-1)L+2, m}=\cdots=E_{qL, m}\\
\hat{P}_{m, q}&\define P_{m, (q-1)L+1}=P_{m, (q-1)L+2}=\cdots=P_{m,qL} 
\eann
for $\natseg{q}{1}{Q}$ and $\natseg{m}{1}{M}$.
With this notation, the I-O relations~\eqref{eq:IO1} and~\eqref{eq:IO2} for the $\mathcal{S}_{m}\to \mathcal{G}_{q}$ links and the $\mathcal{G}_{q}\to \mathcal{D}_{m}$ links can be written as
\bann
\vecr_{q}&=\sum_{m=1}^{M} \hat{E}_{q,m} \vech_{q,m} s_{m} + \vecz_{q},\qquad \natseg{q}{1}{Q}\\
\intertext{and}
y_{m}&=\sum_{q=1}^{Q} \hat{P}_{m,q} \tp{\vecf_{m,q}} \vect_{q} + w_{m},\qquad \natseg{m}{1}{M}
\eann
respectively. Next, we describe the generalization of the protocols P1 and P2 to the case of v-relays making the aspect of cooperation at the relay level explicit. 

\subsubsection{P1 for the Cooperative Case}

Like in the case of single-antenna relays (described in Section~\ref{sec:P1}), we partition the $Q$ v-relay terminals
into $M$ subsets $\mathcal{M}_m \left(\natseg{m}{1}{M}\right)$ with\footnote{For simplicity, we assume that $M$ divides $Q$.}
$\card{\mathcal{M}_m}=Q/M$. The v-relays (each of which has $L$ antenna elements) in $\mathcal{M}_m$ are assumed to assist the $m$th source-destination terminal pair $\{\mathcal{S}_m, \mathcal{D}_m\}$,
and the relay partitioning function $p: [1,Q]\rightarrow [1,M]$ is defined as
\benn
p(q)\define m \Leftrightarrow \mathcal{G}_q \in \mathcal{M}_m.
\eenn

We assume that the $q$th v-relay terminal has perfect knowledge of the phases of the single-input multiple-output backward channel $\mathcal{S}_{p(q)}\rightarrow \mathcal{G}_q$ 
and the phases of the corresponding multiple-input single-output forward channel $\mathcal{G}_q\rightarrow \mathcal{D}_{p(q)}$. This implies that perfect knowledge of the vectors
\bmnn
\tilde{\vech}_{q,p(q)}\define\tp{\left[e^{j\! \arg\left(\ent{\vech_{q,p(q)}}_1\right)}, \,\,\, e^{j\! \arg\left(\ent{\vech_{q,p(q)}}_2\right)},\,\,\cdots\right. \\
 \left. \cdots\,\, , e^{j\! \arg\left(\ent{\vech_{q,p(q)}}_L\right)}\right]}
\emnn
and
\bmnn
\tilde{\vecf}_{p(q),q}\define\tp{\left[e^{j\! \arg\left(\ent{\vecf_{p(q),q}}_1\right)}, \,\,\, e^{j\! \arg\left(\ent{\vecf_{p(q),q}}_2\right)},\,\,\cdots\right.\\
 \left.\cdots\,\, , e^{j\! \arg\left(\ent{\vecf_{p(q),q}}_L\right)}\right]}
\emnn
is available at~$\mathcal{G}_{q}$.
The signal~$\vecr_q$ received at the $q$th v-relay terminal is phase-matched-filtered first w.r.t.\ the assigned backward channel $\mathcal{S}_{p(q)}\rightarrow \mathcal{G}_{q}$ 
and then w.r.t.\ the assigned forward channel $\mathcal{G}_q\rightarrow \mathcal{D}_{p(q)}$ followed by a normalization so that
\be
\vect_q=d_{\mathrm{P1}, q}\,  \conj{\tilde{\bmf}_{p(q),q}} \left(\herm{\tilde{\vech}_{q,p(q)}}\,{\vecr}_q\right)
\label{eq:RelayProc1P1L}
\ee
where\footnote{The quantity~$d_{\mathrm{P1},q}$, used in this section is (for~$L>1$) different from~$d_{\mathrm{P1},k}$ defined in~\eqref{eq:dP1L1}. We use the same symbol for notational simplicity and employ the index~$q$ (instead of~$k$) consistently, in order to resolve potential ambiguities. The same comment applies to other variables redefined in this section.} the choice
\bmnn
d_{\mathrm{P1},q}\define \frac{1}{L}\sqrt{P_{\mathrm{rel}}}
\\\times 
\left[\frac{Q}{M}\sum_{m=1}^M \hat{E}_{q,m}+\frac{ \pi (L-1)Q}{4 M} \hat{E}_{q,p(q)}+ Q \sigma^2 \right]^{-1/2}
\emnn
ensures that the per-v-relay power constraint $\Ex{\vecnorm{\vect_{q}}^2}=P_{\mathrm{rel}}/Q\ \left(\natseg{q}{1}{Q}\right)$ and consequently the total (across v-relays) power constraint $\sum_{q=1}^{Q}\Ex{\vecnorm{{\vect}_q}^2}=P_{\mathrm{rel}}$ is met. As in the single-antenna relay (i.e., noncooperative) case, P1 ensures that the relays $\mathcal{G}_{q}\in\mathcal{M}_m$ forward the 
signal intended for $\mathcal{D}_m$ in a \textquotedblleft doubly coherent\textquotedblright~(w.r.t.\ the assigned backward and forward channel) fashion whereas the signals transmitted 
by the source terminals $\mathcal{S}_{\hat{m}}$ with $\hat{m}\, \neq\, m$ are forwarded to $\mathcal{D}_{m}$ in a \textquotedblleft noncoherent\textquotedblright~fashion  (i.e., phase 
incoherence occurs either on the backward  or the forward link or on both links). From~\eqref{eq:RelayProc1P1L}, we can see that cooperation in groups of $L$ single-antenna relays is realized by phase combining on the backward and forward links of each v-relay. More sophisticated forms of cooperation such as equalization on the backward link and precoding on the forward link are certainly possible, but are beyond the scope of this paper.

\subsubsection{P2 for the Cooperative Case}

Like in the case of single-antenna relays (i.e., the noncooperative case), P2 requires that each relay, in fact here v-relay, knows
the phases of all its $M$ vector-valued backward and forward channels, i.e., $\mathcal{G}_q$ needs knowledge of $\tilde{\vech}_{q,m}$ and $\tilde{\vecf}_{m,q}$, respectively,
for $\natseg{m}{1}{M}$. The relay processing stage in P2 computes
\benn
{\vect}_q=d_{\mathrm{P2},q}\!\left(\sum_{m=1}^M \conj{\tilde{\vecf}_{m,q}} \herm{\tilde{\vech}_{q,m}}\right) \vecr_q 
\eenn
where 
\bmnn
d_{\mathrm{P2},q}\define \frac{1}{L}\sqrt{P_{\mathrm{rel}}}\\
\times 
\lefto[Q\sum_{m=1}^M \hat{E}_{q,m}+\frac{\pi(L-1)Q}{4 M}\sum_{m=1}^M \hat{E}_{q,m}+ M Q\sigma^2 \right]^{-1/2}
\emnn
ensures that the per-v-relay power constraint $\Ex{\vecnorm{\vect_{q}}^2}=P_{\mathrm{rel}}/Q\ \left(\natseg{q}{1}{Q}\right)$ and, consequently, 
the total (across relays) power constraint $\sum_{q=1}^{Q}\Ex{\vecnorm{{\vect}_q}^2}=P_{\mathrm{rel}}$ is met. 

\subsubsection{Ergodic-Capacity Results}
We are now ready to establish
the impact of cooperation at the relay level on the ergodic capacity scaling laws for P1 and P2. Our results
are summarized in Theorems~\ref{thm:P1ergL} and~\ref{thm:P2ergL} below.

\begin{theorem}[Ergodic capacity of P1 with cooperation]
\label{thm:P1ergL}
Suppose that destination terminal $\mathcal{D}_m$
$\left(\natseg{m}{1}{M}\right)$ has perfect knowledge of the mean of the effective channel gain of the $\mathcal{S}_m\rightarrow\mathcal{D}_m$ link, given by $(\pi/4)L^2\!\sum_{q: p(q)=m}  d_{\mathrm{P1}, q}  \hat{P}_{m,q} \hat{E}_{q,m}$.
Then, for any $\epsilon,\delta>0$, there exist $M_0, Q_{0}>0$ s.t.\ for all $M\,\ge\,M_0$ and~$Q\ge Q_{0}$ the per source-destination terminal 
pair capacity achieved by P1 satisfies\footnote{Note that the quantities $\overline{C}_{\mathrm{SN}}$, $\underline{C}$, $\overline{C}$, and $\underline{C}_{\mathrm{SN}}$ used in this section have been defined in Section~\ref{sec:coh}.}
\ba
\label{eq:P1boundsL}
&\frac{1}{2}\log\lefto(1+\frac{\pi^2}{16}\frac{Q L^{2}}{M^3}\frac{\underline{C}^{2}}{\overline{C}_{\mathrm{SN}}^{2}}(1-\epsilon)\right)\le C_{\mathrm{P1}}
\nonumber \\&\ \ \le
\frac{1}{2}\log\lefto(1+\frac{\pi^2}{16}\frac{\max\lefto[Q, M^{2+\delta}\right]L^{2}}{M^3}\frac{\overline{C}^{2}}{\underline{C}_{\mathrm{SN}}^{2}}(1-\epsilon)\right).
\ea
\end{theorem}
\vspace*{2mm}
\begin{theorem}[Ergodic capacity of P2 with cooperation]
\label{thm:P2ergL}
Suppose that destination terminal $\mathcal{D}_m$
$\left(\natseg{m}{1}{M}\right)$ has perfect knowledge of the mean of the effective channel gain of the $\mathcal{S}_m\rightarrow\mathcal{D}_m$ link, given by $(\pi/4)L^2\!\sum_{q=1}^{Q}  d_{\mathrm{P2}, q}  \hat{P}_{m,q} \hat{E}_{q,m}$. 
Then, for any $\epsilon,\delta>0$, there exist $M_0, Q_0$ s.t.\ for all $M\,\ge\,M_0$, $Q\,\ge\,Q_0$ the per source-destination terminal pair capacity 
achieved by P2 satisfies
\ba
\label{eq:P2boundsL}
&\frac{1}{2}\log\lefto(1+\frac{\pi^2}{16}\frac{QL^{2}}{M^2}\frac{\underline{C}^{2}}{\overline{C}_{\mathrm{SN}}^{2}}(1-\epsilon)\right)\le C_{\mathrm{P2}}
\nonumber\\&\ \ \le 
\frac{1}{2}\log\lefto(1+\frac{\pi^2}{16}\frac{\max\lefto[Q,M^{1+\delta}\right]L^{2}}{M^2}\frac{\overline{C}^{2}}{\underline{C}_{\mathrm{SN}}^{2}}(1-\epsilon)\right).
\ea
\end{theorem}

\begin{IEEEproof}[Proof of Theorems~\ref{thm:P1ergL} and~\ref{thm:P2ergL}]
The upper bounds in~\eqref{eq:P1boundsL} and~\eqref{eq:P2boundsL} are again established based on a concentration result for the individual link SINRs and the lower bounds build on the technique summarized in Appendix~\ref{appendix:Medard}. 
The proofs of Theorems~\ref{thm:P1ergL} and~\ref{thm:P2ergL} are almost identical to the proofs of Theorems~\ref{thm:P1erg} and~\ref{thm:P2erg}, respectively, and do not require new techniques.  
There is, however, one important aspect in which Theorems~\ref{thm:P1ergL} and~\ref{thm:P2ergL} differ from Theorems~\ref{thm:P1erg} and~\ref{thm:P2erg}, namely, the appearance of the factor~$L^{2}$ in~\eqref{eq:P1boundsL} and~\eqref{eq:P2boundsL}. To demonstrate where this factor comes from, we provide the proof of the ergodic capacity lower bound for P1 in Appendix~\ref{appendix:ErgP1LThProof}. The proofs of the remaining statements will be omitted for brevity of exposition. 
\end{IEEEproof}

\paragraph*{Discussion of results} Just like in the noncooperative (i.e., single-antenna relay) case, we can conclude that asymptotically in $M$ if $K \propto M^{3+\alpha}$ in P1 and 
$K \propto M^{2+\alpha}$ in P2 with $\alpha\,>\,0$, the network decouples.

The effect of cooperation (through phase matched-filtering) at the relay level manifests itself in the presence of the factor~$L^{2}$ inside the $\log$ in the bounds
for $C_{\mathrm{P1}}$ and $C_{\mathrm{P2}}$ stated in Theorems~\ref{thm:P1ergL} and~\ref{thm:P2ergL}, respectively. We can summarize the
results of Theorems~\ref{thm:P1ergL} and~\ref{thm:P2ergL} as\footnote{Note that we use the $\Theta(\cdot)$ notation only to hide the dependence on $\underline{E}$, $\overline{E}$, $\underline{P}$, and $\overline{P}$. Strictly speaking, as $L$ is finite it should also be hidden under the $\Theta(\cdot)$ notation. However, our goal is to exhibit the impact of cooperation at the relay level on $C_\mathrm{P1}$ and $C_\mathrm{P2}$, which is the reason for making the dependence on $L$ explicit.}
\bann
C_{\mathrm{P1}} &= \frac{1}{2}\log\lefto(1+\Theta\lefto(\frac{QL^{2}}{M^{3}}\right)\right)\nonumber\\
C_{\mathrm{P2}} &= \frac{1}{2}\log\lefto(1+\Theta\lefto(\frac{QL^{2}}{M^{2}}\right)\right).
\eann
We can, therefore, conclude that the per-stream array gain $A$ is given by $A_{\mathrm{P1}}=QL^{2}/M^{3}$ for P1 and $A_{\mathrm{P2}}=QL^{2}/M^{2}$ for P2.
On a conceptual level, the array gain can be decomposed into a contribution due to
distributed array gain, $A_{d}$, and a contribution due to cooperation at the relay level (realized by phase matching on backward and forward links),
$A_{c}$, i.e., $A=A_{d}A_{c}$ with $A_{d,\mathrm{P1}}=QL/M^{3}$, $A_{d,\mathrm{P2}}=QL/M^{2}$, and $A_{c,\mathrm{P1}}=A_{c,\mathrm{P2}}=L$. 
To illustrate the impact of cooperation at the relay level, we compare a network with $K$ noncooperating single-antenna
relays to a network with a total of $K=QL$ single-antenna relays cooperating in groups of $L$ single-antenna relays. 
In the case where there is no cooperation at the relay level, we have
\benn
C^{(nc)}_{\mathrm{P1}}=\frac{1}{2}\log\lefto(1+\Theta\lefto(\frac{K}{M^{3}}\right)\right) 
\eenn
whereas if the relays cooperate in groups of $L$ single-antenna relays, we get
\benn
C^{(c)}_{\mathrm{P1}}=\frac{1}{2}\log\lefto(1+\Theta\lefto(\frac{K L}{M^{3}}\right)\right). 
\eenn
Cooperation at the relay level (realized by phase matched-filtering) in groups of~$L$ single-antenna relays therefore yields an $L$-fold 
increase in the effective per-stream SINR due to additional array gain given by~$A_c=L$. Equivalently, the total number of
single-antenna relays needed to achieve a given per source-destination terminal pair capacity is reduced by a factor of $L$ through cooperation in groups of~$L$ single-antenna
relay elements. The conclusions for P2 are identical. 

As already pointed out above, the network decouples into effectively isolated source-destination pair links for any finite $L\,>\,1$.
Even though a concentration analysis along the lines of Theorems~\ref{thm:P1Conv} and~\ref{thm:P2Conv}
was not performed (for the sake of brevity), it can be shown that for finite $L\,>\,1$ the individual links converge to nonfading links as $M,Q\,\rightarrow\,\infty$, provided that~$Q$ scales supercritically as a function of~$M$.

\begin{figure}
\centering
\includegraphics[width=\figwidth]{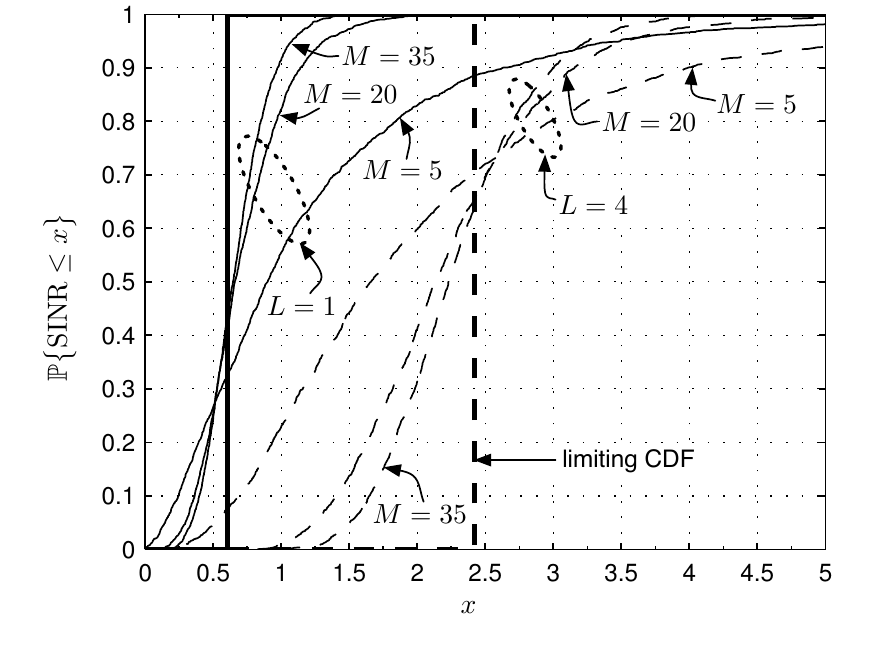}
 \caption{Simulated (Monte-Carlo) SINR CDFs for different values of~$M$ for $QL=M^{3}$ in P1 with $L=1$ and $L=4$.}
      \label{fig:coop}
\end{figure}

\paragraph*{Numerical example} We conclude this section with a numerical example that demonstrates the impact of cooperation at the relay level, where we use the same parameters as in the simulation examples at the end of Section~\ref{sec:crystallization}.
Figure~\ref{fig:coop} shows the SINR CDF for P1 with~$L=4$ and~$QL=M^{3}$ (the case $L=1$ shown in Fig.~\ref{fig:convergenceP1} is included for reference). We observe that, as pointed out above, for increasing $M$, we, indeed, get convergence of the fading link to a nonfading
link. Moreover, we can also see that increasing~$L$ for fixed~$M$ results in higher per source-destination terminal pair capacity, but at the same time
slows down convergence (w.r.t.\ $M$ and hence also~$Q$) of the link SINRs to their deterministic limits.

\section{Noncoherent (AF) Relay Networks}
\label{sec:ncoh}
So far, we have considered coherent relay networks, where each relay terminal
knows its assigned backward and forward channels (P1) or all backward and forward channels (P2) perfectly. In the 
following, we 
relax this assumption and study networks with no CSI at the relay
terminals, i.e., noncoherent relay networks.
In particular, we investigate a simple
AF architecture where the relay terminals, in the second time slot, forward (without additional
processing) a scaled version of the signal received in the first time slot. As already mentioned in Section~\ref{sec:model}, the source terminals do not have CSI. The destination terminals cooperate and perform joint decoding. The assumptions on CSI at the destination terminals will be made specific in Section~\ref{sec:capAF}. 

\subsection{The AF Protocol}
Throughout this section, we assume that $E_{k,m}=P_{m,k}=1$ for all $\natseg{m}{1}{M}$, $\natseg{k}{1}{K}$. This assumption is conceptual as the technique used to derive the main result in this section does not seem to be applicable for general~$E_{k,m}$ and~$P_{m,k}$. On the other hand, the results in this section do not require~$\matH$ and~$\matF$ to have Gaussian entries. Upon reception of~$r_{k}$, the $k$th relay terminal simply scales the received signal to obtain $t_{k}=\bigl(d/\sqrt{K}\bigr) r_{k}$. Choosing $d=\sqrt{P_{\mathrm{rel}}/(1+\sigma^{2})}$ ensures that the per-relay power constraint $\Ex{\abs{t_{k}}^{2}}\le P_{\mathrm{rel}}/K$ and hence the total power constraint $\Ex{\vecnorm{\vect}^{2}}\le P_{\mathrm{rel}}$ is met.

With these assumptions, inserting~\eqref{eq:IO1} into~\eqref{eq:IO2}, we get the following I-O relation
\be
\label{eq:IOAF}
\vecy=\frac{d}{\sqrt{K}} \matF \matH \vecs+ \frac{d}{\sqrt{K}}\matF\vecz+\vecw.
\ee
In the remainder of this section, we assume that the {\it jointly decoding destination terminals} have access to the realizations of~$\matH$ and~$\matF$. In fact, as the analysis below shows, knowledge of~$\matF\matH$ and~$\matF$ is sufficient.

\subsection{Capacity of the AF Protocol}
\label{sec:capAF}
Based on the I-O relation~\eqref{eq:IOAF}, we shall next study the behavior of~$I\lefto(\vecy;\vecs\given\matF\matH,\matF\right)$ when~$M,K\to\infty$ with~$K/M\to\beta$. We start by noting that
\bann
&I\lefto(\vecy;\vecs\given\matF\matH,\matF\right)
\\ &\ =
\log\det\lefto(\matI+\frac{d^{2}}{\sigma^{2} M K}\herm\matH\herm\matF\left(\frac{d^{2}}{K}\matF\herm\matF+\matI\right)^{-1}\matF\matH\right).
\eann

Since the destination terminals perform joint decoding, the ergodic capacity per source-destination terminal pair is given by
\be
\label{eq:CAFsum}
C_{\mathrm{AF}}=\frac{1}{2}\Ex{\frac{1}{M}\sum_{k=1}^{K}\log\lefto(1+\frac{1}{\sigma^{2}}\lambda_{k}\lefto(\frac{1}{M}\matH\herm\matH\matT\right)\right)}
\ee  
where 
\benn
\matT\define \frac{d^{2}}{K}\herm\matF\!\left(\matI+\frac{d^{2}}{K}\matF\herm\matF\right)^{-1}\!\matF
\eenn
and the factor $1/2$ in~\eqref{eq:CAFsum} results from the fact that data is transmitted over two time slots.

\subsection{Asymptotic Capacity Behavior}
To compute $C_{\mathrm{AF}}$ in the $M,K\to\infty$ limit with $K/M\to\beta$, we start by analyzing the corresponding asymptotic behavior of $\lambda_{k}\lefto((1/M)\matH\herm\matH\matT\right)$.
To this end, we define
the empirical spectral distribution (ESD) of a matrix (random or deterministic).
\begin{definition}
Let $\matX\in\complexset^{N\times N}$ be a Hermitian matrix. The ESD of $\matX$ is defined as 
\benn
F_{\matX}^{N}(x)\define\frac{1}{N}\sum_{n=1}^{N} I\lefto[\lambda_{n}(\matX)\le x\right].
\eenn
\end{definition}
For random~$\matX$, the quantity $F_{\matX}^{N}(x)$ is random as well, i.e., it is a RV for each~$x$.
In the following, our goal is to prove the convergence (in the sense defined below), when $M,K\to\infty$ with $K/M\to\beta$ and $\beta\in(0,\infty)$, of $F_{(1/M)\matH\herm\matH\matT}^{K}(x)$ to a deterministic limit and to find the corresponding limiting eigenvalue distribution. 
\begin{definition}
We say that the ESD $F_{\matX}^{N}(x)$ of a random Hermitian matrix $\matX\in\complexset^{N\times N}$ converges almost surely (\as) to a deterministic limiting function $F_{\matX}(x)$, when $N\to\infty$, if
for any~$\epsilon>0$ there exists an $N_{0}>0$ s.t. for all $N\ge N_{0}$ \as
\benn
\sup_{x\in\reals}\abs{F_{\matX}^{N}(x)-F_{\matX}(x)}\le \epsilon.
\eenn
\end{definition}
To prove the convergence of $F_{(1/M)\matH\herm\matH\matT}^{K}(x)$ to a deterministic limiting function, we start by analyzing $F_{\matT}^{K}(x)$.

\begin{lemma}
\label{lemma:ftransform}
For~$M,K\to\infty$ with~$K/M\to\beta$, the ESD $F_{\matT}^{K}(x)$ converges \as to a nonrandom limiting distribution $F_{\matT}(x)$ with corresponding density given by\footnote{Note that~\eqref{eq:density} implies that~$f_{\matT}(x)$ is compactly supported in the interval~$\left[\gamma_1/(1+\gamma_1), \gamma_2/(1+\gamma_2)\right].$}
\ba
\label{eq:density}
&f_{\matT}(x)
=
\frac{\sqrt{(1+\gamma_1)(1+\gamma_2)}}{2\pi d^{2} x (1-x)^{2}}\nonumber\\
&\ {\times}\:\sqrt{\left(\frac{\gamma_2}{1+\gamma_2}-x\right)^{\!\!+}\!\!\left(x-\frac{\gamma_1}{1+\gamma_1}\right)^{\!\!+}}
+\left[1-\frac{1}{\beta}\right]^{\!+}\! \delta(x)
\ea
where $\gamma_1\define d^{2}(1-1/\sqrt{\beta})^{2}$ and $\gamma_2\define d^{2} (1+1/\sqrt{\beta})^{2}$.
\end{lemma}
\begin{IEEEproof}
We start with the singular value decomposition
\benn
\frac{d}{\sqrt{K}}\matF=\matU\matSigma\matV
\eenn
where the columns of $\matU\in \complexset^{M,M}$ are the eigenvectors of the matrix $(d^{2}/K)\matF\herm\matF$, the columns of $\herm\matV\in \complexset^{K,K}$ are the eigenvectors of $(d^{2}/K)\herm\matF\matF$, and the matrix $\matSigma\in \reals^{M,K}$ contains $R=\min(M,K)$ nonzero entries $\Sigma_{11},\Sigma_{22},\ldots,\Sigma_{RR}$, which are the positive square roots of the nonzero eigenvalues of the matrix $(d^{2}/K)\bF\bF^{H}$. Defining $\matLambda\define\matSigma\herm\matSigma\in \reals^{M,M}$, we have
\benn
\matT=\herm\matV\herm\matSigma\left(\matI+\matLambda\right)^{-1}\matSigma\matV.
\eenn
By inspection, it follows that
\bmult
\label{eq:RVtransform}
F^{K}_{\herm\matSigma\left(\matI+\matLambda\right)^{-1}\matSigma}(x)
=\frac{M}{K} F^{M}_{\matLambda}\lefto(\frac{x}{1-x}\right)+\left(1-\frac{M}{K}\right) u(x).
\emult

As $F^{M}_{\matLambda}(x)=F^{M}_{(d^{2}/K) \matF\herm\matF}(x)$, by the Mar\v cenko-Pastur law (see Theorem~\ref{theorem:MP} in Appendix~\ref{appendix:RM}), we conclude that $F^{M}_{\matLambda}(x)$ converges \as to a limiting nonrandom distribution $F_{\matLambda}(x)$ with corresponding density
\be
\label{eq:MPdensity}
f_{\matLambda}(x)=
\frac{\beta}{2\pi x d^{2}}\sqrt{\left(\gamma_2-x\right)^{+}\left(x-\gamma_1\right)^{+}}+
[1-\beta]^{+} \delta(x).
\ee
From~\eqref{eq:RVtransform} we can, therefore, conclude that~$F^{K}_{\herm\matSigma\left(\matI+\matLambda\right)^{-1}\matSigma}(x)$ converges \as to a nonrandom limit
given by
\be
F_{\herm\matSigma\left(\matI+\matLambda\right)^{-1}\matSigma}(x)=\frac{1}{\beta} F_{\matLambda}\lefto(\frac{x}{1-x}\right)+\left(1-\frac{1}{\beta}\right) u(x).
\label{eq:RVtransformLim}
\ee
Taking the derivative  w.r.t.~$x$ on both sides of~\eqref{eq:RVtransformLim}, the density corresponding to~$F_{\herm\matSigma\left(\matI+\matLambda\right)^{-1}\matSigma}(x)$ is obtained as
\bmult
\label{eq:SLSdensity}
f_{\herm\matSigma\left(\matI+\matLambda\right)^{-1}\matSigma}(x)\\
=\frac{1}{\beta} f_{\matLambda}\lefto(\frac{x}{1-x}\right)\frac{1}{(1-x)^{2}}+\left(1-\frac{1}{\beta}\right)\delta(x).
\emult
We obtain the final result in~\eqref{eq:density} now by noting that $f_{\matT}(x)=f_{\herm\matSigma\left(\matI+\matLambda\right)^{-1}\matSigma}(x)$ because of the unitarity of~$\matV$ and by inserting~\eqref{eq:MPdensity} into~\eqref{eq:SLSdensity} and carrying out straightforward algebraic manipulations. 
\end{IEEEproof}

Based on Lemma~\ref{lemma:ftransform}, we can now apply Theorem~\ref{theorem:Silv} (Appendix~\ref{appendix:RM}) to conclude that $F_{(1/M)\matH\herm\matH\matT}^{K}(x)$ converges \as to a deterministic function $F_{(1/M)\matH\herm\matH\matT}(x)$ as $M,K\to\infty$ with $K/M\to\beta$. The corresponding limiting density $f_{(1/M)\matH\herm\matH\matT}(x)$ is obtained through the application of the Stieltjes inversion formula~\eqref{eq:inversion} to the solution of the fixed-point equation
\be
\label{eq:fixed_point_main}
G(z)=\underbrace{\int_{-\infty}^{\infty}\frac{f_{\matT}(x) dx}{x (1-\beta-\beta z G(z))-z}}_{I},\ \  z\in \complexset^{+}
\ee
in the set 
\be
\label{eq:domain}
\left\{G(z)\in\complexset \setgiven -(1-\beta)/z+\beta G(z)\in\complexset^{+}\right\},\ \  z\in \complexset^{+}
\ee
where we used the symbol~$G(z)$ to denote the Stieltjes transform $G_{(1/M)\matH\herm{\matH}\matT}(z)$. In the following, for brevity, we write~$G$ instead of~$G(z)$.
To solve~\eqref{eq:fixed_point_main}, we first compute the integral $I$ on the RHS of~\eqref{eq:fixed_point_main}. We substitute $f_{\matT}(x)$ from~\eqref{eq:density} into~\eqref{eq:fixed_point_main} and define 
\benn
\eta_{1}\define \frac{\gamma_{1}}{1+\gamma_{1}},\ \eta_{2}\define \frac{\gamma_{2}}{1+\gamma_{2}},\ \rho\define \frac{\sqrt{(1+\gamma_1)(1+\gamma_2)}}{2\pi d^{2}}
\eenn
to obtain
\be
\label{eq:TheIntegral}
I=-\frac{1}{z}\left[1-\frac{1}{\beta}\right]^{\!+}\!\!
+\frac{1}{z}
\underbrace{\int_{\eta_{1}}^{\eta_{2}}\!\!\frac{\rho\sqrt{\left(\eta_{2}-x\right)\left(x-\eta_{1}\right)}\, dx}{ x (1-x)^{2}\left(x \left(\frac{1-\beta}{z}-\beta G\right)-1\right)}}_{\hat{I}}.
\ee
The integral $\hat{I}$ is computed in Appendix~\ref{appendix:integral}. Employing the notation introduced in Appendix~\ref{appendix:integral}, we can finally write the fixed point equation~\eqref{eq:fixed_point_main} as
\be
\label{eq:fixed_point1}
G z=-\left[1-\frac{1}{\beta}\right]^{+}\!\! + \chi\left( A_{1} \hat{I}_{1}+ A_{2} \hat{I}_{2}+ A_{3} \hat{I}_{3}+ A_{4} \hat{I}_{4}\right).
\ee
It is tedious, but straightforward, to show that for any~$\beta>0$
\benn
-\left[1-\frac{1}{\beta}\right]^{+}+\chi A_{1} \hat{I}_{1}=-\frac{\beta-1}{2\beta}
\eenn
so that~\eqref{eq:fixed_point1} can be written as
\be
\label{eq:fixed_point2}
G z+\frac{\beta-1}{2\beta}-\chi A_{2} \hat{I}_{2}-\chi A_{3} \hat{I}_{3}=\chi A_{4} \hat{I}_{4}.
\ee
Next, multiplying~\eqref{eq:fixed_point2} by $2 d^2 \beta  (G \beta  z+z+\beta -1)^2$, squaring both sides, introducing the auxiliary variable 
\benn
\hat{G}\define -\frac{1-\beta}{z}+\beta G
\eenn
we obtain after straightforward, but tedious, manipulations that~$\hat{G}$ must satisfy the following quartic equation
\ba
\label{eq:quartic}
 \hat{G}^4+a_{3} \hat{G}^3+a_{2} \hat{G}^2+a_{1} \hat{G}+a_{0}=0
\ea
with the coefficients 
\bann
&a_{3}=\frac{1}{z}(2 z-\beta +1)& &a_{2}=\frac{1}{z} \left(z -\beta  +3 -\frac{\beta}{d^{2}} \right)\\
&a_{1}=\frac{1}{z^{2}} \left(2 z -\beta +1-\frac{\beta}{d^{2}} \right)& &a_{0}=\frac{1}{z^{2}}.
\eann

The quartic equation~\eqref{eq:quartic} can be solved analytically. The resulting expressions are, however, very lengthy, do not lead to interesting insights, and will therefore be omitted.
It is important to note, however, that~\eqref{eq:quartic} has two pairs of complex conjugate roots. The solutions of~\eqref{eq:quartic} will henceforth be denoted as $\hat{G}_{1}, \conj{\hat{G}_{1}}, \hat{G}_{2}$, and~$\conj{\hat{G}_{2}}$. We recall that our goal is to find the unique solution~$G$ of the fixed point equation~\eqref{eq:fixed_point_main} s.t.\ $\hat{G}=-(1-\beta)/z+ \beta G\in\complexset^{+}$ for all $z\in\complexset^{+}$. Therefore, in each point $z\in\complexset^{+}$ we can immediately eliminate the two solutions (out of the four) that have a negative imaginary part. In practice, this can be done conveniently by constructing the functions~$\hat{G}'_{1}\define\Re \hat{G}_{1}+j\!\abs{\Im \hat{G}_{1}}$ and $\hat{G}'_{2}\define\Re \hat{G}_{2}+j\!\abs{\Im \hat{G}_{2}}$, which can be computed analytically, satisfy~\eqref{eq:quartic}, and are in~$\complexset^{+}$ for any $z\in\complexset^{+}$.
Next, note that~\eqref{eq:fixed_point2} has a unique solution in the set~\eqref{eq:domain}, which is also the unique solution of~\eqref{eq:fixed_point_main}. We can obtain this solution~$G(z)$, $z\in\complexset^{+}$, by substituting $G_{1}=(1/\beta)(\hat{G}'_{1}-(\beta-1)/z)$ and $G_{2}=(1/\beta)(\hat{G}'_{2}-(\beta-1)/z)$ into~\eqref{eq:fixed_point2} and checking which of the two satisfies the equation. Unfortunately, it seems that this verification cannot be formalized in the sense of identifying the unique solution of~\eqref{eq:fixed_point2} in analytic form. The primary reason for this is that to check {\it algebraically} if~$G_{1}$ and~$G_{2}$ satisfy~\eqref{eq:fixed_point2}, we have to perform a noninvertible transformation (squaring) of~\eqref{eq:fixed_point2}, which doubles the number of solutions of this equation, and results in~$G_{1}$ and~$G_{2}$ both satisfying the resulting formula. The second reason is that depending on the values of the parameters~$\beta>0, d>0$, the correct solution is either~$G_{1}$ or~$G_{2}$, and the dependence between~$G_{1}$, $G_{2}$, $\beta$, and~$d$ has a complicated structure. Starting from the analytical expressions for~$G_{1}$ and~$G_{2}$, we can identify, however, for any fixed $\beta>0, d>0$, the density function $f_{(1/M)\matH\matH^{H}\matT}(x)=(1/\pi)\lim_{y\to 0^{+}}\Im\lefto[G(x+j y)\right]$ corresponding to the unique solution of~\eqref{eq:fixed_point2} [and hence of~\eqref{eq:fixed_point_main}] numerically.   
This is accomplished as follows.
We know that, for given~$x$, $\lim_{y\to 0^{+}}\Im\left[G(x+j y)\right]$ is either equal to  
\bann
L_{1}(x)&\define\lim_{y\to 0^{+}}\Im\left[G_{1}(x+j y)\right]\\
\intertext{or}
L_{2}(x)&\define\lim_{y\to 0^{+}}\Im\left[G_{2}(x+j y)\right].
\eann
Even though the functions $L_{1}(x)$ and $L_{2}(x)$ can be computed analytically (with the resulting expressions being very lengthy and involved), it seems that for any fixed~$x>0$ the correct choice between the values $L_{1}(x)$ and $L_{2}(x)$ can only be made numerically. The following algorithm constitutes one possibility to solve this problem.
\algobox{Algorithm---Choice of the Limit}{
\label{alg:value}
{\it Input:} $x>0$
		\begin{enumerate}[\setlabelwidth{3a)}]
		 \item
		 \label{alg:step1}
Choose a small enough $y>0$
\item
Substitute $G_{1}(x+j y)$ and $G_{2}(x+j y)$ into~\eqref{eq:fixed_point2}
\item
\label{alg:step3}
{\it If} $G_{1}(x+j y)$ satisfies~\eqref{eq:fixed_point2}, {\it then} 

\hspace{0.5cm}{\it return} $L_{1}(x)$

{\it otherwise} 

\hspace{0.5cm}{\it return} $L_{2}(x)$
\end{enumerate}
}

As any other numerical procedure, this algorithm includes a heuristic element.
The following comments are therefore in order. 
\begin{itemize}
\item
In Step~\ref{alg:step1} of the algorithm, the choice of~$y$ cannot be formalized in the sense of giving an indication of how small it has to be as a function of~$\beta$ and~$d$. On the one hand,~$y$ has to be strictly greater than zero, because~\eqref{eq:fixed_point2} in general holds in~$\complexset^{+}$  only and does not need to hold neither for $G_{1}(x+j0)$ nor for $G_{2}(x+j 0)$. On the other hand,~$y$ should be small enough for $G_{1}(x+j y)$ to be close to $L_{1}(x)$ and $G_{2}(x+j y)$ to be close to~$L_{2}(x)$. The correctness of the output of the algorithm is justified by the fact that $G(z)$ is analytic in~$\complexset^{+}$ (see Definition~\ref{def:stieltjes_tr} in Appendix~\ref{appendix:RM}).
\item
In Step~\ref{alg:step3} the check whether $G_{1}(x+j y)$ satisfies~\eqref{eq:fixed_point2} is performed numerically. Therefore, rounding errors will arise. It turns out, however, that in
practice, unless $\abs{L_{1}(x)-L_{2}(x)}$ is very small (in this case it does not matter which of the two values we choose), the solution of~\eqref{eq:fixed_point2} yields a clear indication of whether $G_{1}(x+j y)$ or $G_{2}(x+j y)$ is the correct choice.      
\item
To compute the density $f_{(1/M)\matH\matH^{H}\matT}(x)$ using the proposed algorithm, we need to run Steps 1--3 for every~$x$. It will be proved below that $f_{(1/M)\matH\matH^{H}\matT}(x)$ is always compactly supported and bounds for its support will be given in analytic form (as a function of~$\beta$ and~$d$). Since the algorithm consists of very basic arithmetic operations only, it is very fast and can easily be run on a dense grid inside the support region of $f_{(1/M)\matH\matH^{H}\matT}(x)$.
\end{itemize}

\begin{figure}
\centering
\label{fig:densitybeta}
\subfigure[]{\includegraphics[width=\figwidth]{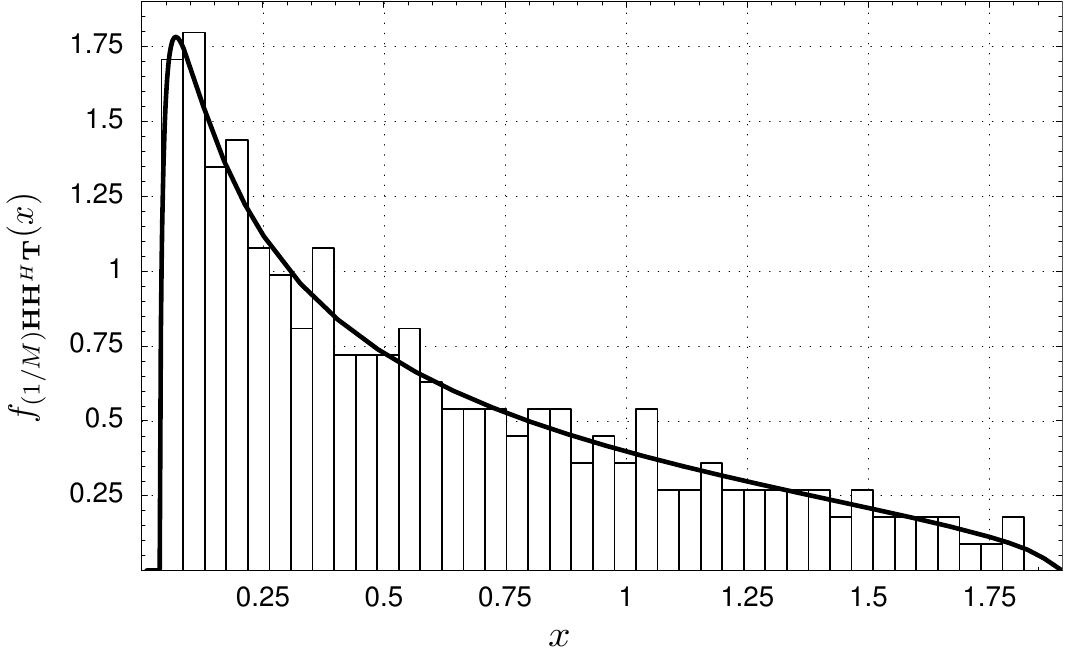}\label{fig:densitybetahist}}
\subfigure[]{\includegraphics[width=\figwidth]{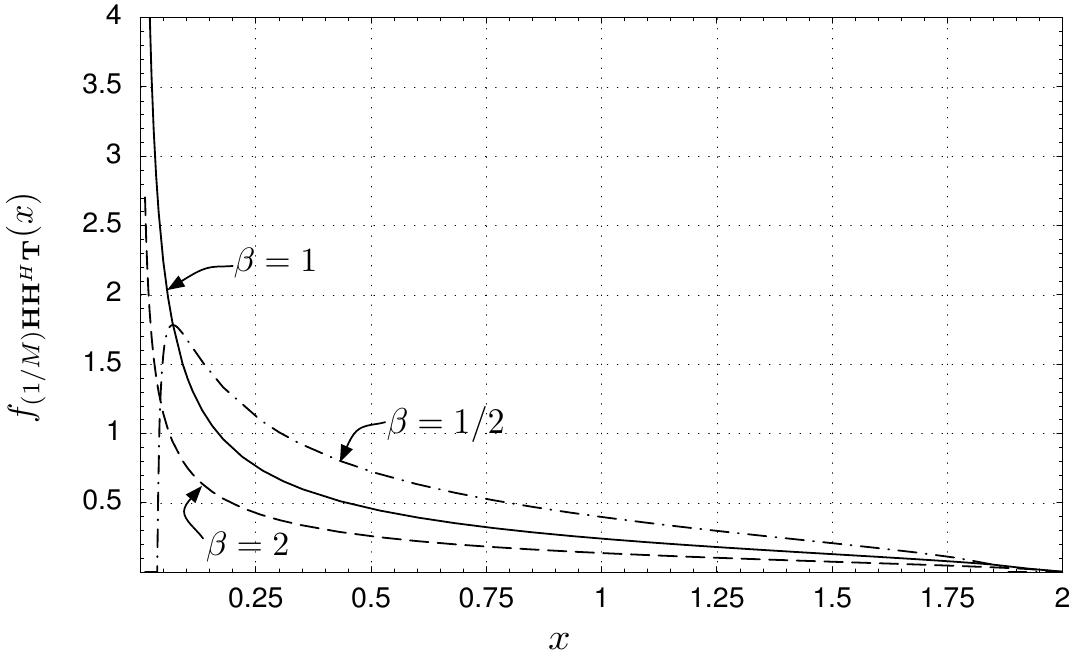}\label{fig:denistybeta3}}
\caption{Limiting density $f_{(1/M)\matH\matH^{H}\matT}(x)$ (a) for $\beta=1/2$ and $d=1$ along with its histogram (Monte-Carlo) and (b) for different values of $\beta=2,1,1/2$ and $d=1$.}
\end{figure}

As an example, for~$d=1$ and $\beta=1/2$, Fig.~\ref{fig:densitybetahist} shows the density $f_{(1/M)\matH\matH^{H}\matT}(x)$ obtained by the algorithm formulated above along with
the histogram of the same density obtained through Monte-Carlo simulation. We can see that the two curves match very closely and that our method allows to obtain a much more refined picture of the limiting density. Fig.~\ref{fig:denistybeta3} shows the density $f_{(1/M)\matH\matH^{H}\matT}(x)$  for~$\beta=2$, $1$, $1/2$ obtained through our algorithm. We can see that the density function is always compactly supported.  

\begin{figure}
\centering
\includegraphics[width=\figwidth]{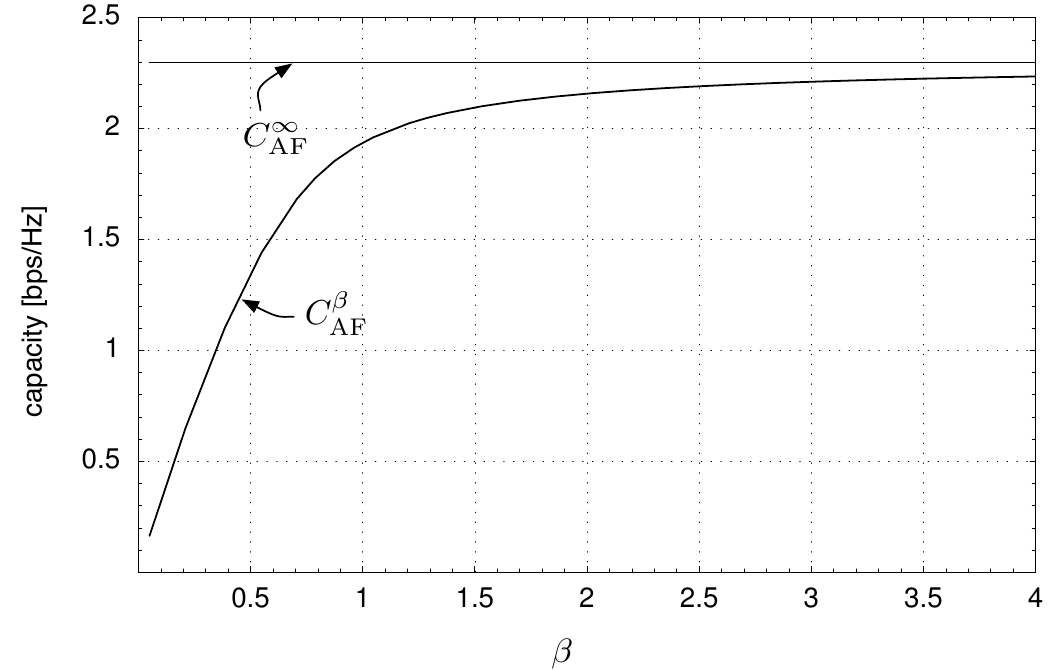}
\caption{Capacity $C_{\mathrm{AF}}^{\beta}$ as a function of $\beta$ for $d=1$ and~$\sigma^{2}=0.01$.}
\label{fig:capacitybeta}
\end{figure}

The final step in computing the asymptotic capacity of the AF relay network is to take the limit $K,M\to\infty$ with $K/M\to\beta$ in~\eqref{eq:CAFsum} and to evaluate the resulting integral
\be
\label{eq:betaC}
C_{\mathrm{AF}}^{\beta}\define\frac{\beta}{2}\int_{0}^{\infty}\log\lefto(1+\frac{x}{\sigma^{2}}\right) f_{(1/M) \matH\matH^{H}\matT}(x)\, dx
\ee 
numerically. The evaluation of~\eqref{eq:betaC} is drastically simplified if we consider that~$f_{(1/M)\matH\matH^{H}\matT}(x)$ is compactly supported. The corresponding interval boundaries (or, more specifically, bounds thereon) can be computed analytically as a function of~$\beta$ and~$d$. We start by noting that the second part of Theorem~\ref{theorem:MP} in Appendix~\ref{appendix:RM} implies that \as 
$\lim_{M\to\infty}\lambda_{\mathrm{max}}\left((1/M)\matH\herm\matH\right)=(1+\sqrt{\beta})^{2}$. From~\eqref{eq:SLSdensity} and Theorem~\ref{theorem:MP}, it follows that \as 
$\lambda_{\mathrm{max}}\lefto(\matT\right)=d^{2}(1+\sqrt{\beta})^{2}/(\beta+d^{2}(1+\sqrt{\beta})^{2})$. For any realization of~$\matH$ and~$\matT$ and any $M,K$, by the submultiplicativity of the spectral norm, we have 
\benn
\lambda_{\mathrm{max}}\lefto((1/M)\matH\herm\matH\matT\right)\le \lambda_{\mathrm{max}}\lefto((1/M)\matH\herm\matH\right)\lambda_{\mathrm{max}}\lefto(\matT\right)
\eenn
which implies that for $M,K\to\infty$ with $K/M\to\beta$ \as
\benn
\lambda_{\mathrm{max}}\lefto((1/M)\matH\herm\matH\matT\right)\le \frac{d^{2}(1+\sqrt{\beta})^{4}}{\beta+d^{2}(1+\sqrt{\beta})^{2}}\define x_{\mathrm{max}}.
\eenn
We can thus conclude that $f_{(1/M)\matH\matH^{H}\matT}(x)$ is compactly supported on the interval\footnote{The actual supporting interval of~$f_{(1/M)\matH\matH^{H}\matT}(x)$ may, in fact, be smaller.} $[0, x_{\mathrm{max}}]$. Consequently, the integral in~\eqref{eq:betaC} becomes 
\benn
C_{\mathrm{AF}}^{\beta}=\frac{\beta}{2}\int_{0}^{x_{\mathrm{max}}}\log\lefto(1+\frac{x}{\sigma^{2}}\right) f_{(1/M)\matH\matH^{H}\matT}(x)\, dx
\eenn
which we can compute numerically, using any standard method for numerical integration and employing the algorithm described above to evaluate~$f_{(1/M)\matH\matH^{H}\matT}(x)$ at the required grid points. Using this procedure, we computed~$C_{\mathrm{AF}}^{\beta}$ as a function of~$\beta$ for~$d=1$ with the result depicted in Fig.~\ref{fig:capacitybeta}.
We can see that for $\beta<1$ (i.e.,~$K<M$), $C_{\mathrm{AF}}^{\beta}$ increases very quickly with~$\beta$, which is because the corresponding  effective MIMO channel matrix builds up rank and hence spatial multiplexing gain. For $\beta>1$ (i.e.,~$K>M$), when the effective MIMO channel matrix is already full rank with high probability, the curve flattens out and for $\beta\to\infty$, the capacity $C_{\mathrm{AF}}^{\beta}$ seems to converge to a finite value. In the next subsection, we prove that $C_{\mathrm{AF}}^{\beta}$ indeed converges to a finite limit as $\beta\to\infty$. This result has an interesting interpretation as it allows to relate the AF relay network to a point-to-point MIMO channel.  

\subsection{Convergence to Point-to-Point MIMO Channel}
In~\cite{bolcskei06-}, it was shown that for finite~$M$, as $K\to\infty$, the two-hop AF relay network capacity converges to half the capacity of a point-to-point MIMO link; the factor~$1/2$ penalty comes from the fact that communication takes place over two time slots. In the following, we demonstrate that the result in~\cite{bolcskei06-} can be generalized to the~$M,K\to\infty$ case. More specifically, we show that for $\beta\to \infty$ the asymptotic ($M,K\to\infty$) capacity of the two-hop AF relay network is equal to half the asymptotic ($M\to\infty$) capacity of a~point-to-point MIMO channel with~$M$ transmit and $M$ receive antennas.
We start by dividing~\eqref{eq:quartic} by $\beta$ and taking the limit\footnote{It is important that first we take the limit $M,K\to\infty$ with $K/M\to\beta$ and afterwards let $\beta\to\infty$.} $\beta\to\infty$, which yields the quadratic equation
\be
z\hat{G}^2+z\! \left(1+\frac{1}{d^{2}}\right)\! \hat{G}+\left(1+\frac{1}{d^{2}}\right)=0.
\label{eq:Quadratic}
\ee
The two solutions of~\eqref{eq:Quadratic} are given by
\ba
&\hat{G}_{1,2}(z)=\frac{-z\!\left(1+\frac{1}{d^{2}}\right)\pm \sqrt{z^{2}\left(1+\frac{1}{d^{2}}\right)^{2}-4 z \left(1+\frac{1}{d^{2}}\right)}}{2 z}.
\label{eq:GLim}
\ea
Applying the Stieltjes inversion formula~\eqref{eq:inversion} to~\eqref{eq:GLim} and choosing the solution that yields a positive density function, we obtain
\ba
&\beta f_{(1/M)\matH\matH^{H}\matT}(x)=\frac{1}{\pi}\lim_{y\to 0^{+}}\Im\left[\beta G(x+j y)\right]\nonumber\displaybreak[0]\\
&\qquad=\frac{1}{\pi}\lim_{y\to 0^{+}}\Im\left[\hat{G}(x+j y)\right]\nonumber\displaybreak[0]\\
&\qquad=\frac{1}{2 \pi x}\sqrt{\left[4 x \left(1+\frac{1}{d^{2}}\right)-x^{2}\left(1+\frac{1}{d^{2}}\right)^{2}\right]^{+}}.
\label{eq:betaf}
\ea
Inserting~\eqref{eq:betaf} into~\eqref{eq:betaC} and changing the integration variable according to $u\define x\!\left(1+1/d^{2}\right)$, we find that $C_{\mathrm{AF}}^{\beta}\xrightarrow{\beta\to\infty} C_{\mathrm{AF}}^{\infty}$, where 
\ba
C_{\mathrm{AF}}^{\infty}\define\frac{1}{4\pi}\int_{0}^{4}\sqrt{\frac{4}{u}-1} \log\lefto(1+\frac{d^{2} }{(d^{2}+1)\sigma^{2}}u\right)  du.
\label{eq:CAFLim}
\ea
Comparing~\eqref{eq:CAFLim} with~\cite[Eq.~(13)]{telatar99-11}, it follows that for $\beta\to\infty$ the asymptotic $M,K\to\infty$ with $K/M\to\beta$ per source-destination terminal pair capacity in the two-hop AF relay network is equal to half the asymptotic $\left(M\to\infty\right)$ per-antenna capacity in a point-to-point MIMO link with~$M$ transmit and~$M$ receive antennas, provided the SNR in the relay case is defined as $\mathrm{SNR}\define d^{2}/\left((d^{2}+1)\sigma^{2}\right)$. For~$M$ and~$K$ large, it is easy to verify that this choice corresponds to the SNR at each destination terminal in the AF relay network.  In this sense, we can conclude that for $\beta\to\infty$ the AF relay network ``converges'' to a point-to-point MIMO link with the same received SNR.

\section{Conclusion}
\label{sec:conclusion}

The minimum rate of growth of the number 
of relays~$K$, as a function of the number of source-destination terminal 
pairs~$M$, for coherent fading interference relay networks to decouple
was shown to be $K\propto M^{3}$ under protocol P1 and $K \propto M^{2}$ 
under protocol P2. P1 requires relay partitioning and the 
knowledge of one backward and one forward fading coefficient at each relay, whereas
P2 does not need relay partitioning, but requires that each relay knows all its
$M$ backward and $M$ forward fading coefficients. The protocols P1 and P2
are thus found to trade off CSI at the relays for the 
required (for the network to decouple) rate of growth of~$K$ as a function of~$M$.

We found that cooperation at the relay level in groups of~$L$ relays, both for P1 and P2, 
results in an $L$-fold reduction of the total number of relays needed to achieve a given
per source-destination terminal pair capacity. An interesting open question in this context
is whether more sophisticated signal processing at the relays (such as equalization for
the backward link and precoding for the forward link) could lead to improved capacity scaling behavior.

It was furthermore shown that the critical growth rates $K \propto M^{3}$ in P1 and $K \propto M^{2}$ in P2 are sufficient
to not only make the network decouple, but to also make the individual source-destination fading links
converge to nonfading links. We say that the {\em network ``crystallizes''}
as it breaks up into a set of {\em effectively isolated ``wires in the air''}. More pictorially, the decoupled links experience increasing distributed spatial (or more specifically relay) diversity.
Consequently, in the large-$M$ limit time diversity
(achieved by coding over a sufficiently long time horizon) is not needed to achieve ergodic capacity.
We furthermore characterized the ``crystallization'' rate (more precisely a guaranteed ``crystallization'' rate as we do not know whether our bounds are tight), 
i.e., the rate (as a function of~$M,K$) at which the decoupled links converge to nonfading links.
In the course of our analysis, we developed a new technique for characterizing the large-deviations behavior of
certain sums of dependent random variables. 

For noncoherent fading interference relay networks with amplify-and-forward relaying and joint decoding 
at the cooperating destination terminals, we 
computed the asymptotic (in~$M$ and~$K$ with $K/M\to\beta$ fixed) network capacity using tools
from large random-matrix theory. To the best of our knowledge, this is the first application of
large random-matrix theory to characterize the capacity behavior of large fading networks. An elegant extension of this approach to the case of multiple layers of relays was recently reported in~\cite{yeh07-06}. We furthermore
demonstrated that for $\beta\to\infty$ the relay network converges to a point-to-point MIMO
link. This generalizes the finite-$M$ result in~\cite{bolcskei06-} and shows that the use
of relays as active scatterers can recover spatial multiplexing gain in poor scattering environments,
even if the number of transmit and receive antennas grows large. More importantly, our result shows
that linear increase in the number of relays as a function of transmit-receive antennas is sufficient
for this to happen.

The large-deviations analysis, along with the notion of decoupling of the network, as carried out in this paper
could serve as a general tool to assess the impact of protocols, processing at the relays, propagation conditions, routing,
and scheduling on network outage and ergodic capacity performance. More specifically, an interesting question is under which
conditions ``crystallization'' can happen in a general network and, if it occurs, what the corresponding
``crystallization'' rate would be. It has to be noted, however, that, in view of the technical difficulties
posed by the basic case analyzed in this paper, it is unclear whether this framework
can yield substantial analytical insights into the above-mentioned questions.

Finally, we note that 
if we interpret our results in terms of per-node throughput, we find that
P1 achieves~$O\lefto(1/n^{2/3}\right)$ whereas P2 realizes~$O\lefto(1/\sqrt{n}\right)$. The scaling law for P2 is exactly the same as the behavior established by Gupta and Kumar 
in~\cite{gupta02-03} and the per-node throughput goes to zero.
On the other hand, it is interesting to observe that we can get an~$O\lefto(1/\sqrt{n}\right)$ throughput
without imposing any assumptions on the path-loss behavior.
General conclusions on the impact of fading on the network-capacity scaling law cannot be drawn
as we are considering a specific setup and specific protocols. 
It was recently shown~\cite{ozgur07-}, however, that under optimistic assumptions on CSI in the network $O\lefto(1\right)$ throughput can be achieved using hierarchical cooperation.

\section*{Acknowledgment}
The authors are indebted to Prof.~O.~Zeitouni for suggesting the application of the truncation
technique to establish the large-deviations behavior of the sums of dependent random variables occuring
in the proofs of Theorems~\ref{thm:P1Conv} and~\ref{thm:P2Conv}. We are furthermore grateful to Prof.~Zeitouni for pointing out an error in an earlier version of Theorem~\ref{lemma:trunc} and for suggesting the correction. Helpful discussions with Prof.~Zeitouni on noncoherent (AF) relay networks are acknowledged as well. We would furthermore
like to thank A.~Dana for pointing out that P2 as introduced in~\cite{dana03-} leads to decoupling of the network.

\appendices
\section{Truncation of Random Variables and Large Deviations}
\label{sec:Trunc} 
We start by recalling the famous Hoeffding inequality along with an important variation that will be central for our developments.
\begin{theorem}[Hoeffding~\cite{hoeffding63-03}]
\label{thm:hoeffding}
Let $X_1,X_2,\ldots,X_N$ be independent real-valued RVs and $A_n\le X_n \le B_n$ for $\natseg{n}{1}{N}$. Let~$S_N=\sum_{n=1}^N X_n$. Then,
\benn
\Prob\lefto\{S_N-\Ex{S_N}\ge N x\right\}\le \exp\lefto(-\frac{2N^2x^2}{\sum_{n=1}^N(B_n-A_n)^2}\right).
\eenn 
\end{theorem}

\begin{theorem}[Maurer~\cite{maurer03-}]
\label{thm:maurer}
Let $X_1,X_2,\ldots,X_N$ be independent real-valued RVs with $X_{n}\ge 0$ and $\Ex{X_{n}^{2}}<\infty$ for $\natseg{n}{1}{N}$. Let~$S_N=\sum_{n=1}^N X_n$. Then,
\benn
\Prob\lefto\{S_N-\Ex{S_N}\le - N x\right\}\le \exp\lefto(-\frac{N^2x^2}{2 \sum_{n=1}^N\Ex{X_{n}^{2}}}\right).
\eenn 
\end{theorem}

The following theorem builds on the Hoeffding inequality (Theorem~\ref{thm:hoeffding}) and constitutes the core of the truncation technique.
\begin{theorem}
\label{lemma:trunc}
Assume on a common probability space:
\begin{itemize}
\item
The real-valued RVs~$X_{1}, X_{2}, \ldots, X_{N}$ (possibly dependent) have marginal distribution functions~$\cdf{X_{n}}(x)$, $\natseg{n}{1}{N}$. The tails of these distributions are exponentially decaying uniformly in~$n$, i.e., there exist $B>0,\ \alpha>0,\ \beta>0$, and $x_{0}>0$ s.t.\ for $x\ge x_0>0$ and $\natseg{n}{1}{N}$
\ba
\Prob\mathopen{}\Bigl\{\abs{X_{n}}\ge x\Bigr\}&=1-\cdf{X_n}(x)+\cdf{X_n}(-x)
\le B e^{-\alpha\, x^\beta}.
\label{eq:DecayCond}
\ea
\item
The real-valued RVs $\phi_1, \phi_{2},\ldots,\phi_{N}$ are jointly independent  
and satisfy 
\benn
-1\le \phi_n \le 1,\quad \Ex{\phi_n}=0,\qquad  \natseg{n}{1}{N}.
\eenn
\item
The real-valued deterministic nonnegative coefficients $A_{1}, A_{2},\ldots, A_{N}$ are uniformly bounded from above, i.e., there exists a constant~$A$ independent of~$n$ s.t.\
\benn
0\le A_n \le A,\qquad  \natseg{n}{1}{N}.
\eenn
\item
The set of RVs $\left\{X_{n}\right\}_{n=1}^{N}$ is independent of the set $\left\{\phi_{n}\right\}_{n=1}^{N}$.
\end{itemize}

Let $S_N=\sum_{n=1}^N A_n X_n \phi_n$.
Then, for all~$N>0$ and~$x>0$ s.t.\ $x\ge x_0^{(2+\beta)/2}$
\bmult
\Prob\lefto\{\abs{S_N}\ge \sqrt{N} x\right\}\\
\le 
2\max\lefto[2,N B\right]\exp\lefto(-\min\lefto[\frac{1}{2 A^{2}},\alpha\right]x^{\frac{2\beta}{2+\beta}}\right).
\label{eq:truncbound}
\emult
\end{theorem}
\begin{IEEEproof}
The proof is based on the idea of truncation 
of the RVs~$X_{n}$. We start by fixing $N$ and choosing $t$ s.t.\ $\left(N t^2\right)^{\gamma}\ge x_0$. The truncation parameter~$0<\gamma<1$ will be chosen later.  Next, we truncate the RVs $X_n$, $\natseg{n}{1}{N}$, 
according to 
\benn
\hat{X}_n\define X_n\, I\lefto[\abs{X_n}\le \left(N t^2\right)^\gamma\right].%
\eenn%
Define $\hat{S}_N\define\sum_{n=1}^N A_n \hat{X}_n \phi_n$. Note that the independence of $\left\{X_{n}\right\}_{n=1}^{N}$ and $\left\{\phi_{n}\right\}_{n=1}^{N}$ and the condition $\Ex{\phi_n}=0$ $(\natseg{n}{1}{N})$ implies that $\Ex{S_{N}}=\Ex{\hat{S}_{N}}=0$. Let $I_{n}$ denote the event that $X_{n}$ is equal to its truncated version, i.e., $I_{n}\define\bigl\{X_{n}=\hat{X}_{n}\bigr\}$ and, $\bar{I}_{n}$ the event that $X_{n}\ne \hat{X}_{n}$, i.e., $\bar{I}_{n}\define\bigl\{X_{n}\ne\hat{X}_{n}\bigr\}$. With these definitions,
distinguishing the events where either all~$X_{n}$ are equal to their truncated version, i.e., $\Intersect_{n=1}^N I_{n}$ and where at least one of the~$X_{n}$ is not equal to its truncated version, i.e., $\Union_{n=1}^N \bar{I}_{n}$, we get 
\ba
&\Prob\lefto\{\abs{S_N}\ge N t\right\}\nonumber\\
&\phantom{\Prob\lefto\{\right.}=\Prob\lefto\{\abs{S_N}\ge N t\setgiven \Intersect_{n=1}^N I_{n}\right\}\Prob\lefto\{\Intersect_{n=1}^N I_{n}\right\}\nonumber\\
&\phantom{\Prob\lefto\{\right.}\relphantom{=}\phantom{\Prob\lefto\{\abs{S_N}\right.}{}+\Prob\lefto\{\abs{S_N}\ge N t\setgiven \Union_{n=1}^N \bar{I}_{n}\right\}\Prob\lefto\{\Union_{n=1}^N \bar{I}_{n}\right\}\nonumber\displaybreak[0]\\
&\phantom{\Prob\lefto\{\right.}=\Prob\lefto\{\abs{\hat{S}_N}\ge N t\right\}\Prob\lefto\{\Intersect_{n=1}^N I_{n}\right\}\nonumber\\
&\phantom{\Prob\lefto\{\right.}\relphantom{=}\phantom{\Prob\lefto\{\abs{S_N}\right.}{}+\Prob\lefto\{\abs{S_N}\ge N t\setgiven \Union_{n=1}^N \bar{I}_{n}\right\}\Prob\lefto\{\Union_{n=1}^N \bar{I}_{n}\right\}\nonumber\displaybreak[0]\\
&\phantom{\Prob\lefto\{\right.}\le\Prob\lefto\{\abs{\hat{S}_N}\ge N t\right\}+\sum_{n=1}^N \Prob\lefto\{\bar{I}_{n}\right\}
\label{eq:TruncTotal}
\ea
where the last step follows by using the trivial bounds
\benn
\Prob\lefto\{\Intersect_{n=1}^N I_{n}\right\}\le 1,\qquad \Prob\lefto\{\abs{S_N}\ge N t\setgiven \Union_{n=1}^N \bar{I}_{n}\right\}\le 1\nonumber
\eenn
and applying the union bound to $\Prob\mathopen{}\bigl\{\Union_{n=1}^N \bar{I}_{n}\bigr\}$. Since $-1\le \phi_{n}\le 1, \natseg{n}{1}{N}$, we obtain the following bounds for the individual terms in $\hat{S}_{N}$
\benn
-A_{n}\!\left(N t^{2}\right)^{\gamma}\le A_n \hat{X}_n \phi_n \le A_{n}\!\left(N t^{2}\right)^{\gamma}, \ \ \natseg{n}{1}{N}.
\eenn
Moreover, owing to the independence of the $\phi_n$, conditioned on the set $\setX\define\bigl\{\hat{X}_{1}, \hat{X}_{2},\ldots, \hat{X}_{N}\bigr\}$, the RVs $A_n \hat{X}_n \phi_n$ are independent.
Therefore, using Bayes' rule and the Hoeffding inequality (Theorem~\ref{thm:hoeffding}), noting that $\Exop\mathopen{}\bigl[\hat{S}_N \given \setX\bigr]=0$,  we can conclude that
\ba
\Prob\lefto\{\abs{\hat{S}_N}\ge N t\right\} &=
\Exop_{\setX}\lefto[\Prob\lefto\{\abs{\hat{S}_N-\Ex{\hat{S}_N \Big| \setX} }\ge N t \setgiven\setX\right\}\right]\nonumber\\
&\le 2\exp\lefto(-\frac{N^2 t^2}{2\sum_{n=1}^N A_n^{2} \left(Nt^2\right)^{2\gamma}}\right)
\nonumber\\
&\le 2\exp\lefto(-\frac{\left(N t^2\right)^{1-2\gamma}}{2A^2}\right).
\label{eq:TruncP1}
\ea
Next, using~\eqref{eq:DecayCond}, and assuming [this will be justified in~\eqref{eq:TresholdCheck}] that~$\left(N t^2\right)^\gamma\ge x_{0}$, we have
\ba
\Prob\lefto\{\bar{I}_{n}\right\}&=\Prob\lefto\{X_{n}\ne\hat{X}_{n}\right\}\nonumber\\
&
=\Prob\lefto\{\abs{X_n}\ge \left(N t^2\right)^\gamma\right\}\le B e^{-\alpha \left(N t^2\right)^{\gamma\beta}}.
\label{eq:TruncP2}
\ea
To get the fastest possible exponential decay in~\eqref{eq:TruncTotal}, we need to choose the free parameter~$\gamma$ s.t.\ it maximizes $\min\lefto[1-2\gamma, \gamma\beta\right]$, which is the solution that makes the exponents of~$t$ in~\eqref{eq:TruncP1} and~\eqref{eq:TruncP2} equal and is given by $\gamma=1/(2+\beta)$. Finally, setting~$t=x/\sqrt{N}$ results in
\be
\label{eq:TresholdCheck}
\left(N t^2\right)^\gamma=x^{2\gamma}=x^{2/(2+\beta)}\ge x_{0}
\ee
as required.
Combining~\eqref{eq:TruncTotal},~\eqref{eq:TruncP1} and~\eqref{eq:TruncP2}, we finally obtain
\bmult
\Prob\lefto\{\abs{S_N}\ge \sqrt{N} x\right\}
\\ \le 
2\exp\lefto(-\frac{1}{2 A^{2}} x^{\frac{2\beta}{2+\beta}}\right)+
N B \exp\lefto(-\alpha\, x^{\frac{2\beta}{2+\beta}}\right).
\label{eq:truncboundtight}
\emult
The final result~\eqref{eq:truncbound} is a trivial upper bound to~\eqref{eq:truncboundtight}.
\end{IEEEproof}

The following corollary is the generalization of Theorem~\ref{lemma:trunc} to the complex-valued case and will be used repeatedly in the proofs of Theorems~\ref{thm:P1Conv} and~\ref{thm:P2Conv}. 
\begin{corollary}
\label{cor:trunc}
Assume on a common probability space:
\begin{itemize}
\item
The absolute values of the complex-valued (possibly dependent) RVs $X_{1}, X_{2}, \ldots, X_{N}$  have marginal distribution functions $\cdf{X_{n}}(x)$, $\natseg{n}{1}{N}$. The tails of these distributions are exponentially decaying uniformly in $n$, i.e., there exist $B>0,\ \alpha>0,\ \beta>0$ and $x_{0}>0$ s.t.\ for $x\ge x_0>0$ and $\natseg{n}{1}{N}$
\be
\Prob\lefto\{\abs{X_{n}}\ge x\right\}=1-\cdf{X_n}(x)\le B e^{-\alpha\, x^\beta}.
\label{eq:DecayCondCor}
\ee
\item
The real-valued RVs $\phi_1, \phi_{2},\ldots,\phi_{N}$ are jointly independent and satisfy $\phi_{n}\sim \uniform\left(-\pi,\pi\right)$ and hence $\Ex{e^{j \phi_n}}=0$ for all $\natseg{n}{1}{N}$.
\item
The real-valued deterministic nonnegative coefficients $A_{1}, A_{2},\ldots, A_{N}$ are uniformly bounded from above, i.e., there exists a constant~$A$ independent of~$n$ s.t.\
\benn
0\le A_n \le A,\qquad  \natseg{n}{1}{N}.
\eenn
\item
The set of RVs $\left\{X_{n}\right\}_{n=1}^{N}$ is independent of the set $\left\{\phi_{n}\right\}_{n=1}^{N}$.
\end{itemize}

Let $S_N=\sum_{n=1}^N A_n X_n e^{j \phi_n}$.
Then, for all~$N>0$ and~$x>0$ s.t.\ $x\ge x_0^{(2+\beta)/2}$
\bmnn
\Prob\lefto\{\abs{S_N}\ge \sqrt{N} x\right\}\\
\le 
4\max\lefto[2,N B\right]\exp\lefto(-\min\lefto[\frac{1}{2 A^{2}},\alpha\right] 2^{-\frac{\beta}{\beta+2}} x^{\frac{2\beta}{\beta+2}}\right).
\emnn
\end{corollary}
\begin{IEEEproof}
Apply Theorem~\ref{lemma:trunc} to $\Re{S_{N}}$ and $\Im{S_{N}}$ separately and combine the two bounds using the Pythagorean union bound (Lemma~\ref{lemma:UBPyth}). 
\end{IEEEproof}

The following corollary is a modification of Theorem~\ref{lemma:trunc} for the case of independent nonnegative RVs and will be used repeatedly in the proofs of Theorems~\ref{thm:P1Conv} and~\ref{thm:P2Conv}.
\begin{corollary}
\label{cor:trunc1}
Assume on a common probability space:
\begin{itemize}
\item
The real-valued nonnegative RVs~$X_{1}, X_{2}, \ldots, X_{N}$ are jointly independent and have marginal distribution functions~$\cdf{X_{n}}(x)$, $\natseg{n}{1}{N}$. The right tails of these distributions are exponentially decaying uniformly in~$n$, i.e., there exist $B>0,\ \alpha>0,\ \beta>0$ and $x_{0}>0$ s.t.\ for all $x\ge x_0>0$ and $\natseg{n}{1}{N}$
\be
\Prob\lefto\{X_{n}\ge x\right\}=1-\cdf{X_n}(x)\le B e^{-\alpha\, x^\beta}.
\label{eq:DecayCondCor2}
\ee
\item
The expectations~$\Ex{X_{n}^{2}}$ are uniformly bounded from above, i.e., there exists a constant~$C$ independent of~$n$ s.t.
\be
\Ex{X_{n}^{2}}\le C,\qquad \natseg{n}{1}{N}.
\label{eq:XnBound}
\ee
\item
The real-valued deterministic nonnegative coefficients $A_{1}, A_{2},\ldots, A_{N}$ are uniformly bounded from above, i.e., there exists a constant~$A$ independent of~$n$ s.t.\
\be
0\le A_n \le A,\qquad \natseg{n}{1}{N}.
\label{eq:AnBounds}
\ee
\end{itemize}

Let $S_N=\sum_{n=1}^N A_n X_n$.
Then, for all~$N>0$ and~$x>0$ s.t.\ $x\ge x_0^{(2+\beta)/2}$
\ba
&\hspace{-2mm}\Prob\lefto\{\abs{S_N-\Ex{S_N}}\ge \sqrt{N} x\right\}\nonumber \\
&\le 
3\max\lefto[1,N B\right]\exp\lefto(\!-\min\lefto[\frac{2}{A^{2}},\alpha, \frac{1}{2 A^{2} C} \right]\! x^{\frac{2\beta}{\beta+2}}\right).
\label{eq:truncbound2}
\ea
\end{corollary}
\begin{IEEEproof}
The proof idea of this corollary is similar to that used in Theorem~\ref{lemma:trunc}. However, there are several technical details, which do not occur in the proof of Theorem~\ref{lemma:trunc}. We have, therefore, decided to present the full version of the proof of Corollary~\ref{cor:trunc1}. 

Unlike in the proof of Theorem~\ref{lemma:trunc}, here we have~$\Ex{S_{N}}\ne 0$. To obtain an upper bound on $\Prob\mathopen{}\bigl\{\abs{S_N-\Ex{S_N}}\ge \sqrt{N} x\bigr\}$, we establish an upper bound on $\Prob\mathopen{}\bigl\{S_N\ge \Ex{S_N}+\sqrt{N} x\bigr\}$ and on $\Prob\mathopen{}\bigl\{S_N\le \Ex{S_N}-\sqrt{N} x\bigr\}$ and use the union bound to combine the results. 

We start by deriving an upper bound on $\Prob\mathopen{}\bigl\{S_N\ge \Ex{S_N}+\sqrt{N} x\bigr\}$.
Following the same steps as in the proof of Theorem~\ref{lemma:trunc}, we define the truncation parameter~$0<\gamma<1$, which will be chosen later. Fix $N$ and choose $t$ s.t.\ $\left(N t^2\right)^{\gamma}\ge x_0$. We truncate the RVs $X_n$ $\left(\natseg{n}{1}{N}\right)$ 
according to 
\benn
\hat{X}_n\define X_n\, I\lefto[X_n\le \left(N t^2\right)^\gamma\right]
\eenn
and define $\hat{S}_N\define\sum_{n=1}^N A_n \hat{X}_n$. It is easily seen that $\Ex{S_{N}}\ge \Ex{\hat{S}_{N}}$ and therefore
\be
\Prob\mathopen{}\Bigl\{S_N\ge \Ex{S_N}+N t\Bigr\}\le \Prob\lefto\{S_N\ge \Ex{\hat{S}_N}+N t\right\}.
\label{eq:hatSbound2}
\ee
Let $I_{n}$ denote the event that $X_{n}$ is equal to its truncated version, i.e.,~$I_{n}\define\bigl\{X_{n}=\hat{X}_{n}\bigr\}$, and $\bar{I}_{n}$ the event that~$X_{n}\ne \hat{X}_{n}$, i.e., $\bar{I}_{n}\define\bigl\{X_{n}\ne\hat{X}_{n}\bigr\}$. With these definitions,
distinguishing the events where either all~$X_{n}$ are equal to their truncated version, i.e., $\Intersect_{n=1}^N I_{n}$ and where at least one of the~$X_{n}$ is not equal to its truncated version, i.e., $\Union_{n=1}^N \bar{I}_{n}$, we get 
\ba
&\Prob\lefto\{S_N\ge \Ex{\hat{S}_N}+N t\right\}\nonumber\\
&\phantom{\Prob\lefto\{\right.}=\Prob\lefto\{S_N\ge \Ex{\hat{S}_N}+N t\setgiven \Intersect_{n=1}^N I_{n}\right\}\Prob\lefto\{\Intersect_{n=1}^N I_{n}\right\}\nonumber\\
&\phantom{\Prob\lefto\{\right.}\relphantom{=}{+}\:\Prob\lefto\{S_N\ge \Ex{\hat{S}_N}+N t\setgiven \Union_{n=1}^N \bar{I}_{n}\right\}\Prob\lefto\{\Union_{n=1}^N \bar{I}_{n}\right\}\nonumber \displaybreak[0]\\
&\phantom{\Prob\lefto\{\right.}=\Prob\lefto\{\hat{S}_N\ge \Ex{\hat{S}_N}+N t\right\}\Prob\lefto\{\Intersect_{n=1}^N I_{n}\right\}\nonumber\\
&\phantom{\Prob\lefto\{\right.}\relphantom{=}{+}\:\Prob\lefto\{S_N\ge \Ex{\hat{S}_N}+N t\setgiven \Union_{n=1}^N \bar{I}_{n}\right\}\Prob\lefto\{\Union_{n=1}^N \bar{I}_{n}\right\}\nonumber\\
&\phantom{\Prob\lefto\{\right.}\le 
\Prob\lefto\{\hat{S}_N\ge \Ex{\hat{S}_N}+N t\right\}+\sum_{n=1}^N \Prob\lefto\{\bar{I}_{n}\right\}
\label{eq:TruncTotal2}
\ea
where the last step is obtained by using the trivial bounds
\benn
\Prob\lefto\{\Intersect_{n=1}^N I_{n}\right\}\le 1,\qquad \Prob\lefto\{S_N\ge \Ex{\hat{S}_N}+N t\setgiven \Union_{n=1}^N \bar{I}_{n}\right\}\le 1\nonumber
\eenn
and applying the union bound to $\Prob\mathopen{}\bigl\{\Union_{n=1}^N \bar{I}_{n}\bigr\}$. The individual terms in~$\hat{S}_{N}$ are bounded according to
\benn
0\le A_n \hat{X}_n \le A_{n}\!\left(N t^{2}\right)^{\gamma}, \ \ \natseg{n}{1}{N}.
\eenn
Using Bayes' rule and the Hoeffding inequality (Theorem~\ref{thm:hoeffding}), we can conclude that
\ba
\Prob\lefto\{\hat{S}_N\ge \Ex{\hat{S}_N}+N t\right\}
&\le \exp\lefto(-\frac{2 N^2 t^2}{\sum_{n=1}^N A_n^{2} \left(Nt^2\right)^{2\gamma}}\right)\nonumber\\
&\le \exp\lefto(-\frac{2 \left(N t^2\right)^{1-2\gamma}}{A^2}\right).
\label{eq:TruncP12}
\ea
Next, using~\eqref{eq:DecayCondCor2}, and assuming [this will be justified in~\eqref{eq:TresholdCheck2}] that $\bigl(N t^2\bigr)^\gamma\ge x_{0}$, we have
\ba
\Prob\lefto\{\bar{I}_{n}\right\}&=\Prob\lefto\{X_{n}\ne\hat{X}_{n}\right\}\nonumber\\
&
=\Prob\lefto\{X_n\ge \left(N t^2\right)^\gamma\right\}\le B e^{-\alpha \left(N t^2\right)^{\gamma\beta}}.
\label{eq:TruncP22}
\ea
To get the fastest possible exponential decay in~\eqref{eq:TruncTotal2}, we need to choose the free parameter~$\gamma$ s.t.\ it maximizes $\min\mathopen{}\bigl[1-2\gamma, \gamma\beta\bigr]$, which is the solution that makes the exponents of~$t$ in~\eqref{eq:TruncP12} and~\eqref{eq:TruncP22} equal and is given by $\gamma=1/(2+\beta)$. Finally, setting $t=x/\sqrt{N}$ results in
\be
\label{eq:TresholdCheck2}
\left(N t^2\right)^\gamma=x^{2\gamma}=x^{2/(2+\beta)}\ge x_{0}
\ee
as required.
Combining~\eqref{eq:hatSbound2}-\eqref{eq:TruncP22}, we obtain
\bmult
\Prob\lefto\{S_N\ge \Ex{S_N}+\sqrt{N} x\right\}\\
\le 
\exp\lefto(-\frac{2}{A^{2}} x^{\frac{2\beta}{2+\beta}}\right)+
N B \exp\lefto(-\alpha\, x^{\frac{2\beta}{2+\beta}}\right).
\label{eq:truncboundtight2}
\emult

It remains to establish an upper bound on $\Prob\mathopen{}\bigl\{S_N\le \Ex{S_N}-\sqrt{N} x\bigr\}$. From Theorem~\ref{thm:maurer} it follows that
\benn
\Prob\lefto\{S_N\le \Ex{S_N}-\sqrt{N} x\right\}\le \exp\lefto(-\frac{N x^{2}}{2 \sum_{n=1}^N \Ex{A_{n}^{2} X_{n}^{2}}}\right)
\eenn
which, using~\eqref{eq:XnBound} and~\eqref{eq:AnBounds}, can be further upper-bounded as
\be
\Prob\lefto\{S_N\le \Ex{S_N}-\sqrt{N} x\right\}\le \exp\lefto(-\frac{x^{2}}{2 A^{2} C}\right).
\label{eq:SNlower}
\ee

Combining~\eqref{eq:truncboundtight2} and~\eqref{eq:SNlower} and using the union bound, we obtain
\bmult
\Prob\lefto\{\abs{S_N-\Ex{S_N}}\ge \sqrt{N} x\right\}
\le \exp\lefto(\!-\frac{2}{A^{2}} x^{\frac{2\beta}{2+\beta}}\!\right)\\+
N B \exp\lefto(\!-\alpha\, x^{\frac{2\beta}{2+\beta}}\!\right)+\exp\lefto(\!-\frac{x^{2}}{2 A^{2} C}\!\right).
\label{eq:truncboundtightfin}
\emult

The final result~\eqref{eq:truncbound2} is a trivial upper bound to~\eqref{eq:truncboundtightfin}.
\end{IEEEproof}

\section{Union Bounds}
\label{sec:union}
In this appendix, as a reference, we present several variations of union bounds for probability that we use frequently throughout the paper. \
\begin{lemma}[Union bound for sums]
\label{lemma:UBsums}
Assume the complex-valued RVs $X_{1}, X_{2}, \ldots,X_{N}$ are s.t.\ 
\benn
\Prob\mathopen{}\Bigl\{\abs{X_n}\ge C_n\Bigr\}\le P_n,\qquad \natseg{n}{1}{N}
\eenn
where $C_{1}, C_{2},\ldots, C_{N}$ and $P_{1}, P_{2},\ldots, P_{N}$ are fixed positive constants. Then,
\benn
\Prob\lefto\{\abs{\sum_{n=1}^{N} X_n}\ge\sum_{n=1}^{N}C_n\right\}\le \sum_{n=1}^{N} P_n.
\eenn
\end{lemma}
\begin{IEEEproof}
Let $A_n$ denote the event that $\abs{X_n}\ge C_n,\, \natseg{n}{1}{N}$. Let $B$ denote the event that $\abs{\sum_{n=1}^{N} X_n}\ge \sum_{n=1}^{N} C_n$. By inspection, it follows that $B \Rightarrow \Union_{n=1}^{N} A_n$, which implies $
\Prob\lefto\{B\right\}\le \sum_{n=1}^{N} \Prob\{A_{n}\}$.
\end{IEEEproof}
The proofs of the remaining union bounds follow exactly the same pattern as the proof of Lemma~\ref{lemma:UBsums} and will hence be omitted.

\begin{lemma}[Pythagorean union bound]
\label{lemma:UBPyth}
Assume the complex-valued RV~$X$ is s.t.\ 
\benn
\Prob\mathopen{}\Bigl\{\abs{\Re X}\ge C_{\mathrm{R}}\Bigr\}\le P_{\mathrm{R}}
\text{ and }
\Prob\mathopen{}\Bigl\{\abs{\Im X}\ge C_{\mathrm{I}}\Bigr\}\le P_{\mathrm{I}}
\eenn
where $C_{\mathrm{R}}, C_{\mathrm{I}}, P_{\mathrm{R}},$ and $P_{\mathrm{I}}$ are fixed positive constants. Then,
\benn
\Prob\mathopen{}\Bigl\{\abs{X}\ge\sqrt{C_{\mathrm{R}}^{2}+C_{\mathrm{I}}^{2}}\Bigr\}\le P_{\mathrm{R}}+P_{\mathrm{I}}.
\eenn
\end{lemma}

\begin{lemma}[Union bound for mixed sums]
\label{lemma:UBsumsMixed}
Assume that the complex-valued RVs~$X_{1}, X_{2}, \ldots,X_{N}$ are s.t.\ 
\benn
\Prob\mathopen{}\Bigl\{\abs{X_n}\ge C_n\Bigr\}\le P_n,\qquad \natseg{n}{1}{N}
\eenn
where $C_{1}, C_{2},\ldots, C_{N}$ and $P_{1}, P_{2},\ldots, P_{N}$ are fixed positive constants;
then, the following statements hold:
\begin{enumerate}
\item
If the real-valued RVs $X_{1}', X_{2}', \ldots,X_{N'}'$ are s.t.\ 
\benn
\Prob\mathopen{}\Bigl\{X'_n\le C_n'\Bigr\}\le P_n',\qquad \natseg{n}{1}{N}
\eenn
where $C'_{1}, C'_{2},\ldots, C'_{N}$ and $P'_{1}, P'_{2},\ldots, P'_{N}$ are fixed positive constants,
then
\bann
\Prob\lefto\{\abs{\sum_{n=1}^{N} X_n+\sum_{n=1}^{N'} X'_n}\le\max\lefto[0, \sum_{n=1}^{N'} C'_n - \sum_{n=1}^{N} C_n\right]\right\}\\
\quad\le \sum_{n=1}^{N} P_n+\sum_{n=1}^{N'} P'_n.
\eann
\item
If the real-valued RVs $X_{1}', X_{2}', \ldots,X_{N'}'$ are s.t.\ 
\benn
\Prob\mathopen{}\Bigl\{X'_n\ge C_n'\Bigr\}\le P_n',\qquad \natseg{n}{1}{N}
\eenn
then,
\bmnn
\Prob\lefto\{\abs{\sum_{n=1}^{N} X_n+\sum_{n=1}^{N'} X'_n}\ge\sum_{n=1}^{N'} C'_n + \sum_{n=1}^{N} C_n\right\}\\
\le 
\sum_{n=1}^{N} P_n+\sum_{n=1}^{N'} P'_n.
\emnn
\end{enumerate}
\end{lemma}

\begin{lemma}[Union bound for products]
\label{lemma:UBprods}
Assume the complex-valued RVs $X_{1}, X_{2}, \ldots,X_{N}$ are such that 
\benn
\Prob\mathopen{}\Bigl\{\abs{X_n}\ge C_n\Bigr\}\le P_n,\ \natseg{n}{1}{N}
\eenn
where $C_{1}, C_{2},\ldots, C_{N}$ and $P_{1}, P_{2},\ldots, P_{N}$ are fixed positive constants. Then,
\benn
\Prob\lefto\{\abs{\prod_{n=1}^{N} X_n}\ge\prod_{n=1}^{N}C_n\right\}\le \sum_{n=1}^{N} P_n.
\eenn
\end{lemma}

\begin{lemma}[Union bound for fractions]
\label{lemma:UBfrac}
If for real-valued positive RVs $X_{1}$ and $X_{2}$ and positive constants $C_{1}, C_{2}$ and $P_{1}, P_{2}$
\benn
\Prob\mathopen{}\Bigl\{X_1\ge C_1\Bigr\}\le P_1\text{ and } \Prob\mathopen{}\Bigl\{X_2\le C_2\Bigr\}\le P_2
\eenn
then 
\benn
\Prob\mathopen{}\Bigl\{X_{1}/X_{2}\ge C_{1}/C_{2}\Bigr\}\le P_1+P_2.
\eenn
If, in turn,
\benn
\Prob\mathopen{}\Bigl\{X_1\le C_1\Bigr\}\le P_1 \text{ and } \Prob\mathopen{}\Bigl\{X_2\ge C_2\Bigr\}\le P_2
\eenn
then 
\benn
\Prob\mathopen{}\Bigl\{X_{1}/X_{2}\le C_{1}/C_{2}\Bigr\}\le P_1+P_2.
\eenn
\end{lemma}

\section{Proof of Theorem~\ref{thm:P1Conv}}
\label{appendix:P1ConvProof}
We start by recalling that we want to establish a concentration result for~$\mathrm{SINR}^{\mathrm{P1}}_{m}$, given by~\eqref{eq:SINRP1Sums}, using the truncation technique throughout. As already mentioned, this entails establishing the large-deviations behavior of~$S^{(1)}, S^{(2)}, S^{(3)}$, and~$S^{(4)}$. For~$S^{(3)}$, this has already been done in Section~\ref{sec:AppTrunc}. It remains to establish the corresponding (based on the truncation technique) concentration results for~$S^{(1)}$, $S^{(2)}$, and $S^{(4)}$ defined by~\eqref{eq:S1def}, \eqref{eq:S2def}, and~\eqref{eq:S4def}, respectively. 
\begin{figure*}[!t]
\normalsize
\setcounter{MYtempeqncnt}{\value{equation}}
\setcounter{equation}{136}
\ba
\label{eq:POutP1App}
P_{\mathrm{P1}}^{\mathrm{U}}& \define  6\, \frac{K}{M} e^{-\Delta^{\!(1)} x_{1}^{2/3}}
+8\, \frac{K (M-1)}{M} e^{-\Delta^{\!(2)} x_{2}^{2/3}}
+6\, (M-1) K e^{-\Delta^{\!(31)} x_{31}^{2/5}}
\nonumber\\&\qquad\quad{+}\: 
64\, \frac{(K-1)K(M-1)^{2}}{M}  e^{-\Delta^{\!(32)} x_{321}^{2/7}}
+64\, \frac{(K-1)K(M-1)}{M}  e^{-\Delta^{\!(32)} x_{322}^{2/7}}
+3 K\! e^{-\Delta^{\!(4)} x_{4}^{2/3}}\\
\label{eq:UP1N}
\hat{U}_{\mathrm{P1}}^{\mathrm{N}}&\define \left(1+\frac{4}{\overline{C} \pi}\sqrt{\frac{M}{K}}\, x_{1}+\frac{4}{\overline{C} \pi}\sqrt{\frac{M (M-1)}{K}}\, x_{2}\right)^{2}\\
\label{eq:UP1D}
\hat{U}_{\mathrm{P1}}^{\mathrm{D}}&\define \max\lefto[0,\frac{\underline{C}^{2}}{\underline{C}_{\mathrm{SN}}^{2}}\frac{M-1}{M}-\frac{1}{\underline{C}_{\mathrm{SN}}^{2}}\frac{M-1}{M\sqrt{K}}\, x_{31}
-\frac{1}{\underline{C}_{\mathrm{SN}}^{2}}\sqrt{\frac{(K-1)(M-1)^{2}}{K M^{3}}}\, x_{321}\right.
\nonumber\\&\qquad\quad{-}\:
\left.\frac{1}{\underline{C}_{\mathrm{SN}}^{2}}\sqrt{\frac{(K-1)(M-1)}{K M^{3}}}\, x_{322}\right]
+
\frac{\sigma^{2}}{\underline{C}_{\mathrm{SN}}^{2}}\max\lefto[0,\underline{c}^{2}-\frac{1}{\sqrt{K}}\, x_{4}\right]+\frac{\sigma^{2}}{\underline{C}^{2}_{\mathrm{SN}}}
\ea
\setcounter{equation}{\value{MYtempeqncnt}}
\hrulefill
\vspace*{-5pt}
\end{figure*}

\subsection{Analysis of $S^{(1)}$}
The sum~$S^{(1)}$ can be written as
\be
\label{eq:S1P1DefMod}
S^{(1)}=\sum_{k: p(k)=m} C_{\mathrm{P1}, k}^{m,m} Z_{k}^{(1)}
\ee
with
\benn
Z_{k}^{(1)}\define \abs{f_{m,k}} \abs{h_{k,m}}.
\eenn
For any~$\natseg{k}{1}{K}$ s.t.\ $p(k)=m$, we have~$\Ex{Z_{k}^{(1)}}=\pi/4$ and $\Ex{\bigl(Z_{k}^{(1)}\bigr)^{2}}=1$. Application of the union bound for products yields
\benn
\Prob\lefto\{Z_{k}^{(1)} \ge x\right\}\le 2 e^{-x},\qquad x\ge 0.
\eenn
Noting that the sum~$S^{(1)}$ contains~$K/M$ terms, which are jointly independent, taking into account~\eqref{eq:P1Cbound}, and using Corollary~\ref{cor:trunc1}, we get for~$x\ge 0$ and~$K/M\ge 1$
\bann
&\Prob\lefto\{\abs{S^{(1)}-\frac{\pi}{4}\sum_{k: p(k)=m}C_{\mathrm{P1}, k}^{m,m}}\ge \sqrt{\frac{K}{M}} x\right\}
\le 
6 \frac{K}{M} e^{-\Delta^{\!(1)} x^{2/3}}
\eann
with $\Delta^{\!(1)}=\min\mathopen{}\bigl[1,1/\big(2\,\overline{C}^{\,2}\big) \bigr]$.
Finally, using~\eqref{eq:P1Cbound}, it follows that
\be
\Prob\lefto\{S^{(1)}\ge \frac{\pi}{4}\overline{C} \frac{K}{M} + \sqrt{\frac{K}{M}}\, x\right\}\le
6\, \frac{K}{M} e^{-\Delta^{\!(1)} x^{2/3}}
\label{eq:S1P1u}
\ee
and
\be
\Prob\lefto\{S^{(1)}\le \frac{\pi}{4}\underline{C} \frac{K}{M} - \sqrt{\frac{K}{M}}\, x\right\}\le
6 \frac{K}{M} e^{-\Delta^{\!(1)} x^{2/3}}
\label{eq:S1P1l}
\ee
for any~$x\ge 0$ and~$K/M\ge 1$.
\subsection{Analysis of $S^{(2)}$}
The sum $S^{(2)}$ can be written as
\benn
S^{(2)}=\sum_{k: p(k)\ne m} C_{\mathrm{P1}, k}^{m,m}\, Z_{k}^{(2)}
\eenn
with
\benn
Z_{k}^{(2)}\define\conj{\tilde{f}_{p(k)\!,k}}\, f_{m,k}\, \conj{\tilde{h}_{k,p(k)}}\, h_{k,m}.
\eenn
For any $\natseg{k}{1}{K}$ s.t.\ $p(k)\ne m$, we have $\Ex{Z_{k}^{(2)}}=0$. Application of the union bound for products yields
\benn
\Prob\lefto\{\abs{Z_{k}^{(2)}} \ge x\right\}\le 2 e^{-x},\qquad x\ge 0.
\eenn
Noting that the sum~$S^{(2)}$ contains~$K (M-1)/M$ terms, which are jointly independent, taking into account~\eqref{eq:P1Cbound}, and using Corollary~\ref{cor:trunc}, we get for $x\ge 0$ and $K(M-1)/M\ge 1$
\be
\Prob\lefto\{\abs{S^{(2)}}\ge \sqrt{\frac{K (M-1)}{M}}\, x\right\}
\le 
8 \frac{K (M-1)}{M} e^{-\Delta^{\!(2)} x^{2/3}}
\label{eq:S2P1lu}
\ee
with $\Delta^{\!(2)}=2^{-\frac{1}{3}} \min\mathopen{}\bigl[1,1/\big(2\, \overline{C}^{\,2}\big) \bigr]$.
\begin{figure*}[!t]
\normalsize
\setcounter{MYtempeqncnt}{\value{equation}}
\setcounter{equation}{141}
\ba
\label{eq:LP1N}
\hat{L}_{\mathrm{P1}}^{\mathrm{N}}&\define \max\lefto[0,1-\frac{4}{\underline{C} \pi}\sqrt{\frac{M}{K}}\, x_{1}-\frac{4}{\underline{C} \pi}\sqrt{\frac{M (M-1)}{K}}\, x_{2}\right]^{2}\\
\label{eq:LP1D}
\hat{L}_{\mathrm{P1}}^{\mathrm{D}}&\define \frac{\overline{C}^{2}}{\overline{C}_{\mathrm{SN}}^{2}}\frac{M-1}{M}+\frac{1}{\overline{C}_{\mathrm{SN}}^{2}}\frac{M-1}{M\sqrt{K}}\, x_{31}+\frac{1}{\overline{C}_{\mathrm{SN}}^{2}}\sqrt{\frac{(K-1)(M-1)^{2}}{K M^{3}}}\, x_{321}
\nonumber\\&\qquad\qquad\qquad\qquad\qquad\qquad\qquad\qquad\quad{+}\:\frac{1}{\overline{C}_{\mathrm{SN}}^{2}}\sqrt{\frac{(K-1)(M-1)}{K M^{3}}}\, x_{322}
+\frac{\sigma^{2}}{\overline{C}_{\mathrm{SN}}^{2}}\left(\overline{c}^{2}+\frac{1}{\sqrt{K}}\, x_{4}\right)+\frac{\sigma^{2}}{\overline{C}^{2}_{\mathrm{SN}}}\\
\setcounter{equation}{145}
\label{eq:fm}
\bar{F}_{m}&=\frac{\pi}{4} \frac{L^{2}}{\sqrt{Q}}\sum_{q: p(q)=m} C_{\mathrm{P1},q}^{m,m}\\
\label{eq:vfm}
\Var{\tilde{F}_{m}}&=\frac{L^{2}}{Q}\sum_{q:p(q)\ne m} \left(C_{\mathrm{P1},q}^{m,m}\right)^{2}+
\frac{\left(L+(\pi/4) (L-1)L\right)^{2}-(\pi^{2}/16)L^{4}}{Q}\sum_{q:p(q)= m} \left(C_{\mathrm{P1},q}^{m,m}\right)^{2}\\
\label{eq:wm}
\Var{W_{m}}&=\frac{L^{2}+(\pi/4) L^{2}(L-1)}{Q M}\sum_{\hat{m}\ne m}\sum_{q:p(q)=m} \left(C_{\mathrm{P1},q}^{m,\hat{m}}\right)^{2}+\frac{L^{2}+(\pi/4)L^{2}(L-1)}{Q M}\sum_{\hat{m}\ne m}\sum_{q:p(q)=\hat{m}}\left(C_{\mathrm{P1},q}^{m,\hat{m}}\right)^{2}
\nonumber\\&\quad
{+}\:\frac{L^{2}}{Q M}\sum_{\hat{m}\ne m}\sum_{q:p(q)\ne m\atop p(q)\ne\hat{m}}\!\!\! \left(C_{\mathrm{P1},q}^{m,\hat{m}}\right)^{2}+\frac{L^{2}+(\pi/4) L^{2}(L-1)}{Q}\sigma^{2}\!\!\!\sum_{q:p(q)=m}\!\!\left(C_{\mathrm{P1},q}^{m}\right)^{2}
+\frac{L^{2}}{Q}\sigma^{2}\!\!\sum_{q:p(q)\ne m}\!\!\! \left(C_{\mathrm{P1},q}^{m}\right)^{2}+\sigma^{2}
\ea
\setcounter{equation}{\value{MYtempeqncnt}}
\hrulefill
\vspace*{-5pt}
\end{figure*}
\subsection{Analysis of $S^{(4)}$}
\label{sec:S4P1}
The sum $S^{(4)}$ can be written as
\benn
S^{(4)}=\sum_{k=1}^{K} \left(C_{\mathrm{P1}, k}^{m}\right)^{2} Z_{k}^{(4)}
\eenn
with
\bann
Z_{k}^{(4)}=\abs{f_{m,k}}^{2}.
\eann
Since $Z_{k}^{(4)}$ is exponentially distributed with parameter $\lambda=1$, we have
\benn
\Prob\lefto\{Z_{k}^{(4)} \ge x\right\}\le e^{-x},\qquad \natseg{k}{1}{K},\ x\ge 0.
\eenn
Noting that the sum $S^{(4)}$ contains~$K$ jointly independent terms, taking into account~\eqref{eq:P1cbound} and using $\Ex{Z_{k}^{(4)}}=1$ and $\Ex{\bigl(Z_{k}^{(4)}\bigr)^{2}}=2$ $\left(\natseg{k}{1}{K}\right)$, we get for $x\ge 0$ and $K\ge 1$ 
\benn
\Prob\lefto\{\abs{S^{(4)}-\sum_{k=1}^{K}\left(C_{\mathrm{P1}, k}^{m}\right)^{2}}\ge \sqrt{K} x\right\}\le 3 K e^{-\Delta^{\!(4)} x^{2/3}}
\eenn
with $\Delta^{\!(4)}=\min\lefto[1,1/\big(4\, \overline{c}^{\, 4}\big) \right]$.
Therefore, using~\eqref{eq:P1cbound}, it follows that
\be
\Prob\lefto\{S^{(4)}\ge K\,\overline{c}^2+  \sqrt{K}x\right\}\le 3 K e^{-\Delta^{\!(4)} x^{2/3}}
\label{eq:S4P1u}
\ee
and
\be
\Prob\lefto\{S^{(4)}\le K\,\underline{c}^2-  \sqrt{K} x\right\}\le 3 K e^{-\Delta^{\!(4)} x^{2/3}}.
\label{eq:S4P1l}
\ee

We are now ready to carry out the final Step~\ref{step:sum5} of the program outlined in the first paragraph of Section~\ref{sec:convergence}.  
The concentration result for $\mathrm{SINR}^{\mathrm{P1}}_{m}$ is expressed in terms of upper bounds on $\Prob\mathopen{}\bigl\{\mathrm{SINR}^{\mathrm{P1}}_{m}\ge \hat{U}_{\mathrm{P1}}\bigr\}$ and $\Prob\mathopen{}\bigl\{\mathrm{SINR}^{\mathrm{P1}}_{m}\le \hat{L}_{\mathrm{P1}}\bigr\}$, where the exact form of $\hat{U}_{\mathrm{P1}}$ and $\hat{L}_{\mathrm{P1}}$ is specified below.

To establish an upper bound on $\Prob\mathopen{}\bigl\{\mathrm{SINR}^{\mathrm{P1}}_{m}\ge \hat{U}_{\mathrm{P1}}\bigr\}$, we proceed as follows:
\begin{enumerate}
\item
\label{step:1th1}
Apply Part 2 of Lemma~\ref{lemma:UBsumsMixed} to~\eqref{eq:S1P1u} and~\eqref{eq:S2P1lu} to establish a stochastic upper bound\footnote{For a RV~$X$, a ``stochastic upper bound'' in this context means a bound of the form $\Prob\lefto\{X\ge A\right\}\le P$.} for $\abs{S^{(1)}+S^{(2)}}$. 
\item
\label{step:2th1}
Apply Part 1 of Lemma~\ref{lemma:UBsumsMixed} to~\eqref{eq:S31Boundl}, \eqref{eq:S321Bound}, and~\eqref{eq:S322Bound} to establish a stochastic lower bound\footnote{For a RV~$X$, a ``stochastic lower bound'' in this context means a bound of the form $\Prob\mathopen{}\bigl\{X\le A\bigr\}\le P$.} for $\abs{S^{(3)}}$.
\item
\label{step:3th1}
Apply Part 1 of Lemma~\ref{lemma:UBsumsMixed} to the result from Step~\ref{step:2th1}) and~\eqref{eq:S4P1l} to establish a stochastic lower bound for $S^{(3)}+\sigma^{2} M S^{(4)}+K M \sigma^{2}$.
\item
Apply the union bound for fractions (Lemma~\ref{lemma:UBfrac}) to the stochastic upper bound from Step~\ref{step:1th1} and to the stochastic lower bound from Step~\ref{step:3th1} to establish the final result:
\be
\label{eq:PUpperP1}
\Prob\lefto\{\mathrm{SINR}^{\mathrm{P1}}_{m}\ge \hat{U}_{\mathrm{P1}}\right\}\le P_{\mathrm{P1}}^{\mathrm{U}}
\ee
with
\be
\label{eq:LUP1}
\hat{U}_{\mathrm{P1}}\define\frac{\pi^{2}}{16}\frac{\overline{C}^{2}}{\underline{C}_{\mathrm{SN}}^{2}}\frac{K}{M^{3}}\frac{\hat{U}_{\mathrm{P1}}^{\mathrm{N}}}{\hat{U}_{\mathrm{P1}}^{\mathrm{D}}}
\ee
and $P_{\mathrm{P1}}^{\mathrm{U}}$, $\hat{U}_{\mathrm{P1}}^{\mathrm{N}}$ and $\hat{U}_{\mathrm{P1}}^{\mathrm{D}}$ defined at the top of the page in \eqref{eq:POutP1App}, \eqref{eq:UP1N} and \eqref{eq:UP1D}, respectively.
\addtocounter{equation}{3}

\end{enumerate}
An upper bound on $\Prob\mathopen{}\bigl\{\mathrm{SINR}^{\mathrm{P1}}_{m}\le \hat{L}_{\mathrm{P1}}\bigr\}$ can be obtained as follows:
\begin{enumerate}
\item
\label{step:12th1}
Apply Part 1 of Lemma~\ref{lemma:UBsumsMixed} to~\eqref{eq:S1P1l} and~\eqref{eq:S2P1lu} to establish a stochastic lower bound for $\abs{S^{(1)}+S^{(2)}}$.
\item
\label{step:22th1}
Apply Part 2 of Lemma~\ref{lemma:UBsumsMixed} to~\eqref{eq:S31Boundu}, \eqref{eq:S321Bound}, and~\eqref{eq:S322Bound} to establish a stochastic upper bound for $\abs{S^{(3)}}$.
\item
\label{step:32th1}
Apply Part 2 of Lemma~\ref{lemma:UBsumsMixed} to the result from Step~\ref{step:22th1} and to~\eqref{eq:S4P1u} to establish a stochastic upper bound for $S^{(3)}+\sigma^{2} M S^{(4)}+K M \sigma^{2}$.
\item
Apply the union bound for fractions to the stochastic lower bound from Step~\ref{step:12th1} and to the stochastic upper bound from Step~\ref{step:32th1} to establish the final result:
\be
\label{eq:PLowerP1}
\Prob\lefto\{\mathrm{SINR}^{\mathrm{P1}}_{m}\le \hat{L}_{\mathrm{P1}}\right\}\le P_{\mathrm{P1}}^{\mathrm{U}}
\ee
with  
\be
\hat{L}_{\mathrm{P1}}\define \frac{\pi^{2}}{16}\frac{\underline{C}^{2}}{\overline{C}_{\mathrm{SN}}^{2}}\frac{K}{M^{3}}\frac{\hat{L}_{\mathrm{P1}}^{\mathrm{N}}}{\hat{L}_{\mathrm{P1}}^{\mathrm{D}}}
\ee
and $\hat{L}_{\mathrm{P1}}^{\mathrm{N}}$ and $\hat{L}_{\mathrm{P1}}^{\mathrm{D}}$ defined at the top of the page in~\eqref{eq:LP1N} and~\eqref{eq:LP1D}, respectively.
\addtocounter{equation}{2}

\end{enumerate}

The result presented in Theorem~\ref{thm:P1Conv} is a simpler and slightly weaker form of the bounds~\eqref{eq:PUpperP1} and~\eqref{eq:PLowerP1}. To obtain this simplification we proceed as follows. Set $x_{1}=x_{2}=x_{31}=x_{4}=x_{321}=x_{322}=x$ in~\eqref{eq:POutP1App}, \eqref{eq:UP1N}, \eqref{eq:UP1D}, \eqref{eq:LP1N} and \eqref{eq:LP1D}. Note that in this case $L_{\mathrm{P1}}(x)\le \hat{L}_{\mathrm{P1}}(x)$ and $U_{\mathrm{P1}}(x)\ge \hat{U}_{\mathrm{P1}}(x)$ and therefore
\ba
\label{eq:ProbUApp}
&\Prob\mathopen{}\Bigl\{\mathrm{SINR}^{\mathrm{P1}}_{m}\ge U_{\mathrm{P1}}\Bigr\}\le \Prob\mathopen{}\Bigl\{\mathrm{SINR}^{\mathrm{P1}}_{m}\ge \hat{U}_{\mathrm{P1}}\Bigr\}\le P_{\mathrm{P1}}^{\mathrm{U}}\\
\label{eq:ProbLApp}
&\Prob\mathopen{}\Bigl\{\mathrm{SINR}^{\mathrm{P1}}_{m}\le L_{\mathrm{P1}}\Bigr\}\le \Prob\mathopen{}\Bigl\{\mathrm{SINR}^{\mathrm{P1}}_{m}\le \hat{L}_{\mathrm{P1}}\Bigr\}\le P_{\mathrm{P1}}^{\mathrm{U}}.
\ea
Finally, combine the bounds~\eqref{eq:ProbUApp} and~\eqref{eq:ProbLApp} according to 
\bmnn
\Prob\lefto\{\left(\mathrm{SINR}^{\mathrm{P1}}_{m}\ge U_{\mathrm{P1}}\right)\Union \left(\mathrm{SINR}^{\mathrm{P1}}_{m}\le L_{\mathrm{P1}} \right)\right\}\\
\le
\Prob\mathopen{}\Bigl\{\mathrm{SINR}^{\mathrm{P1}}_{m}\ge U_{\mathrm{P1}}\Bigr\}+ \Prob\mathopen{}\Bigl\{\mathrm{SINR}^{\mathrm{P1}}_{m}\le L_{\mathrm{P1}}\Bigr\}\le 2 P^\mathrm{U}_\mathrm{P1}
\emnn
and note that $2 P^\mathrm{U}_\mathrm{P1}$ is upper bounded by the RHS of~\eqref{eq:POutP1}.
\hspace*{\fill}~\IEEEQED

\section{Proof of Lower Bound in Theorem~\ref{thm:P1ergL}}
\label{appendix:ErgP1LThProof}
As already mentioned in the main body of the paper, the proof of the lower bound in~\eqref{eq:P1boundsL} is based on the technique summarized in Appendix~\ref{appendix:Medard}. After straightforward algebra, it follows that the I-O relation of the SISO channel between the terminals $\mathcal{S}_m$ and $\mathcal{D}_m$ $\left(\natseg{m}{1}{M}\right)$ is given by
\benn
y_{m}=\left(\bar{F}_{m}+\tilde{F}_{m}\right) s_{m}+W_{m}
\eenn
where
{\allowdisplaybreaks
\bann
\bar{F}_{m}&\define \frac{1}{\sqrt{Q}}\sum_{q=1}^{Q} \Ex{a_q^{m,m}}\\
\tilde{F}_{m}&\define \frac{1}{\sqrt{Q}}\sum_{q=1}^{Q} \left(a_q^{m,m}-\Ex{a_q^{m,m}}\right)\\
W_{m}&\define \sum_{\hat{m}\ne m} s_{\hat{m}}\frac{1}{\sqrt{Q}}\sum_{q=1}^Q a_q^{m,\hat{m}}+
\frac{1}{\sqrt{Q}}\sum_{q=1}^{Q} b_q^{m} \herm{\tilde{\vech}}_{q,p(q)} \vecz_q+w_m
\eann}%
and
{\allowdisplaybreaks
\bann
a_q^{m,\hat{m}}&\define C_{\mathrm{P1}, q}^{m,\hat{m}}\left(\herm{\tilde{\vecf}}_{p(q),q} \vecf_{m,q}\right)\left(\herm{\tilde{\vech}}_{q,p(q)} \vech_{q,\hat{m}}\right)\\
\label{eq:P1bDefL}
b_q^{m}&\define C_{\mathrm{P1},q}^{m}\left(\herm{\tilde{\vecf}}_{p(q),q} \vecf_{m,q}\right)\\
C_{\mathrm{P1},q}^{m,\hat{m}}&\define \sqrt{Q}\, d_{\mathrm{P1}, q}  \hat{P}_{m,q} \hat{E}_{q,\hat{m}}\\
C_{\mathrm{P1},q}^{m}&\define \sqrt{Q}\, d_{\mathrm{P1}, q} \hat{P}_{m,q}.%
\eann}%
It is not difficult, but tedious, to verify that $\bar{F}_{m}$, $\Var{\tilde{F}_{m}}$ and $\Var{W_{m}}$ are given by~\eqref{eq:fm}, \eqref{eq:vfm} and~\eqref{eq:wm}, respectively.
\addtocounter{equation}{3}
Using~\eqref{EPbounds}, we lower-bound $\bar{F}_{m}$ and upper-bound $\Var{\tilde{F}_{m}}$ and $\Var{W_{m}}$, substitute the resulting bounds 
into~\eqref{eq:Medard}, and obtain 
\be
\label{eq:Th5MIbound}
I(y_{m};s_{m})\ge \frac{1}{2}\log\lefto(1+\frac{\pi^{2}}{16}\frac{Q}{M^{3}}\underline{f}(M,L)\right)
\ee
where
\bann
&\underline{f}(M,L)\\&\quad=\frac{\underline{P}\ \underline{E}\, P_{\mathrm{rel}}L^{2}}{\left(\overline{E}+\frac{\pi (L-1)}{4M}\overline{E}+\sigma^{2}\right)\left(\epsilon(M,L)+\overline{C}^{2}+\sigma^{2}\overline{c}^{2}+\sigma^{2}\right)}
\eann
with
\bmnn
\epsilon(M,L)=\frac{\,\overline{C}^{2}}{M}+ \frac{\left(1+(\pi/4) (L-1)\right)^{2}}{M^{2}} \overline{C}^{2}
\\
+\frac{\left(1+(\pi/4)(L-1)\right)\left(2\overline{C}^{2}+\sigma^{2}\overline{c}^{2}\right)}{M}.
\emnn
Finally, since~$L$ is finite, it follows by inspection that $\lim_{M\to\infty}\epsilon(M,L)=0$ and, therefore,
\benn
\lim_{M\to\infty} \underline{f}(M,L)=\frac{L^{2} \underline{C}^{2}}{\overline{C}_{\mathrm{SN}}^{2}}
\eenn
which, together with~\eqref{eq:Th5MIbound}, concludes the proof. 
\hspace*{\fill}~\IEEEQED

\section{Lower Bound on Channel Capacity with Imperfect Channel Knowledge}
\label{appendix:Medard}
The following Lemma is obtained by 
recognizing that the expression in~\cite[Eq. (66)]{lapidoth02-05} is trivially a lower bound to $I(X; Y)$ in~\eqref{eq:Medard} below. For completeness, we present the result in the form needed in this paper. For the proof of the (general) statement the interested reader is referred to~\cite{lapidoth02-05}.
\begin{lemma}
\label{lemma:Medard}
Consider a SISO channel with I-O relation
\benn
Y=F X + W
\eenn
where $X\sim\circnorm\lefto(0, \sigma_{X}^{2}\right)$, $W$ is zero-mean noise\footnote{In contrast to~\cite[Section~III]{medard00-05}, the noise is not necessarily Gaussian.} with variance $\sigma_{W}^{2}$, $F$ is the random channel gain with variance $\sigma_{F}^{2}$, and $Y$ is the output of the channel. Assume that~$F$ can be decomposed as
\benn
F=\bar{F}+\tilde{F}
\eenn
where~$\bar{F}=\Ex{F}$ is known at the receiver and $\tilde{F}$ with $\Ex{\tilde{F}}=0$ is not known at the receiver. Assume that~$X$ is statistically independent\footnote{In~\cite[Section~III]{medard00-05}, it is assumed that $X$, $F$, and~$W$ are statistically independent. The condition required here is weaker: $F$ and~$W$ need not be statistically independent.} of both~$F$ and~$W$. Then, the mutual information~$I(X;Y)$ can be lower-bounded as follows
\be
\label{eq:Medard}
I(X;Y)\ge \log\lefto(1+\frac{\bar{F}^{2}\sigma_{X}^{2}}{\sigma_{F}^{2}\sigma_{X}^{2}+\sigma_{W}^{2}}\right).
\ee 
\end{lemma}

\section{Some Essentials from Large Random-Matrix Theory}
\label{appendix:RM}
In this section, we briefly summarize the basic definitions and results from large random-matrix theory used in this paper. An excellent tutorial on this subject is~\cite{tulino04}. 

\begin{definition}[Stieltjes transform]
\label{def:stieltjes_tr}
Let $F(x)$ be a distribution function with density $f(x)$. The analytic function
\benn
G_{F}(z)\define\int \frac{f(x)}{x-z}\,dx,\quad z\in\complexset^{+}
\eenn
is called the Stieltjes transform of $F(x)$.
\end{definition}

\begin{lemma}[Inversion formula]
Let $G_{F}(z)$ be the Stieltjes transform of a distribution function $F(x)$. The corresponding density function can be obtained as
\be
\label{eq:inversion}
f(x)=\frac{1}{\pi}\lim_{y\to 0+}\Im\left[G_{F}(x+j y)\right].
\ee
\end{lemma}

\begin{theorem}[Silverstein~\cite{silverstein95-11}]
\label{theorem:Silv}
Define the following quantities on a common probability space:
\begin{itemize}
\item
The random matrix $\matX\in\complexset^{N\times N'}$ has i.i.d.\ zero-mean entries with variance one.
\item
The random matrix $\matY\in\complexset^{N\times N}$ is Hermitian nonnegative definite with $F_{\matY}^{N}(x)$, for $N\to\infty$,  converging on $[0,\infty)$ \as to a nonrandom distribution function $F_{\matY}(x)$ with corresponding density $f_{\matY}(x)$.
\end{itemize}
Assume that the matrices $\matX$ and $\matY$ are statistically independent.
Then, for $N,N'\to\infty$ with $N/N'\to\beta$, 
\benn
F_{(1/N')\matX\herm\matX \matY}^{N}(x)\xrightarrow{\mathrm{a.s.}} F_{(1/N')\matX\herm\matX \matY}(x)
\eenn 
with its Stieltjes transform $G_{F_{(1/N')\matX\herm\matX \matY}}(z)$ satisfying 
\bmnn
G_{F_{(1/N')\matX\herm\matX \matY}}(z)\\=\int_{-\infty}^{\infty}\frac{f_{\matY}(x) dx}{x (1-\beta-\beta z\,G_{F_{(1/N')\matX\herm\matX \matY}}(z))-z},\ z\in \complexset^{+}.
\emnn
The solution of this fixed-point equation is unique in the set 
\bmnn
\left\{G_{F_{(1/N')\matX\herm\matX \matY}}(z)\in\complexset \setgiven\right. \\
\left. -\frac{1-\beta}{z}+\beta\, G_{F_{(1/N')\matX\herm\matX \matY}}(z)\in\complexset^{+}\right\}.
\emnn
\end{theorem}

We shall furthermore use the Mar\v cenko-Pastur law as stated in~\cite{bai99-}.
\begin{theorem}[Mar\v cenko-Pastur~\cite{marcenko67}]
\label{theorem:MP}
Assume that the matrix $\matX\in \complexset^{N\times N'}$ has i.i.d.\ zero-mean entries with variance~$d^2$. Then, for $N,N'\to\infty$ with $N'/N\to\beta$, the ESD of $(1/N') \matX\herm\matX$ converges \as to a limiting distribution function with density
\bmnn
f_{(1/N')\matX\herm\matX}(x)=
\frac{\beta}{2\pi x d^2}\sqrt{\left(\gamma_{2}-x\right)^{+}\left(x-\gamma_{1}\right)^{+}}\\+
[1-\beta]^{+} \delta(x)
\emnn
where $\gamma_{1}=d^2(1-1/\sqrt{\beta})^{2}$ and $\gamma_{2}=d^2(1+1/\sqrt{\beta})^{2}$.

Under the same assumptions as in the first statement, if, in addition, the entries of $\matX$ have finite fourth moments, then \as
\bann
&\lim_{N'\to\infty} \lambda_{\mathrm{min}}\lefto(\frac{1}{N'}\matX\herm\matX\right)=\gamma_{1}\\
&\lim_{N'\to\infty} \lambda_{\mathrm{max}}\lefto(\frac{1}{N'}\matX\herm\matX\right)=\gamma_{2}.
\eann
\end{theorem}

\section{Computation of the Integral~$\hat{I}$ in~\eqref{eq:TheIntegral}}
\label{appendix:integral}
In the following, we detail the computation of the integral
\benn
\hat{I}\define\rho\int_{\eta_{1}}^{\eta_{2}}\frac{\sqrt{\left(\eta_{2}-x\right)\left(x-\eta_{1}\right)}\, dx}{ x (1-x)^{2}\left(x \left(\frac{1-\beta}{z}-\beta G\right)-1\right)}
\eenn
on the RHS of~\eqref{eq:TheIntegral}.
With the change of variables
\benn
t=\sqrt{\frac{x-\eta_{1}}{\eta_{2}-x}}
\eenn
and the notation
{\allowdisplaybreaks
\bsann
&\mu_{1}\define 1-\eta_{1}\qquad\qquad &\nu_{1}\define \eta_{1}\left(\frac{1-\beta}{z}-\beta G\right)-1\\
&\mu_{2}\define 1-\eta_{2}\qquad\qquad &\nu_{2}\define \eta_{2}\left(\frac{1-\beta}{z}-\beta G\right)-1
\esann}%
the integral~$\hat{I}$ can be written as
\bann
\hat{I}=2(\eta_{2}-\eta_{1})^{2}\rho\int_{0}^{\infty}\frac{t^{2} (t^{2}+1)dt}{(\eta_{2}t^{2}+\eta_{1})(\mu_{2}t^{2}+\mu_{1})^{2}(\nu_{2}t^{2}+\nu_{1})}.
\eann
To simplify further, we introduce the notation
\bann
\kappa_{1}\define -\frac{\eta_{1}}{\eta_{2}},\ \ \kappa_{2}\define -\frac{\mu_{1}}{\mu_{2}},\ \ \kappa_{3}\define -\frac{\nu_{1}}{\nu_{2}},\ \  \chi\define \frac{2(\eta_{2}-\eta_{1})^{2}}{\eta_{2}\,\mu_{2}^{2}\, \nu_{2}}\rho
\eann
so that
\be
\hat{I}=\chi\int_{0}^{\infty}\frac{t^{2} (t^{2}+1)dt}{(t^{2}-\kappa_{1})(t^{2}-\kappa_{2})^{2}(t^{2}-\kappa_{3})}.
\label{eq:Int2}
\ee
Upon partial fraction expansion of the integrand in~\eqref{eq:Int2}, we obtain
\benn
\hat{I}=\chi(A_{1}\hat{I}_{1}+A_{2}\hat{I}_{2}+A_{3}\hat{I}_{3}+A_{4}\hat{I}_{4})
\eenn
where
\bsa
&\hat{I}_{1}\define \int_{0}^{\infty}\frac{dt}{t^{2}-\kappa_{1}}\quad\quad\quad &\hat{I}_{2}\define \int_{0}^{\infty}\frac{dt}{(t^{2}-\kappa_{2})^{2}}\nonumber\\
&\hat{I}_{3}\define \int_{0}^{\infty}\frac{dt}{t^{2}-\kappa_{2}}\quad\quad\quad &\hat{I}_{4}\define \int_{0}^{\infty}\frac{dt}{t^{2}-\kappa_{3}}
\label{eq:4Ints}
\esa
with
\ba
\label{eq:A1}
&A_{1}= \frac{\kappa_{1}(\kappa_{1}+1)}{(\kappa_{1}-\kappa_{2})^{2}(\kappa_{1}-\kappa_{3})}\\
\label{eq:A2}
&A_{2}= \frac{\kappa_{2}(\kappa_{2}+1)}{(\kappa_{2}-\kappa_{1})(\kappa_{2}-\kappa_{3})}\\
\label{eq:A3}
&A_{3}= \frac{-\kappa_{2}^{2}-\kappa_{1}\kappa_{2}^{2}+\kappa_{1}\kappa_{3}+2\kappa_{1}\kappa_{2}\kappa_{3}-\kappa_{2}^{2}\kappa_{3}}{(\kappa_{2}-\kappa_{1})^{2}(\kappa_{2}-\kappa_{3})^{2}}\\
\label{eq:A4}
&A_{4}= \frac{\kappa_{3}(\kappa_{3}+1)}{(\kappa_{3}-\kappa_{1})(\kappa_{3}-\kappa_{2})^{2}}.
\ea
The integrals in~\eqref{eq:4Ints} can be evaluated resulting in 
{\allowdisplaybreaks
\ba
\label{eq:I11}
\hat{I}_{1}&=\left.\frac{1}{\sqrt{-\kappa_{1}}}\arctan\frac{t}{\sqrt{-\kappa_{1}}}\right|_{0}^{\infty}=\frac{\pi}{2 \sqrt{-\kappa_{1}}}\\
\label{eq:I12}
\hat{I}_{2}&=\left.-\frac{t}{2\kappa_{2}(t^{2}-\kappa_{2})}\right|_{0}^{\infty}-\left.\frac{1}{2\kappa_{2}\sqrt{-\kappa_{2}}}\arctan{\frac{t}{\sqrt{-\kappa_{2}}}}\right|_{0}^{\infty}\nonumber\\
&=-\frac{\pi}{4 \kappa_{2}\sqrt{-\kappa_{2}}}\\
\label{eq:I13}
\hat{I}_{3}&=\left.\frac{1}{\sqrt{-\kappa_{2}}}\arctan \frac{t}{\sqrt{-\kappa_{2}}}\right|_{0}^{\infty}=\frac{\pi}{2 \sqrt{-\kappa_{2}}}\\
\label{eq:I14}
\hat{I}_{4}&=\left.\frac{1}{\sqrt{-\kappa_{3}}}\arctan \frac{t}{\sqrt{-\kappa_{3}}}\right|_{0}^{\infty}=\frac{\pi}{2 \sqrt{-\kappa_{3}}}.
\ea}%
The quantity~$\kappa_{3}$ is complex-valued, and the~$\arctan$ and square root in~\eqref{eq:I13} are understood as the principal values of these functions in~$\complexset$ as defined in~\cite{abramowitz64}.

Finally, by inspection, combining~\eqref{eq:I11}--\eqref{eq:I14} with~\eqref{eq:A1}--\eqref{eq:A4} and resubstituting the values of the parameters $\kappa_{1}$, $\kappa_{2}$, $\kappa_{3}$, $\chi$, $\rho$, $\mu_{1}$, $\mu_{2}$, $\eta_{1}$, $\eta_{2}$, $\nu_{1}$, $\nu_{2}$, $\gamma_{1}$, and $\gamma_{2}$, after straightforward but tedious simplifications, we find 
\bann
\chi A_{1} \hat{I}_{1}&=\frac{\left(\sqrt{\beta}+1\right) \left|\sqrt{\beta }-1\right|}{2 \beta }\\
\chi A_{2} \hat{I}_{2}&=-\frac{z}{\sqrt{\beta} (G \beta  z+z+\beta -1)}\\
\chi A_{3} \hat{I}_{3}&=
-\frac{z d^2 \left(\sqrt{\beta }-1\right)^2 (G \beta z +z+\beta -1)}{2 d^2 \beta  (G \beta  z+z+\beta -1)^2}\\
&\relphantom{=}\qquad\qquad\qquad {+}\:\frac{z \beta  (G \beta z+\beta -1)}{2 d^2 \beta  (G \beta  z+z+\beta -1)^2}\\
\chi A_{4} \hat{I}_{4}&=-\frac{(G \beta z +\beta -1) }{2 d^2 \beta  (G \beta  z+z+\beta -1)^2}\nonumber\\
&\phantom{=
-}{\times}\: \sqrt{\frac{d^2 (G \beta  z+z+\beta -1) \left(\sqrt{\beta}-1\right)^2+z \beta }{d^2 (G \beta  z+z+\beta -1) \left(\sqrt{\beta}+1\right)^2+z \beta }}
\nonumber\\&\phantom{=
-}{\times}\: 
\left(d^2 (G \beta  z+z+\beta -1) \left(\sqrt{\beta}+1\right)^2+z \beta \right).
\eann

\bibliographystyle{IEEEtran}

\newpage
\begin{IEEEbiography}
Veniamin I. Morgenshtern was born in Leningrad, Russia on June 23, 1982. From 1999 to 2004 he studied Mathematics at St.~Petersburg State University, St.~Petersburg, Russia, where he received his M.S. degree.

Since 2004 he has been a research assistant at ETH Zurich, Switzerland, working towards the Dr.~sc. degree. His current research interests are in communication and information theory.
\end{IEEEbiography}

\begin{IEEEbiography}
Helmut B\"{o}lcskei was born in Austria on May 29, 1970, and received the Dipl.-Ing. and Dr. techn. degrees in electrical engineering/communication theory from Vienna University of Technology, Vienna, Austria, in 1994 and 1997, respectively. From 1994 to 1998 he was with Vienna University of Technology. From 1999 to 2001 he was a postdoctoral researcher in the Information Systems Laboratory, Department of Electrical Engineering, Stanford University, Stanford, CA. He was in the founding team of Iospan Wireless Inc., a Silicon Valley-based startup company (acquired by Intel Corporation in 2002) specialized in multiple-input multiple-output (MIMO) wireless systems for high-speed Internet access. From 2001 to 2002 he was an Assistant Professor of Electrical Engineering at the University of Illinois at Urbana-Champaign. He has been with ETH Zurich since 2002, where he is Professor of Communication Theory. He was a visiting researcher at Philips Research Laboratories Eindhoven, The Netherlands, ENST Paris, France, and the Heinrich Hertz Institute Berlin, Germany. His research interests include communication and information theory with special emphasis on wireless communications, signal processing and quantum information processing.

He received the 2001 IEEE Signal Processing Society Young Author Best Paper Award, the 2006 IEEE Communications Society Leonard G. Abraham Best Paper Award, the ETH ÒGolden OwlÓ Teaching Award, and was an Erwin Schr\"{o}dinger Fellow (1999-2001) of the Austrian National Science Foundation (FWF). He was a plenary speaker at several IEEE conferences and served as an associate editor of the IEEE Transactions on Signal Processing, the IEEE Transactions on Wireless Communications and the EURASIP Journal on Applied Signal Processing. He is currently on the editorial board of "Foundations and Trends in Networking", serves as an associate editor for the IEEE Transactions on Information Theory and is TPC co-chair of the 2008 IEEE International Symposium on Information Theory.
\end{IEEEbiography}

\end{document}